\documentclass{aa}  

\usepackage{graphicx}
\usepackage{txfonts}
\usepackage{color}

\begin{document}

   \title{Galaxy populations in the Hydra I cluster from the VEGAS survey}
   \subtitle{II. The ultra-diffuse galaxy population}

   \author{Antonio La Marca\inst{1,2,3} \fnmsep\thanks{\email{antoniolamarca46@gmail.com}}
          \and Enrichetta Iodice\inst{1}
          \and Michele Cantiello\inst{4}
          \and Duncan A. Forbes \inst{5}
          \and Marina Rejkuba \inst{6}
          \and Michael Hilker  \inst{6}
          \and Magda Arnaboldi \inst{6}
          \and Laura Greggio \inst{7}
          \and Chiara Spiniello \inst{1,9}
          \and Steffen Mieske \inst{10}
          \and Aku Venhola \inst{11}
          \and Marilena Spavone \inst{1}
          \and Giuseppe D'Ago \inst{12}
          \and Maria Angela Raj \inst{13}
          \and Rossella Ragusa \inst{1,3}
          \and Marco Mirabile \inst{3}
          \and Roberto Rampazzo \inst{8}
          \and Reynier Peletier\inst{2}
          \and Maurizio Paolillo \inst{1,3}
          \and Nelvy Choque Challapa \inst{2}
          \and Pietro Schipani\inst{1}}

           \institute{INAF $-$ Astronomical Observatory of Capodimonte, Salita Moiariello 16, I-80131, Naples, Italy
            \and
            Kapteyn Institute, University of Groningen, Landleven 12, 9747 AD Groningen, the Netherlands
            \and
            University of Naples ``Federico II'', C.U. Monte Sant'Angelo, Via Cinthia, 80126, Naples, Italy
            \and
            INAF $-$ Astronomical Observatory of Abruzzo, Via Maggini, 64100, Teramo, Italy
            \and
            Centre for Astrophysics \& Supercomputing, Swinburne University of Technology, Hawthorn VIC 3122, Australia
            \and
            European Southern Observatory, Karl$-$Schwarzschild-Strasse 2, 85748 Garching bei München, Germany
            \and
            INAF $-$ Osservatorio Astronomico di Padova, Vicolo dell’Osservatorio 5, I-35122 Padova, Italy
            \and
            INAF $-$ Astronomical Observatory of Padova, Via dell’Osservatorio 8 , I-36012, Asiago (VI), Italy
            \and 
            Department of Physics, University of Oxford, Denys Wilkinson Building, Keble Road, Oxford OX1 3RH, UK
            \and 
            European Southern Observatory, Alonso de Cordova 3107, Vitacura, Santiago, Chile
            \and
            Space Physics and Astronomy Research Unit, University of Oulu, P.O. Box 3000, FI$-$90014, Oulu, Finland
            \and
            Instituto de Astrofísica, Facultad de Física, Pontificia Universidad Católica de Chile, Av. Vicuña Mackenna 4860, 7820436 Macul, Santiago, Chile
            \and
            INAF $-$ Astronomical Observatory of Rome, Via Frascati, 33, 00078, Monte Porzio Catone, Rome, Italy
             }

   \date{Received 04 October 2021; accepted 13 June 2022}

% \abstract{}{}{}{}{} 
% 5 {} token are mandatory
 
\abstract 
{In this work we extend the catalog of low-surface brightness (LSB) galaxies, 
including Ultra-Diffuse Galaxy (UDG) candidates, within $\approx 0.4R_{vir}$ of the \object{Hydra\,I} cluster of galaxies, based on deep images from the {\it VST Early-type GAlaxy Survey (VEGAS).} 
The new galaxies are found by applying an automatic detection tool and carrying out additional visual inspections of $g$ and $r$ band images.
This led to the detection of 11 UDGs and 8 more LSB galaxies. 
For all of them, the cluster membership has been assessed using the color-magnitude relation derived for early-type giant and dwarf galaxies in \object{Hydra\,I}.
The UDGs and new LSB galaxies found in Hydra I span a wide range of central surface brightness ($22.7 \lesssim \mu_{0,g} \lesssim 26.5$~mag/arcsec$^2$), effective radius ($0.6 \lesssim R_e \lesssim 4.0$~kpc) and color ($0.4 \leq g-r \leq 0.9$~mag), and have stellar masses in the range $\sim 5\times 10^6 - 2\times 10^8$~M$_{\odot}$. 
The 2D projected distribution of both galaxy types is similar to the spatial distribution of dwarf galaxies, with over-densities in the cluster core and north of the cluster centre.  
They have similar color distribution and comparable stellar masses to the red dwarf galaxies. 
Based on photometric selection, we identify a total of 9 globular cluster candidates associated to the UDGs and 4 to the LSB galaxies, with the highest number of candidates in an individual UDG being three. 
We find that there are no relevant differences between dwarfs, LSB galaxies and UDGs: the structural parameters (that is surface brightness, size, colors, n-index) and GCs content of the three classes have similar properties and trends. 
This finding is consistent with UDGs being the extreme LSB tail of the size-luminosity distribution of dwarfs in this environment.
}


\maketitle

%-------------------------------------------------------------------

\section{Introduction}\label{sec:intro}

The study of the low-surface brightness (LSB) galaxies in the universe represents a crucial step to map the bottom-up assembly processes of galaxies in all environments, and thus constraining their formation within the Lambda-Cold Dark Matter (LCDM) paradigm \citep[][]{springel2005}. 
In this framework, the ultra-diffuse galaxies (UDGs) have a special role. 
They are considered the extreme tail of the size-luminosity distribution of LSB galaxies, being faint ($\mu_{0,g} \geq 24$ mag arcsec$^{-2}$) and diffuse ($R_e \geq 1.5$~kpc) objects, with stellar masses similar to dwarf galaxies, $M_* \sim 10^7 - 10^8$ M$_{\odot}$ \citep[][]{vanDokkum2015}. 
Given the extremely low baryonic mass density, UDGs are particularly suitable laboratories to test the dark-matter (DM) theories \citep[e.g.,][]{Silk2019,Sales2020}.

The detection and analysis of UDGs is however challenging due to their LSB nature. 
Deep imaging surveys brought attention to an ever increasing detection of UDGs in clusters 
\citep[][]{Yagi2016,vanderBurg2017,Zaritsky2019,Lee2020,Janssens2017,Janssens2019,Venhola2017,Mancera-pina2018udg,Mancera-pina2019udg} and groups \citep[][]{Merritt2016a,Trujillo2017,Forbes2019,Forbes2020b, Habas2020}, as well as in less-dense environments \citep[][]{Leisman2017,Prole2019a,Roman2019,Habas2020,Venhola2021}.
Based on their observed color distribution, it seems that two populations of UDGs are found: 
the red and quenched UDGs, located primarily in clusters of galaxies, but also recently in low density environments \citep{Marleau2021matlas}, and a blue population of UDGs, which are mostly found in low density regions \citep[e.g.,][]{Leisman2017,Roman2017a,Marleau2021matlas}.

Because of their LSB nature, there are only few spectroscopic observations of UDGs, which strongly limits firm conclusions about the nature of their stellar populations. 
The few available spectroscopic studies reveal the existence of both metal-poor ($-0.5 \leq [M/H] \leq -1.5$ dex) and old systems \citep[about 9 Gyr, e.g.][]{Ferre-Mateu2018,Fensch2019,Pandya2018,Muller2020}, and younger star-forming UDGs \citep{Martin-Navarro2019}. 
Based on the few kinematical measurements of UDGs in groups and clusters, only dispersion-dominated systems are known \citep[][and references therein]{Emsellem2019}. 

A particular interest is devoted to the globular cluster (GC) systems in UDGs. 
Observations suggest that there is clearly a large number of GCs in some UDGs. 
For massive galaxies, the GCs specific frequency $S_N$, i.e. the number of GCs ($N_{GC}$) per unit of $V$-band luminosity \citep[$S_N = N_{GC} 10^{0.4 [M_v+15]}$,][]{Harris1981Sn}, is used as a tracer of the host galaxy's halo virial mass \citep[e.g.,][]{Georgiev2010,Hudson2014,Burkert2020}.
The validity of the GC vs. halo mass relation in the LSB regime is still to be confirmed {  \citep[][]{Burkert2020}}. 

Nevertheless, assuming that it holds also for UDGs, in the Coma cluster of galaxies, $S_N$ appears to be larger in UDGs than in dwarf galaxies of similar luminosity \citep[][]{Lim2020,Forbes2020a}, which could indicate the presence of DM halos more massive for this class of objects than for similar luminosity dwarfs ($M_h \geq 10^{11}$~M$_{\odot}$). 
On the other hand, UDGs with $S_N$ similar to dwarf galaxies are found in other clusters of galaxies {  \citep[][]{Prole2019b,Saifollahi2021,Saifollahi2022,Marleau2021matlas}}.
In this respect, the DM content of the UDGs is one of the open and most debated issues. 
Using also spectroscopy, a larger DM amount was inferred for a number of UDGs \citep[e.g.,][]{Toloba2018,vanDokkum2019,Forbes2020a,Gannon2021,Forbes2021} while, at odds, some others with very low DM content 
have also been discovered \citep[e.g.,][]{vanDokkum2018,Collins2021}.

The wide range of observed properties for UDGs, listed above, does not fit in a single formation scenario \citep[see][for a recent summary of various formation processes]{Jones2021}.
To date, there is a general consensus that different formation channels can result in galaxies with UDG-like properties.
To account for the high DM content and large effective radii of the first discovered UDGs, comparable to normal Milky Way (MW) like galaxies, \citet[][]{vanDokkum2015} termed these objects as “failed” galaxies, assuming that this new class of galaxies might have lost gas supply at an early epoch, which prevented the formation of normal, higher surface brightness systems.
UDGs formation was also connected to different internal processes such as 
{\it i)} repeated episodes of star formation feedback, which induce kinematical heating of their stars \citep[i.e. ][]{diCintio2017}, or
{\it ii)} anomalously high spins of DM halos, which prevent an efficient collapse of the gas to form a dense structure \citep{Amorisco2016,Rong2017,Tremmel2019}. 
The latter two proposed mechanisms form UDGs with stellar masses and DM fraction 
comparable with those of dwarf galaxies.

Gravitational interactions and merging between galaxies have been also invoked as possible formation mechanisms for UDGs.
Merging of low-mass galaxies at an early-epoch could result in a present day UDG in the field \citep[][]{Wright2021}.
UDGs may also have a tidal origin.  
A high-velocity galaxy collisions could induce the formation of DM-free UDGs and, simultaneously, their star clusters \citep[][]{Lee2021}.
Diffuse, DM-free, tidal dwarf galaxies (TDGs) are known to form in the tidal tails of strongly interacting galaxies \citep[][]{Lelli2015,Duc2014,Ploeckinger2018}.
The tidal interaction of a dwarf satellite galaxy in the potential well of a major galaxy might puff it up and form a UDG-like system \citep[][]{Conselice2018,Carleton2019,Carleton2021}.
Recently, based on deep optical images of the Fornax cluster, \citet[][]{Venhola2021} found that the UDGs and LSB dwarfs spatial distributions are more concentrated in the cluster core than the normal dwarfs, and show more morphological signs of tidal interactions than other galaxies.
There are few cases of UDGs potentially associated with faint stellar streams, detected in the outskirts of galaxies in a group environment, suggesting that UDGs might also form from material released during a weak tidal interaction \citep[][]{Bennet2018,Muller2019,Montes2020}.
Finally, \citet[][]{Poggianti2019} suggested that the DM-free UDGs might form from ram pressure stripped (RPS) gas clumps in the extended tails of infalling cluster galaxies, {  depending on their subsequent dynamical evolution.} 
A UDG candidate that might have formed from the RPS gas clumps has been recently discovered in NGC~3314A, a \object{Hydra\,I} cluster member, by \citet{Iodice2021}.
Confirmation of the nature of this UDG through follow-up spectroscopy could be the missing observational evidence of this formation channel.

High resolution hydro-dynamical cosmological simulations have been recently developed to resolve the internal structure of low mass galaxies in clusters, and used to trace the formation and evolution of LSB galaxies.
Based on the IllustrisTNG simulations, \citet[][]{Sales2020} also proposed two different formation channels for UDGs. 
A population of “genuine” LSB galaxies, with UDG properties, could form in the field and later enter the cluster environment. 
The so-called tidal-UDGs (T-UDGs), stem from luminous galaxies and evolve into UDGs due to cluster tidal forces that remove their DM fraction and puff up their stellar component. 
The T-UDGs populate the centre of the clusters and, at a given stellar mass, have lower velocity dispersion, higher metallicity and lower DM fraction with respect to the genuine UDGs. 
Using the RomulusC simulations, \citet[][]{Tremmel2020} suggested that UDGs could form from dwarf galaxies, where the gas was removed and the star formation was halted by the ram pressure acting during their motions through the cluster.
Recently, \citet[][]{Benavides2021} proved that quenched and isolated UDGs are formed as backsplash galaxies, which were satellites of a group or a cluster halo in an early epoch, and were today found a few Mpc away from them. 
This kind of interaction would remove the gas and tidally strip the outskirts of the DM halo.

As concluding remark, both the observational and theoretical works on UDGs cited above show that more than one formation channel might exist for this class of objects, which lead to different structural properties (e.g. colors, stellar populations, DM fraction) also depending on the environment where they reside.
Therefore, to shed light on this aspect, it is needed to collect deep imaging and spectroscopy for complete samples of UDGs across different environments, in order to systematically explore their properties from the densest regions in the core of clusters, out to their less populated outskirts, and in the field.
Due to their LSB nature, which could usually prevent their detection, a complete sample of UDGs is meant to be the number density of UDGs expected for the given virial mass of the host environment, based on the relation published by \citet[][]{vanderBurg2017} and \citet[][]{Janssens2019}. 
With the present paper we provide a complete sample of UDGs in the \object{Hydra\,I} cluster and analyse their structural photometric properties. 
This work builds on the previous studies of \object{Hydra\,I} cluster by \citet[][hereafter Paper I]{LaMarca2021} 
which presented a new catalog of 317 dwarf galaxies fainter that $M_r\geq-18.5$~mag, and by \citet[][]{Iodice2020c}. 
The latter provided the first sample of 20 LSB galaxies, with $\mu_{0,g}\geq
23.5$ mag/arcsec$^2$ and $R_e\geq 0.6$ kpc, of which 12 have been classified as UDGs in this cluster. 
Nevertheless, in this work we consider as UDGs only those galaxies which rigorously respect the van Dokkum UDG definition.
Therefore, we count as UDGs in Hydra I only 9 out of the 20 LSB galaxies presented by \citet[][]{Iodice2020c}.
In addition, one more UDG in the Hydra I cluster has been discovered in the stellar filaments of \object{NGC3314}A \citep[hereafter called UDG32,][]{Iodice2021}, bringing the total number of known UDGs in this cluster to 10 up to date.

The \object{Hydra\,I} cluster has a virial mass of $M_{200}=2.1\times10^{14}h^{-1}M_{\odot}$, as estimated from X-ray observations of the intracluster medium \citep[][]{Tamura2000}. 
It is located in the South hemisphere at a distance of $51\pm6$~Mpc  \citep{Christlein2003}.
Recent studies show that the core of the cluster reveals ongoing interactions that trace the extended mass assembly around the brightest cluster member \object{NGC\,3311} \citep[see][and references therein]{Barbosa2018}. 
As found in Paper I, most of the dwarf galaxies in the cluster are concentrated in the core and around a sub-group of galaxies in the North. 
These are the two densest regions of the cluster, which show signs of galaxy interactions and presence of intra-cluster diffuse light (Iodice et al. in preparation).
Therefore, the \object{Hydra\,I} cluster offers an exquisite opportunity to analyse this class of LSB galaxies as function of the environment and relate them to the mass assembly processes.

The paper is organised as follow. 
The \object{Hydra\,I} imaging data used in this paper are introduced in Sect. \ref{sec:data}, while the detection of new UDG candidates is presented in Sect. \ref{sec:detection}. 
In section \ref{sec:GCs} we explain how we detected and selected possible GCs for our sample of UDGs, and in Sect. \ref{sec:result} we report the results of our analysis: the spatial distribution of UDGs, their photometric properties, and their GC candidates.
Finally, in Sect. \ref{sec:disc} we discuss the properties of the UDGs in the \object{Hydra\,I} cluster, while conclusions are provided in Sec.~\ref{sec:concl}.

%-------------------------------------------------------------------

\section{Data: deep images of the Hydra~I Cluster}\label{sec:data}

The imaging data for the Hydra~I cluster presented in this work were collected in the $g$ and $r$ bands, with the European Southern Observatory (ESO) VLT Survey Telescope (VST), as part of the {\it VST Early-type Galaxy Survey} (VEGAS\footnote{See \url{http://www.na.astro.it/vegas/VEGAS/Welcome.html}}).
{ 
The optical camera OmegaCAM \citep{kujiken2011} on the VST records wide field images spanning $1\times1$ deg$^2$ field of view, with a pixel scale of 0.21 arcsec/pixel.
}
The observing strategy and data reduction were presented by \citet{Iodice2020c} and in Paper I.
Briefly, images were acquired in dark time, with total integration times of 2.8 and 3.22 hours in the $g$ and $r$ bands, respectively.
The VST mosaic covers an area of $1^{\circ} \times 2^{\circ}$ ($0.9\times1.8$ Mpc, at the distance of the cluster) around the cluster core.
As described by \citet[][]{Iodice2020c} and in Paper I, the observing strategy and data reduction, coupled with the long integration times and the large covered area, allow us to study the \object{Hydra\,I} cluster down to the LSB regime out to $\approx 0.4 R_{vir}$. 
In particular, to reduce the scattered light from the bright (7th-magnitude) foreground star, which is on the NE side of the cluster core, during the data acquisition this star was always put in one of the two wide OmegaCam gaps. 
The residual light from this bright star has been modelled and subtracted from the reduced mosaic, in both bands.
Also the light distribution of the second brightest star in the field, located SE the core, is accounted, by modelling and subtracting it from the parent image \citep[see Fig.1 in][]{Iodice2020c}.
We estimate surface brightness depths of $\mu_g=28.6\pm0.2$~mag/arcsec$^2$, and $\mu_r=28.1\pm0.2$~mag/arcsec$^2$ for the final stacked images.
These depths are derived as the flux corresponding to $5\sigma$, with $\sigma$ averaged over {  1 arcsec$^2$} of empty area.

The automatic detection of LSB galaxies in \object{Hydra\,I} has been improved and described in Paper I, where we compiled a new catalog counting 317 dwarf galaxies fainter than $M_r\geq-18.5$ mag. 
\footnote{As detailed in Paper I, at $M_B\simeq -18$ mag a separation between dwarfs and giants ellipticals 
emerges in the $\langle \mu\rangle -L$ plane. The early-type dwarfs fainter than $M_B\simeq -18$ mag in general do not share the surface brightness-luminosity relation defined by giant ellipticals \citep[][]{binggeli1991}. 
Hence, considering the typical $B-r$ colors for early-type dwarfs $\geq0.5$ mag, we assumed $M_r=-18.5$ mag as limit for dwarf galaxies.}
A dwarf galaxies catalog on a smaller area was already presented by \citet[][]{Misgeld2008}, {containing galaxies with $M_V\geq-17$ mag.}
In this paper, we focus on the new LSB galaxies, including UDG candidates, detected in the cluster.

All magnitudes and colors are corrected for Galactic extinction using values from \citet[][]{Schlafly2011}.

\section{Discovery of new LSB galaxies and UDG candidates}\label{sec:detection}

As detailed in Paper I, we run the automatic tool {\sc SExtractor} \citep[][]{Bertin1996} on the VST mosaic to detect the LSB galaxies in the \object{Hydra\,I} cluster.
We have tuned {\sc SExtractor} parameters by setting the {\it THRESHOLD} to five times the background noise standard deviation, the minimum area\footnote{The minimum area parameter corresponds to the number of connected pixels above the threshold to have a detection.} to 25 pixels and the background model grid size to 256$\times$256 pixels. 
In this work, to improve the detection of LSB galaxies, we have adopted the following steps: 
{\it i)} the light distribution of the brightest cluster members has been modelled and subtracted from the parent image (Iodice et al. in preparation); 
{\it ii)} on the residual image we made another {\sc SExtractor} run, setting the $THRESHOLD$ parameter to one time the background noise standard deviation; 
{\it iii)} a complementary visual inspection has then been performed on the images, revealing new LSB galaxies, which were not detected by {\sc SExtractor};
{\it iv)} the cluster membership of detected galaxies is inferred using the color-magnitude relation (CMR) for early-type giant and dwarf galaxies in \object{Hydra\,I} \citep[][]{Misgeld2008}.
This CMR is converted from the Johnson $V$ and $I$ bands to the SDSS $g$ and $r$ photometric bands using the conversion factors given by \citet[][]{Kostov2018}.
The band-conversion equations are:
\begin{equation}
    V = r -0.017 + 0.492 \cdot (g-r),
\end{equation}
\begin{equation}
    V-I = 0.27 + 1.26\cdot(g-r).
\end{equation}

Only the LSB galaxies consistent with the assumed CMR, within the errors, are kept in our final sample.
Galaxies with colors redder than $2\sigma$ from the CMR derived by \citet[][]{Misgeld2008} are excluded. 
We consider highly unlikely the possibility that the LSB galaxies are members of other foreground/background structures and only by chance projected along the line of sight towards Hydra I.
Indeed, as also discussed in Paper I, there is evidence of a 40-50 Mpc under-dense region in front and behind the Hydra I cluster, along the line of sight \citep[][]{Richter1982,Richter1987}.

\begin{figure*}
    \centering
    \includegraphics[width=0.89\textwidth]{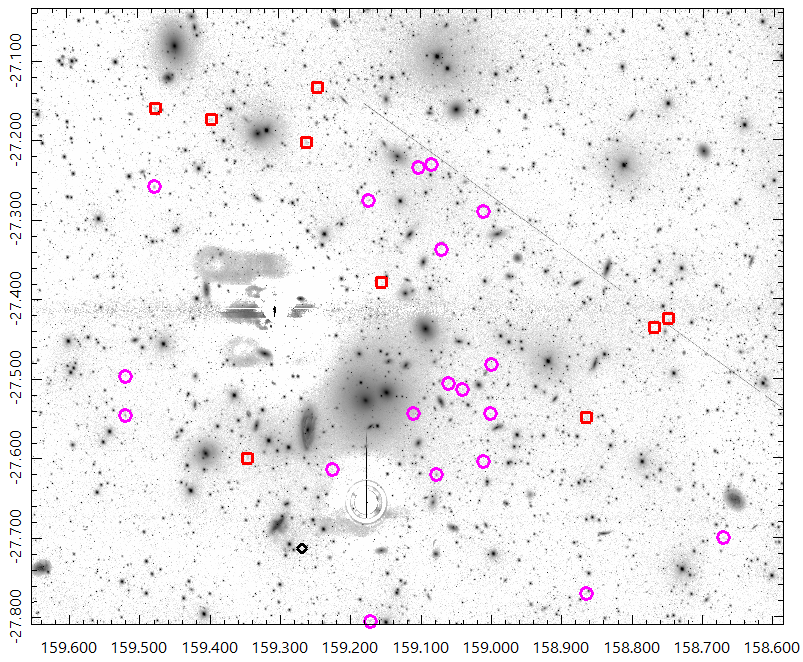}
    \caption{
    VST mosaic ($56.7\arcmin \times 46.55$~\arcmin $\sim0.8 \times 0.7$~Mpc) of the \object{Hydra\,I} cluster in the $g$ band, corresponding to $\approx 0.4 R_{vir}$.
    The 19 new galaxies detected in this work are marked as magenta circles. 
	Red boxes show the first sample of UDGs presented by \citet[][]{Iodice2020c} and the black diamond marks the position of UDG32 reported by \citet[][]{Iodice2021}.
	North is up and East to the left.}
    \label{fig:mosaic}
\end{figure*}

As result of our new search, 19 new LSB galaxies have been added to the catalog of LSB candidates recently presented in Paper I. % by \citet[][Paper I]{LaMarca2021}.
They are listed in Tab.~\ref{tab:UDGsample} and Tab.~\ref{tab:LSBsample}.
Their projected location in the cluster is shown in Fig.~\ref{fig:mosaic}.
For all the new candidates, we have performed the surface photometry to derive the structural parameters as described below. 

In the cutout extracted from the VST mosaic around each new LSB galaxies, we have identified all the background and foreground objects (stars and galaxies) brighter than the $2\sigma$ background level.
The most extended ones (i.e. $\geq 2$ times the LSB in the cutout) are modelled by adopting an isophote fit 
(where all parameters are left free to vary) and then subtracted from the image. 
All the remaining objects are carefully masked.
Once all bright objects are masked or modelled and subtracted, for each LSB galaxy of the sample we have performed the isophote fit, by using the {\sc astropy} package {\sc photutils} \citep[][]{larry_bradley_2020_4044744}, in the $g$ and $r$ bands.
The ellipticity of the ishophotes is also fitted rather than using a fixed value.
From the azimuthally averaged surface brightness profiles, a local background value is estimated for each object \citep[see][for details]{Iodice2021}.
In short, the local background is computed in an outer annulus where the light blends into the average scatter of the residual sky level. 

For all candidates in the new sample, the structural parameters (i.e. central surface brightness $\mu_0$, effective radius $R_e$) 
are derived by fitting the $g$-band azimuthally averaged profile, which sums up the light at the faintest levels, with  
a single S\'ersic function \citep[][]{Sersic}. 
The best fit is obtained by adopting the $\chi ^2$ minimisation method\footnote{The $\chi ^2$ is defined as $\sum_{\substack{ 0\le i \le N} }\sqrt{[f(x)-D]^{2}/N}$, where $f(x)$ is the Sérsic law used to fit the observed values ($D$), and $N$ is the number of the data-points. The goodness of the fit is provided by the reduced $\Tilde{\chi ^2}$, given by $\chi^{2}/(N-N_{param}) \sim 1$, where $N_{param}$ is the number of fitted parameters (i.e. $\mu_0$, $R_e$, $n$).},
where a weighted-fit is performed, taking into account the error on each data-point of the surface brightness profile.
Therefore, the data points corresponding to the central regions of the galaxies, with their small uncertainties have 
considerable weight in determining the best-fit solution. Since we are interested in classifying UDGs, this approach 
can reasonably exclude 
spurious large values of effective radii, which might bias the selection process.
The error estimate on the structural parameters, $R_e$ and $\mu_0$, include the uncertainties on the fitting, 
which are about 0.2\% and 6\%, respectively, and on the sky removal for $\mu_0$ {  \citep[$\sim 1 \%$, see][]{Iodice2020c}}, 
and on the distance ($\sim 12 \%$) for $R_e$. 
The structural parameters derived by the best fit and the reduced $\Tilde{\chi ^2}$, which indicates the goodness of the fit, 
are listed in Tab.~\ref{tab:UDGsample} and Tab.~\ref{tab:LSBsample}.

For all the newly detected LSB galaxies in the \object{Hydra\,I} cluster, we have derived the growth curve from the
isophote fit. These are shown in App.~\ref{sec:UDGimages} and App.~\ref{sec:LSBimages}. From the growth curve, we 
have computed the total integrated magnitudes in the $g$ and $r$ bands, 
and the effective radius $R_e^{50}$, as the distance from the galaxy centre containing half of the light. 
For all galaxies of the sample, the values of $R_e^{50}$ 
(see Fig.~\ref{fig:growUDG_1} and Fig.~\ref{fig:growUDG_2}, and Fig.~\ref{fig:growLSB_1} and Fig.~\ref{fig:growLSB_2}) 
are consistent with the values of $R_e$ estimated by the best fit of the surface brightness profiles.
This is particularly important for those galaxies where the fit appears quite poor ($\Tilde{\chi ^2} \geq 2$), 
as described in Sec.~\ref{sec:UDG}, 
since this provides a robust estimate of the $R_e$, which is independent from any fitting law.

The $g{-}r$ color, obtained by the integrated magnitudes, and the absolute magnitude in the $r$ band (M$_r$) are then used to derive the stellar mass for all LSB galaxies, by using the relation given by \citet{Into2013}, where $log(M/L_r)=1.373\times(g-r)-0.596$.

Nine of the new galaxies were found only by visual inspection.
These are the smallest (semi-major axis radius $R \leq 8$ arcsec $\sim2$~kpc) and/or most diffuse objects of the sample, with effective g-band surface brightness $\mu_e$ ranging from $\sim$ 26 to 28.5 mag/arcsec$^2$ (see Table~\ref{tab:UDGsample} and Table~\ref{tab:LSBsample}). 
Most of the visually detected galaxies do not show a steep increase of the surface brightness toward the galaxy centre, as observed for all those found by SExtractor, which have $25\lesssim \mu_e \lesssim 28.5$~mag/arcsec$^2$, and are also the most extended objects of the sample ($R_e$ up to 16 arcsec $\simeq$ 4 kpc).

\subsection{New UDGs in the Hydra~I cluster}\label{sec:UDG}

As stated in Sec.~\ref{sec:intro}, \citet[][]{vanDokkum2015} empirically defined the UDGs to be faint, with $\mu_{0,g}\geq 24$~mag/arcsec$^2$ and diffuse, with $R_e \geq 1.5$~kpc. 
Other criteria have been proposed in literature, which used
different cuts in size and/or surface brightness limits \citep[][]{Koda2015,Mihos2005,Yagi2016,vanderBurg2017}.
Recently, \citet[][]{Lim2020} adopted a new selection method to classify UDGs in the Virgo cluster, based on the scaling relations between photometric and structural properties of LSB galaxies (e.g. total luminosity versus $R_e$ and $\mu_e$). 
According to this selection method, UDGs are identified as "outliers" in the scaling relationships, being $2.5 \sigma$ fainter and larger than the average distribution of the parent LSB sample. 

In this work, in order to be consistent with the previous UDGs study in the Hydra I cluster \citep[][]{Iodice2020c,Iodice2021}, and to compare the UDGs properties with those published by several other papers (see Sec.\ref{sec:concl}), we decided to adopt the empirical definition proposed by \citet[][]{vanDokkum2015}, i.e. $R_e\geq 1.5$~kpc and $\mu_0\geq24$~mag/arcsec$^2$ in $g$ band, derived by the fit of the light distribution using a Sérsic law. 
Hence, from the whole sample of new 19 LSB galaxies, the selection process led to 11 UDGs in the \object{Hydra\,I} cluster (see Fig.~\ref{fig:mosaic}). 
They are listed in Tab.~\ref{tab:UDGsample}, which includes the structural parameters and the detection method (i.e. SExtractor or visual inspection).

\begin{figure*}
    \centering
    \includegraphics[width=0.49\textwidth]{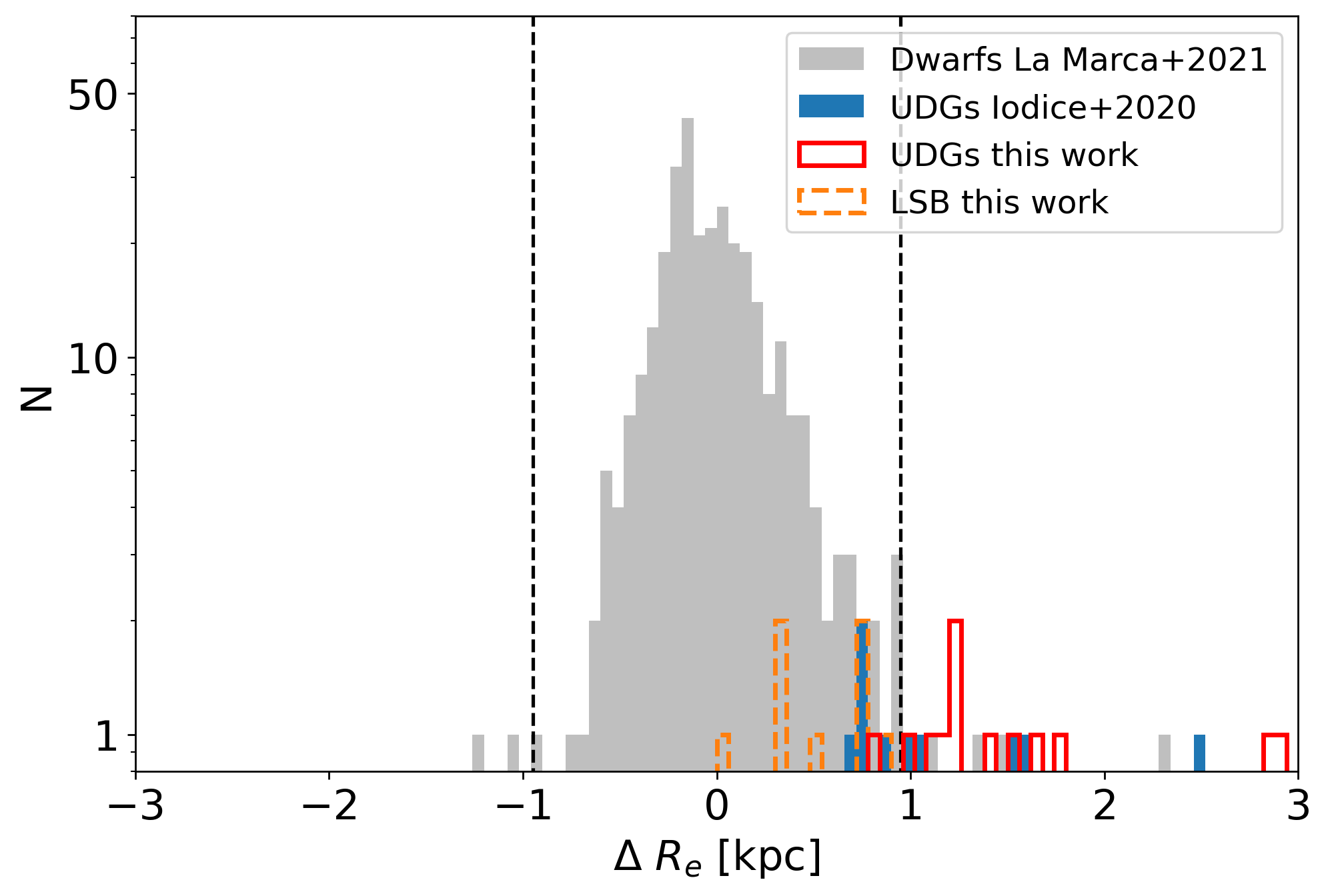}
    \includegraphics[width=0.49\textwidth]{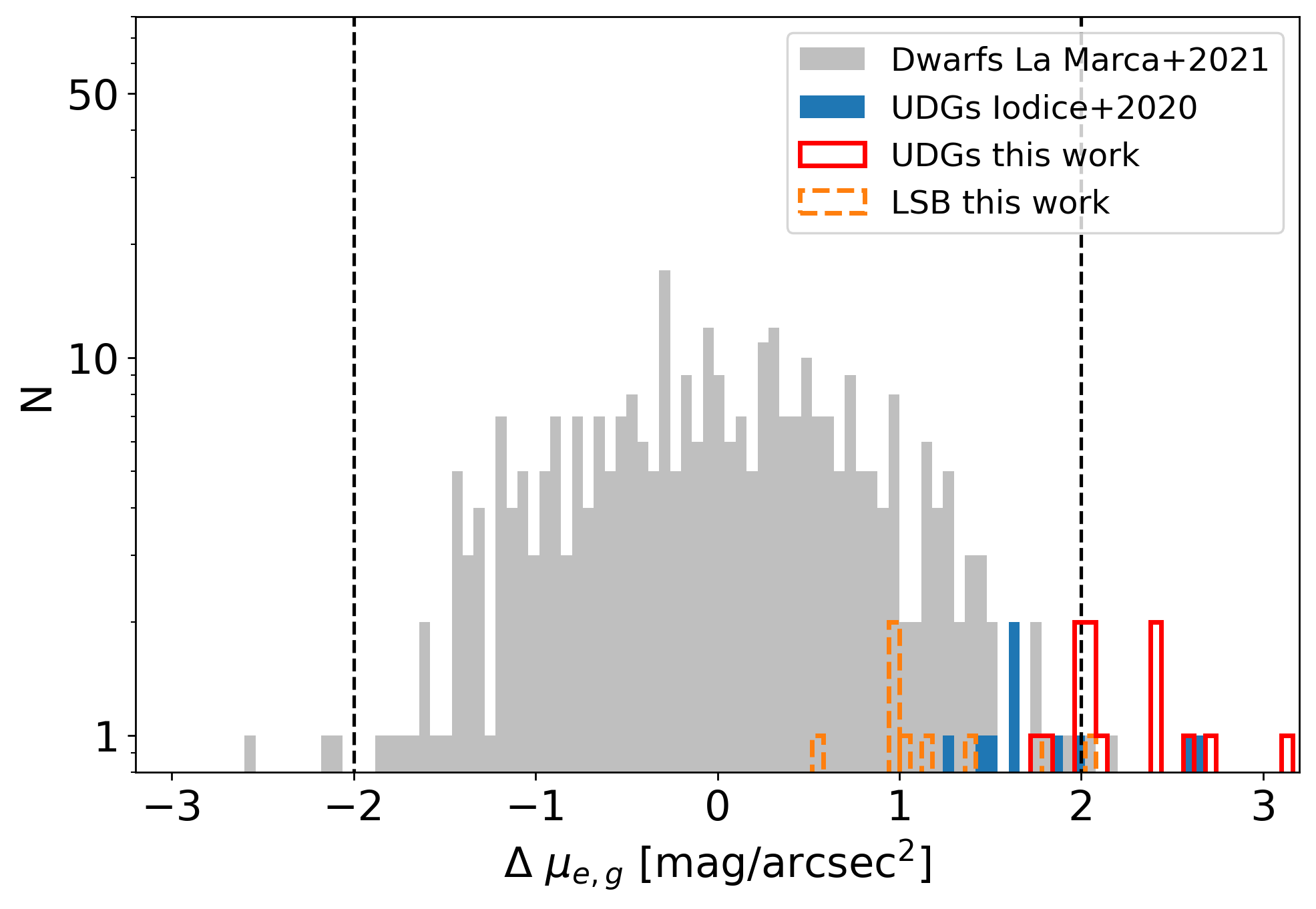}
    \caption{
    Deviations from the mean relation between luminosity and effective radius ({\it left panel}), and from the mean relation between luminosity and effective surface brightness ({\it right panel}). Vertical dashed lines show the 2.5$\sigma$ deviations from the mean relations.
    }
    \label{fig:Lim2020}
\end{figure*}

Using the selection method proposed by \citet[][]{Lim2020} we found a comparable number of UDGs.
In Fig.~\ref{fig:Lim2020} we show the distribution of the deviations from the mean relations between luminosity and effective radius and surface brightness of dwarf galaxies ($\Delta R_e$ and $\Delta\mu_e$) for the whole sample of dwarf and LSB galaxies in the Hydra I cluster \citep[from Paper~I,][and this work]{Iodice2020c,Iodice2021}. 
25 galaxies are found to deviate more than 2.5$\sigma$ for at least one of the two mean scaling relations.
Of these 25 galaxies, 17 are $2.5 \sigma$ fainter and larger than both the average fitted distributions. 
Therefore, according to \citet[][]{Lim2020} 17 galaxies would be defined as {\it primary} UDGs, and 8 as {\it secondary} UDGs. 
13 of 17 {\it primary} UDGs also respect the \citet[][]{vanDokkum2015} empirical definition.

In the sample of 11 new UDGs in Hydra I, three of them (UDG~13, UDG~15 and UDG~23) are poorly fitted with a single Sérsic function, and a second component would be required. 
This is not done here, since we aim at consistently compare the structural properties of all UDGs in Hydra I 
with those of UDGs in other clusters of galaxies, which are obtained by adopting a single Sérsic function.
Nevertheless, for the three UDGs listed above, 
the $R_e^{50}$ derived by the growth curve is larger than 1.5 kpc, and it is consistent with the
$R_e$ estimated by the best fit. 
Therefore, they can be safely identified as UDGs.
It is worth nothing that these UDGs might host a diffuse and faint halo in the galaxy outskirts.
Something similar has been recently observed by \citet[][]{Gannon2021} for a UDG in a loose group of galaxies.

\begin{figure*}
	\centering
% 	\vspace{-9cm}
	\includegraphics[width=0.8\textwidth]{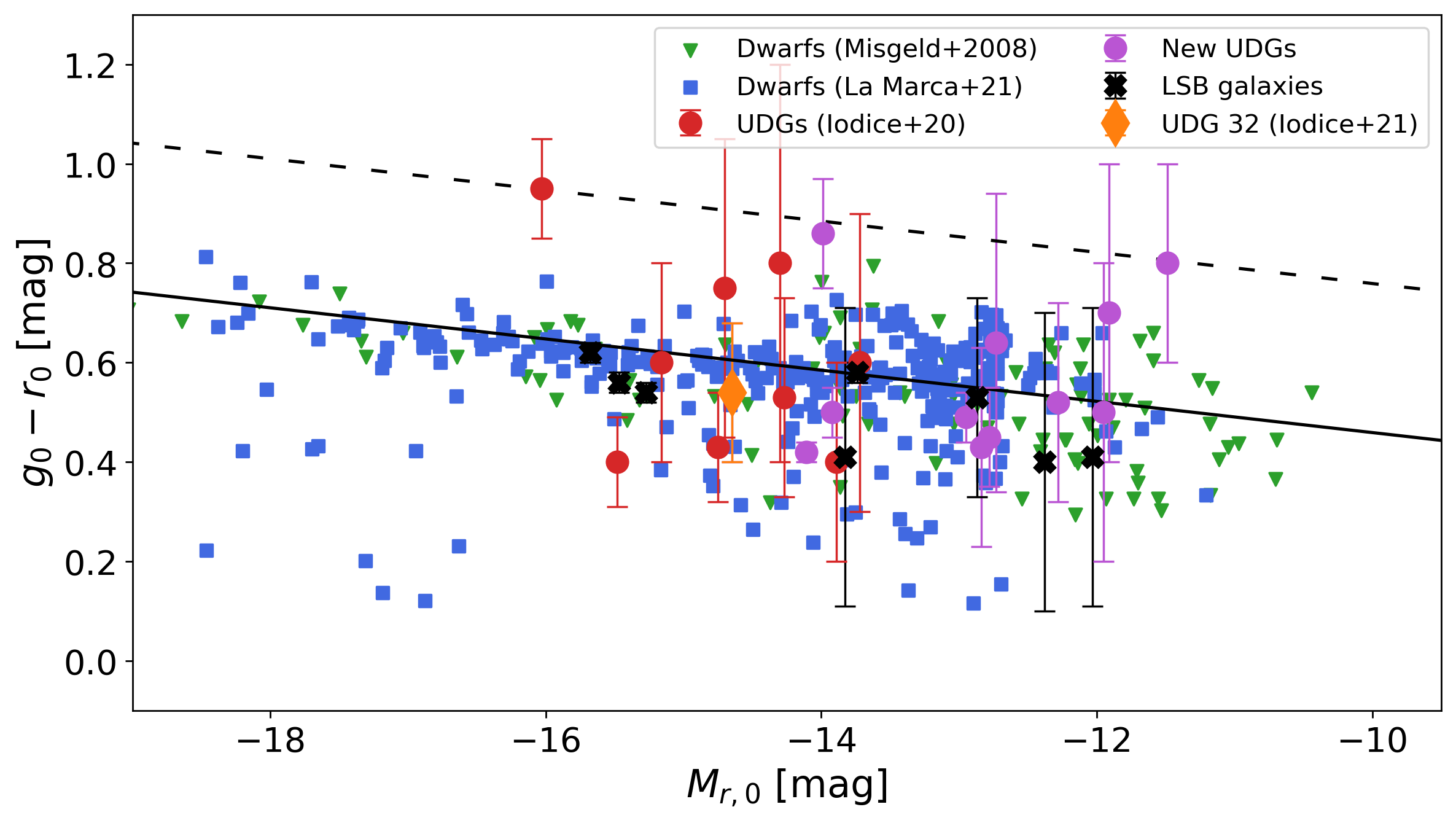}
	\caption{
	color-magnitude diagram for the full sample of dwarfs, LSB galaxies 
	and UDGs detected in the VST \object{Hydra\,I} mosaic.
	The newly detected UDGs are shown as magenta filled circles, while black crosses indicate the new LSB galaxies.
	Blue filled squares are dwarf galaxies presented in Paper I, red points are the UDGs found by \citet{Iodice2020c}, and the green triangles are the \object{Hydra\,I} galaxies presented by \citet{Misgeld2008}.
	The solid black line is the CMR for the \object{Hydra\,I} cluster early-type galaxies, with its upper 2$\sigma$ scatter limit (dashed line), as  derived by \citet[][]{Misgeld2008}: $(g-r)=-0.0314\cdot M_r+0.145$, $\sigma=0.15$. 
	}

	\label{fig:CMD}
\end{figure*}

The sample of new UDGs is displayed on the CMR in Fig. \ref{fig:CMD}.
In Fig.~\ref{fig:corrRe} we show the photometric parameters for the new UDG candidates as a function of their effective radius. 
In both plots the new UDGs are compared with those from the previous sample of 10 UDGs detected in \object{Hydra\,I}, and with the sample of Hydra\,I dwarf galaxies \citep[Paper\,I and ][]{Misgeld2008}.
In Appendix~\ref{sec:UDGimages} we show the individual UDG images and the azimuthally averaged surface brightness profile in the $g$ band, including the best fit model.

The green triangle in the UDG area of the $\mu_0-R_e$ plane, is HCC~087, originally classified as an early-type dwarf in the Hydra Cluster catalog by \citet[][]{Misgeld2008}. 
Its large $R_e$ and low surface brightness are result of strong tidal features detected around a dwarf spheroidal galaxy that is ongoing tidal disruption \citep[][]{Koch2012}. 
We decided not to include this target in the final UDG sample because of its conspicuous S-shape and clear signs of tidal disruption, which is quite different from other UDG candidates that have more round shapes (see Appendix~\ref{sec:UDGimages}).

The new UDGs have $(g-r)_0$ color in the range $0.4 - 0.86$~mag, with an average value of $(g-r)_0=0.57 \pm 0.15$~mag. 
The stellar mass ranges from $\sim 6\times 10^{6}$ to $\sim 1.2 \times 10^{8}$~M$_{\odot}$ (see Tab.~\ref{tab:UDGsample}).

%----------------

\subsection{New LSB galaxies in the Hydra~I cluster}\label{sec:LSB}

\begin{figure*}
	\centering
	\includegraphics[width=0.75\textwidth]{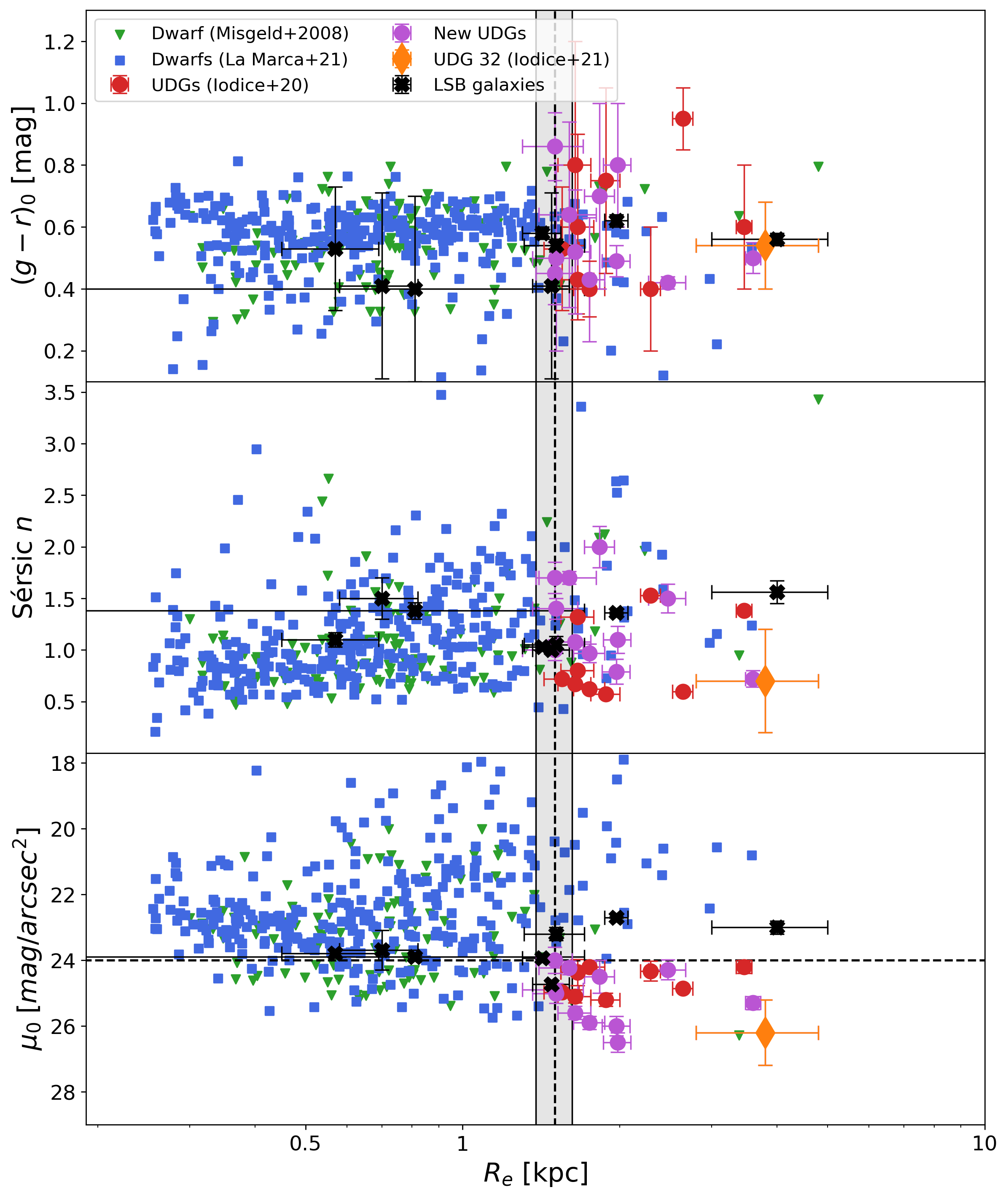}
	\caption{
	Structural and photometric parameters for the newly discovered UDGs (filled magenta circles) as a function of the effective radius. 
	Symbols are the same as Fig. \ref{fig:CMD}.
    The UDG definition criteria, $R_e\geq1.5$~kpc and $\mu_0\geq24$~mag/arcsec$^2$ \citep{vanDokkum2015}, are shown by the dashed lines. 
	The vertical shaded region indicates the range of uncertainty on $R_e$ given by the uncertainties on the cluster distance. 
    }
	\label{fig:corrRe}
\end{figure*}

The structural parameters of the remaining 8 newly detected LSB galaxies are listed in Tab.~\ref{tab:LSBsample}.
They show a wide range of values in effective radii, $0.6\lesssim R_e \lesssim 4.0$ kpc, and surface brightness, 
$22.7\lesssim \mu_{0,g}\lesssim 24.73$ mag/arcsec$^2$.
Six of them have larger size ($R_e \geq 1.2$~kpc) and fainter central surface brightness than typical dwarf galaxies of similar luminosity, hence resembling large LSB dwarfs. 
Three of them are located in the transition region between the dwarf galaxies and the UDG (that is $\mu_0\simeq24$~mag/arcsec$^2$ and $R_e\simeq1.5$~kpc, see Fig.~\ref{fig:corrRe}). 
Taking into account the $1\sigma$ error on the structural parameters, which are $\sim 0.03-0.6$~mag/arcsec$^2$ for $\mu_0$ and $\sim 0.12$~kpc for $R_e$, these galaxies might fall in the region where UDGs are found or being consistent with the typical values observed for dwarfs.
This means that a definitive conclusion on the nature of these galaxies cannot be provided based on photometry alone. 
Spectroscopic data are necessary to reduce the uncertainties on $R_e$, once redshift is estimated, and, to constraint the stellar population content.

The $g$-band images, including surface brightness profiles and their best fit, are provided in Appendix~\ref{sec:LSBimages}.
The surface brightness profiles of two objects (LSB~6 and LSB~7) show a bright central component (with $R_e \sim 0.5-1.7$~kpc, see Fig. \ref{fig:LSB_2} and Fig.~\ref{fig:LSB_3}), resembling a small bulge, in addition to a shallower and extended component at larger radii. 
Therefore, for these objects, the best fit of the light distribution would be obtained with the superposition of two 
S\'ersic profiles.
However, to be consistent with the selection method adopted in this paper, we have used the structural parameters derived by fitting a single S\'ersic law.

Compared to the UDGs, these LSB galaxies have similar range of $(g-r)_0$ color, with $0.41 \leq (g-r)_0 \leq 0.62$~mag, and a fully consistent average value of $(g-r)_0=0.51 \pm 0.08$~mag. The stellar masses are also mostly similar with those derived for UDGs, in the range $\sim 5\times 10^{6}$ to $\sim 4 \times 10^{7}$~M$_{\odot}$, except for three objects, which are one order of magnitude more massive ($\gtrsim 10^{8}$~M$_{\odot}$, see Tab.~\ref{tab:LSBsample}).

%------ Table 1: UDGs
\begin{table*}
\setlength{\tabcolsep}{2pt}
\begin{center}
\small 
%%\scriptsize
\caption{Parameters of the new UDG candidates in the \object{Hydra\,I} cluster.} 
\label{tab:UDGsample}
%%\vspace{13pt}
{
\begin{tabular}{lcccccccccccc}
\hline\hline
Object & R.A. & DEC & detection & $M_{r,0}$ & $(g-r)_0$ & $M/L_r$ & M$_*$ & $\mu_{e}$ &  $\mu_{0}$ & $R_{e}$ & $n$ & $\chi^2$\\ 
 &[deg]& [deg] & & [mag] & [mag] &  & [$10^{7}$~M$_\odot]$ &[mag/arcsec$^{2}$] & [mag/arcsec$^{2}$] & [kpc]& &\\
 (1) & (2) & (3) & (4) & (5) & (6) & (7) & (8) & (9) & (10) & (11) & (12) & (13)\\
\hline\\
%\vspace{-7pt}\\
UDG 13 &159.060384&	-27.507390&	V & -12.73&	0.64$\pm$0.30& 2.00 $\pm$ 1.00 & 2.0$\pm$1.0 & 27.35$\pm$0.11 & 24.23$\pm$0.20 & 1.60$\pm$0.20 & 1.70$\pm$0.06 & 6.3\\
UDG 14 & 159.000949 & -27.544729& A & -11.91&	0.70$\pm$0.30 & 2.00 $\pm$ 2.00 & 1.1$\pm$0.8 & 28.50$\pm$0.04 & 24.50$\pm$0.50 & 1.83$\pm$0.12& 2.0 $\pm$ 0.2 & 1.1\\
UDG 15 & 159.010624& -27.605435 & V & -11.95&	0.50$\pm$0.30 & 1.20 $\pm$ 0.80 & 0.6$\pm$0.4 & 27.70$\pm$0.20 & 25.01$\pm$0.30 & 1.51$\pm$0.15 & 1.40$\pm$0.10 & 11.0 \\
UDG 16 & 159.10242 & -27.235641 & V & -12.84&	0.43$\pm$0.20 & 1.00 $\pm$ 0.50 & 1.1$\pm$0.5 & 27.63$\pm$0.11& 25.90$\pm$0.20& 1.75$\pm$0.12& 0.97$\pm$0.09 & 1.0 \\
UDG 17 & 159.173845 & -27.277078 & V & -13.99&	0.86$\pm$0.11 & 4.00 $\pm$ 1.00 & 12$\pm$3.0 & 26.70$\pm$0.08 & 24.90$\pm$0.10 & 1.50$\pm$0.20 & 1.00$\pm$0.10 & 1.6 \\
UDG 18 & 159.070065 & -27.338011 & V & -12.28 &	0.52$\pm$0.20 & 1.30 $\pm$ 0.60 & 0.9$\pm$0.4 & 27.60$\pm$0.20 & 25.56$\pm$0.19& 1.64$\pm$0.12& 1.08$\pm$0.03 & 1.7 \\
UDG 19 & 159.010682 & -27.290000 & V & -11.49 &	0.80$\pm$0.20 & 3.00 $\pm$ 1.00 & 1.0$\pm$0.5 & 28.50$\pm$0.20 & 26.50$\pm$0.30 & 1.98$\pm$0.12 & 1.10$\pm$0.13 & 1.1 \\
UDG 20 & 159.518455 & -27.497272 & A & -12.95 &	0.49$\pm$0.05& 1.20 $\pm$ 0.14 & 1.4$\pm$0.2 & 27.33$\pm$0.15 & 26.03$\pm$0.33 & 1.97$\pm$0.12 & 0.79$\pm$0.12 & 1.4 \\
UDG 21 & 159.225691 & -27.615297 & A & -12.78 &	0.45$\pm$0.10& 1.00 $\pm$ 0.20 & 1.1$\pm$0.3 & 27.30$\pm$0.50 & 24.00$\pm$0.40 & 1.50$\pm$0.12 & 1.70$\pm$0.15 & 1.5 \\
UDG 22 & 158.670717 & -27.700909 & A & -13.92 &	0.50$\pm$0.05 & 1.23 $\pm$ 0.14 & 3.6$\pm$0.4 & 26.50$\pm$0.14 & 25.26$\pm$0.18& 3.60$\pm$0.12& 0.72$\pm$0.08 & 1.0 \\
UDG 23 & 158.865410 & -27.771273 & A & -14.11&	0.42$\pm$0.02 & 0.96 $\pm$ 0.04 & 3.4$\pm$0.2 & 27.20$\pm$0.20 & 24.29$\pm$0.30 & 2.47$\pm$0.20 & 1.5$\pm$0.14 & 6.0\\
% UDG $32^*$ &	159.26775&	-27.715428&	V & -14.65&	0.54$\pm$0.14& 1.40 $\pm$ 0.40 & 8.0$\pm$3.0 & 27.50$\pm$1.50 & 26.2$\pm$1.0& 3.80$\pm$1.00 &0.70$\pm$0.50\\
\hline
\end{tabular}
\tablefoot{Column 1 reports the name of the UDG candidate. 
In columns 2 and 3 we list the coordinates of the UDGs in J2000. 
In column 4 we report the kind of detection, i.e. visual 
inspection (flag ''V'') or by the automatic tool (flag ''A'').
In columns 5 and 6 we report the total $r$-band magnitude and the average $g-r$  color. 
Columns 7 and 8 give the stellar mass-to-light ratio derived in the $r$ band and stellar mass, respectively.
Columns 9 to 12  list the structural parameters derived from the 2D fit in the $g$ band: the effective and 
central surface brightness, the effective radius in kpc and the $n$ exponent of the S{\'e}rsic law, respectively. 
In Column 13 we report the value of $\Tilde{\chi ^2}$, which indicates the goodness of the fit of the surface brightness profiles.
Magnitudes and  colors are corrected for Galactic extinction using values from \citet{Schlafly2011}.}}
\end{center}
\end{table*}

%------ Table 2: transition LSB
\begin{table*}
\setlength{\tabcolsep}{2pt}
\begin{center}
\small 
{
%%\scriptsize
\caption{Parameters of the new LSB galaxies in the \object{Hydra\,I} cluster.} 
\label{tab:LSBsample}
%%\vspace{13pt}
\begin{tabular}{lcccccccccccc}
\hline\hline
Object & R.A. & DEC & detection & $M_{r,0}$ & $(g-r)_0$ & $M/L_r$ & M$_*$ & $\mu_{e}$ &  $\mu_{0}$ & $R_{e}$ & $n$ & $\chi^2$ \\ 
 &[deg]& [deg] & & [mag] & [mag] &  & [$10^{7}$~M$_\odot]$ &[mag/arcsec$^{2}$] & [mag/arcsec$^{2}$] & [kpc]& \\
 (1) & (2) & (3) & (4) & (5) & (6) & (7) & (8) & (9) & (10) & (11) & (12) & (13)\\
\hline\\
%\vspace{-7pt}\\
LSB 1 &159.000120	& -27.482826 & V & -12.38& 0.40$\pm$0.30& 0.90$\pm$0.60 & 0.60$\pm$0.40 & 26.59$\pm$0.13 & 23.90$\pm$0.20 & 0.81$\pm$0.90& 1.38$\pm$0.08 & 16.0 \\
LSB 2 &	159.040202&-27.514335& V & -12.87& 0.53$\pm$0.20& 1.30$\pm$0.60  & 1.50$\pm$0.70 & 25.84$\pm$0.07 & 23.80$\pm$0.14& 0.57$\pm$0.12& 1.10$\pm$0.07 &  1.4\\
LSB 3 &	159.110634&	-27.544892&	V & -12.03&	0.41$\pm$0.30& 0.90$\pm$0.60 & 0.50$\pm$0.30 & 26.60$\pm$0.80 &  23.69$\pm$0.60 & 0.70$\pm$0.12 & 1.50$\pm$0.20 & 2.0 \\
LSB 4 & 159.084460& -27.231646& A & -13.83& 0.41$\pm$0.30& 0.90$\pm$0.60& 2.50$\pm$1.80 & 26.55$\pm$0.06 & 24.73$\pm$0.09 & 1.48$\pm$0.12 & 1.00$\pm$0.03 & 2.0 \\
LSB 5 &	159.171918&	-27.805707&	A & -13.74&	0.58$\pm$0.02& 1.59$\pm$0.07 & 4.00$\pm$0.20 & 25.81$\pm$0.01 & 23.93$\pm$0.03 & 1.42$\pm$0.12 & 1.03$\pm$0.01 & 1.6\\
LSB 6 &	159.519470&	-27.545830&	A & -15.47&	0.56$\pm$0.02& 1.49$\pm$0.07 & 18.40$\pm$0.90 & 26.02$\pm$0.08 & 23.00$\pm$0.20& 4.00$\pm$1.00 & 1.56$\pm$0.11 & 1.1 \\
LSB 7 &	159.077907&	-27.621647&	A & -15.68&	0.62$\pm$0.02& 1.80$\pm$0.08 & 25.89$\pm$0.02 & 25.30$\pm$0.01 & 22.70$\pm$0.02 & 1.97$\pm$0.10 & 1.36$\pm$0.01 &  1.5 \\
LSB 8 &	159.476947&	-27.258645&	A & -15.27&	0.54$\pm$0.02& 1.40$\pm$0.06 & 14.40$\pm$0.70 & 25.12$\pm$0.11 & 23.20$\pm$0.20& 1.51$\pm$0.20& 1.05$\pm$0.08 & 1.5\\
\hline
\end{tabular}
\tablefoot{Same as Table~\ref{tab:UDGsample}, but for the new LSB galaxies of the sample.}}
\end{center}
\end{table*}

\section{Detection of globular clusters}\label{sec:GCs}

With our data, i.e. deep $g$- and $r$-band imaging, namely two optical and relatively close passbands, it is hard to clearly identify the GC populations because of the expected large fore- and back-ground contamination. Nevertheless, as detailed in the previous VEGAS papers, the large format of the OmegaCam allows to robustly constrain the background contamination from the analysis of regions at large angular distances from the target(s). 

To detect GCs around all the newly detected cluster members, both UDGs and  LSB galaxies, we adopted the same procedure described by \citet[][]{Iodice2020c}, and more extensively presented in \citet{cantiello2018,Cantiello2020}. 
Briefly, we run SExtractor on $g$ and $r$ cutouts centred on each galaxy, with size $\sim20 R_e\times 20R_e$. 
To improve source detection down to the faintest magnitude levels, we subtracted the galaxy model derived from \texttt{GALFIT}. 
The GCs at the distance of \object{Hydra\,I}, in images with a FWHM$\sim0.8$~arcsec seeing, are by all means point sources. 
For each source and in each band we derived the following quantities:
\begin{itemize}
\item the automated aperture magnitude ($MAG\; AUTO$) to estimate the total magnitude of the source;
\item the aperture magnitude ($MAG\;APER$) within 4, 6 and 8 pixels diameter;
\item the $g{-}r$ color\footnote{ 
    The $g$ and $r$ aperture magnitudes are not aperture corrected. The  aperture corrections in the two passbands are nearly identical ($\sim$0.35 mag), so they compensate each other when deriving the $g{-}r$ color, while the total error we estimated on the combined apertures is of the order of $\sim$0.04 mag. In other words, we do not apply any correction as the correction is smaller than its uncertainty.} as:
\begin{equation*}
    MAG\_APER_g [6_{pix}] {-}MAG\_APER_r [6_{pix}];
\end{equation*}
\item the concentration index as:
\begin{equation*}
    CI_X=MAG\_APER_X [4_{pix}] {-}MAG\_APER_X [8_{pix}],
\end{equation*}
which is an indicator of source compactness in the $X$ band \citep{peng11}.
\end{itemize}
At the adopted distance of Hydra~I the peak of the GC luminosity function (GCLF), the so called turn-over magnitude (TOM), is $\mu_{g,TOM}\simeq 26.0$ mag {  \citep[we assumed $M_g^{TOM}=-7.5$, from ][]{villegas10}}.
{  Adopting} a median $\langle g{-}r\rangle = 0.6$ mag for the GC population \citep{Cantiello2018a}, the $r$-band TOM is $\mu_{r,TOM}\simeq 25.4$ mag.
To identify GC candidates, we select sources with:
{\it i)} $g$-band magnitude $23.5\leq m_g\leq 26.0$ mag, the expected range between the TOM of the GCLF and {  3$\sigma_{GCLF}$ mag brighter \citep[$3\sigma_{GCLF} = 2.5$ mag,][]{villegas10};}
{\it ii)} color $0.25\leq g{-}r \leq 1.25$~mag\footnote{Based on models predictions from \url{http://cosmic.yonsei.ac.kr/YEPS.htm}, for ages older than 8 Gyr, \citet[][]{chung2020}, and empirical results from \citet{Cantiello2018a}. Similar color limits were adopted also by \citet[][]{Carlsten2022}};
{\it iii)} SExtractor star-galaxy classifier: {$CLASS\_STAR\geq0.4$}, we assumed a low CLASS\_STAR value for star/galaxy separation, as the classifier gets less reliable toward the faint magnitudes we are interested in;
{\it iv)} elongation (i.e. major-to-minor axis ratio from SExtractor) $\leq2$ in both bands;
{\it v)} $CI_X$ within $\pm0.1$ mag of the local sequence of point sources. 
The choices of 4 and 8 pixels, and $\Delta CI_x=\pm0.1$ mag, were made after several tests, as a compromise between the completeness and contamination \citep[see also][where the same compactness parameter is adopted]{Saifollahi2021,Saifollahi2022}.

With these criteria we identify the GC candidates around each galaxy in the sample, and estimated their local density $\Sigma_{GC}^{UDG}$. 
To correct for background contamination, we need to characterise the expected number of foreground MW stars, background galaxies and possible intra-cluster GCs (ICGC). 
The stars and background galaxies contamination is most effectively derived in regions far from the virial radius of the cluster. 
The local effect of ICGCs is instead better constrained around the UDG or  LSB galaxy itself. 
Hence, we obtained two independent characterisations of the background. 
A {\it local background} is derived by running the GC selection criteria on the catalogs of sources detected in annuli of $5R_e \leq r \leq 10R_e$ around each galaxy\footnote{The radius $5R_e$ is considered a upper limit for bound systems \citep{Kartha2014,Forbes2017,Caso2019}. 
The upper limit to $10R_e$ is  motivated by the need of having an area enclosing a number of GC candidates of $\sim 20$, sufficiently large for constraining the local background.}. 
We find that the number of GC candidates in such areas is on average $\Sigma_{GC}^{bkgr}=5.0\pm1.5$ per square arcmin, ranging from a minimum  of 1.3~arcmin$^{-2}$ to a maximum of 11.5~arcmin$^{-2}$. 

Furthermore, we estimated a {\it global background} from five different fields with size  $\sim10\arcmin \times10\arcmin$, outside the Hydra~I virial radius, free from bright foreground MW stars, and obvious Hydra~I cluster galaxies, obtaining a background density of  $\Sigma_{GC}^{bkgr}=1.35\pm0.17$~arcmin$^{-2}$.

We keep both background estimates as the global background is statistically more robust, but it possibly fails to accurately sample the ICGCs contamination in UDGs closer to the cluster core; the local background, on the other hand, can better constrain the local ICGCs population but it suffers for larger statistical uncertainty.

The physical extent of the GC system around UDGs and LSB galaxies is still subject to debate \citep[][]{Forbes2017,Muller2021,Caso2019,Kartha2014}. 
Hence, we derived the total number of GCs ($N_{GC}$) within 1.5$R_e$, as in \citet[][]{vanDokkum2016}, and also within 3 and $5R_e$. 
Since the photometry roughly reaches the TOM peak (the median photometric uncertainty is $\Delta m_g\simeq 0.2$ mag at $m_g=26$ mag for point sources) and assuming the LSB galaxies GCLF is a Gaussian \citep[e.g.][]{Rejkuba2012,vanDokkum2016}, we derived $N_{GC}$ as twice the background corrected GC density over the 3 and $5 R_e$ area of the galaxy, times the area: $N_{GC}=2\times (\Sigma_{GC}^{LSB}{-}\Sigma_{GC}^{bkgr})\times Area$\footnote{The factor of two correction is strictly valid if the galaxy lies at the adopted distance modulus, at $m_g^{TOM}\sim26$ mag. 
However, if the galaxy is in fore/background with respect to the adopted cluster distance, the TOM would be brighter/fainter, hence the correction factor would be under/over-estimated, respectively.}.
When estimating the $N_{GC}$ within 1.5 $R_e$, we assumed that half of the GC population is within 1.5 $R_e$, thus the total population is four times the background corrected GC density over 1.5 $R_e$ area, times the area \citep{vanDokkum2016}.

To estimate the uncertainty on $N_{GC}$, we combined the $rms$ of the global background, the Poissonian scatter on the over-density, ($\Sigma_{GC}^{UDG}{-}\Sigma_{GC}^{bkgr})\times Area$, plus a further $20\%$ error on the adopted scaling factors (i.e., two for $N_{GC}$ at  3 and 5$R_e$, and four at $1.5R_e$, respectively). 
For the $N_{GC}$ values based on local background decontamination, we propagated Poissonian uncertainties for both the background and the over-density.

We choose as reference value the $N_{GC}$ estimated within 1.5 $R_e$, locally corrected for background contamination. 
In Table \ref{tab:GC_1.5} we report the GC raw counts, the background-subtracted GC counts and the completeness corrected GC counts, within this aperture, including also the counts for the UDG 32 \citep[][]{Iodice2021}.
We also obtain $N_{GC}$ for two larger apertures (3 and 5 $R_e$), with both types of background corrections (local and global). 
We present them as supplementary material, listing the $N_{GC}$ measurements in Table~\ref{tab:UDG_GC}.

The values derived using the local and the global background corrections agree with each other within the quoted uncertainties. 
With a few exceptions, most of our galaxies have a GCs over-density consistent with zero, independently from the background and the $R_e$ analysis radius adopted. 
This is consistent with previous work by \citet[][]{Lim2018}, which also found many UDGs without GCs.

\begin{figure*}
    \centering
    \includegraphics[width=0.9\textwidth]{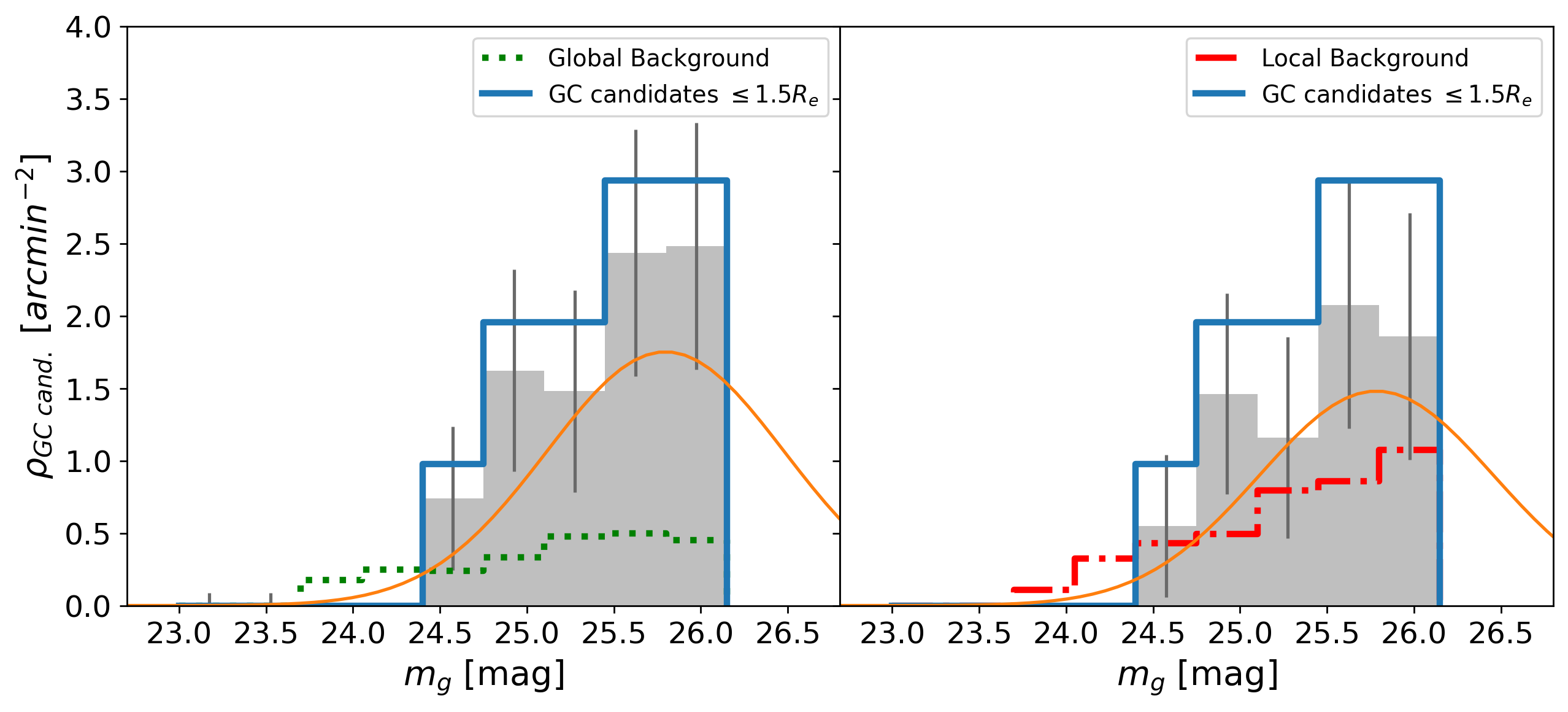}
    \caption{{\it Left}: The total luminosity function of the GC candidates over the joint sample of UDGs and LSB galaxies (Tab. \ref{tab:GC_1.5}) with $N_{GC}\ge 0$ within $1.5Re$, normalised to the total galaxies area (solid blue line). The density distribution for the global background is shown with green-dotted line. The grey shaded histogram shows the residuals of the two. The solid line shows the best fit Gaussian function of the GC luminosity function (grey bins). The best fit parameter are: $\mu_{g,TOM} = 25.8 \pm 0.3$ mag, $\sigma_{GCLF} = 0.69 \pm 0.12$ mag. 
    {\it Right}: As left panel, but using the local background (red dot-dashed line). The best fit parameter are: $\mu_{g,TOM} = 25.8 \pm 0.2$ mag, $\sigma_{GCLF} = 0.66 \pm 0.14$ mag.
    The error bars show the uncertainties on the residual GC density, calculated as detailed in Sect. \ref{sec:GCs}.}
    \label{fig:GC_density_dist}
\end{figure*}

Figure \ref{fig:GC_density_dist} shows the GCLF normalised to the total area, summed over all the galaxies listed in Table \ref{tab:GC_1.5} with $N_{GC}>0$, within 1.5 $R_e$.
We also plot the global/local background luminosity function (left/right panel) and the residual luminosity function (grey shaded histogram). 
The best Gaussian fit parameters for the GCLF locally corrected are: $\mu_{g,TOM} = 25.8 \pm 0.2$ mag, $\sigma_{GCLF} = 0.66 \pm 0.14$ mag.
The g-band turn-over magnitude, $\mu_{g,TOM}$, is consistent with the assumptions discussed above.
The estimated width, $\sigma_{GCLF}$, agrees nicely with the predictions by \citet[][see their Fig. 5]{villegas10} for objects with total magnitude similar to ours.
Similar results are obtained adopting the global background correction (Fig. \ref{fig:GC_density_dist}, left panel).
The diagram reveals several features: 
$i$) the GC candidates luminosity function (blue solid line) shows the expected steep increase when going to fainter magnitude, while approaching the TOM; 
$ii)$ the global background has a relatively stable distribution, with larger densities toward fainter magnitudes; 
$iii$) the local background luminosity function suffers from a slightly larger scatter and a higher density of sources compared to the case of the global background, which is expected because of the local contribution of ICGCs; 
$iv$) both panels reveal the presence of a residual GC over-density (gray histograms), consistent with the expected half-Gaussian shape.
These results, while supporting the evidence that a small yet non-negligible GC population is found in these UDGs, highlight the need for new observational data, either spectroscopy or near-IR photometry, to better constrain the bulk GCs population.

%------------- Table 2

\begin{table}
    \centering
    \caption{Globular clusters candidates numbers and their specific frequencies associated with the new UDGs and LSB galaxies in the Hydra I cluster.}
    \label{tab:GC_1.5}
    \begin{tabular}{lcccc}
    \hline \hline
    Object & Raw GCs & Back. sub. GCs & $N_{GC}$ & $S_N$ \\
    \hline
    UDG 13 & 0 & -1.4 & - & - \\
    UDG 14 & 1 & 0.04 & 0$\pm$1 & 0$\pm$23 \\
    UDG 15 & 1 & 0.46 & 2$\pm$3 & 41$\pm$75 \\
    UDG 16 & 0 & -0.33 & - & - \\
    UDG 17 & 1 & 0.73 & 3$\pm$3 & 11$\pm$14 \\
    UDG 18 & 3 & 2.8 & 11$\pm$7 & 168$\pm$153 \\
    UDG 19 & 0 & -0.24 & - & - \\
    UDG 20 & 0 & -0.69 & - & - \\
    UDG 21 & 0 & -0.6 & - & - \\
    UDG 22 & 0 & -0.24 & - & - \\
    UDG 23 & 0 & -0.45 & - & - \\
    UDG 32 & 3 & 1.8 & 7$\pm$4 & 12$\pm$10\\
\hline
     &  &  &  &  \\
    LSB 1 & 0 & -0.42 & - & - \\
    LSB 2 & 0 & -0.36 & - & - \\
    LSB 3 & 1 & -0.44 & - & - \\
    LSB 4 & 2 & 1.9 & 8$\pm$3 & 28$\pm$16\\
    LSB 5 & 0 & -0.09 & - & - \\
    LSB 6 & 1 & 0.46 & 2$\pm$1 & 2$\pm$1 \\
    LSB 7 & 0 & -0.63 & - & - \\
    LSB 8 & 0 & -0.36 & - & - \\
    \hline
    
    \end{tabular}
    
    \tablefoot{
    For each galaxy in our sample we show the raw GC candidates counts within a 1.5$R_e$ aperture (second column), the local background subtracted counts (third column), the completeness corrected total GC counts ($N_{GC}$, fourth column), and the respective specific frequency (last column).
    }
    
\end{table}

%----------------------------------------------------------------

\section{Results}\label{sec:result}

In this section we explore the main properties we derived for all UDGs over the \object{Hydra\,I} cluster area, including also the first 9 candidates published by \citet[][]{Iodice2020c}.
We focus on the 2-dimensional (2D) projected distribution of the UDGs inside the cluster and on how the structural parameters and colors vary as a function of the cluster centric projected distance. 
In addition, we discuss how the structural properties of the UDGs compare with those derived for the dwarf galaxies in the cluster and, in particular, for the new LSB galaxies also presented in this work.
Finally, we show how the GCs specific frequency is related to the other physical properties and to the clustercentric distance.
The implications of these results on the nature of UDGs in \object{Hydra\,I} cluster will be discussed in Sec.~\ref{sec:disc}.

\subsection{2-dimensional distribution of UDGs inside the cluster}\label{sec:2D_distr}

\begin{figure*}
	\includegraphics[width=9cm]{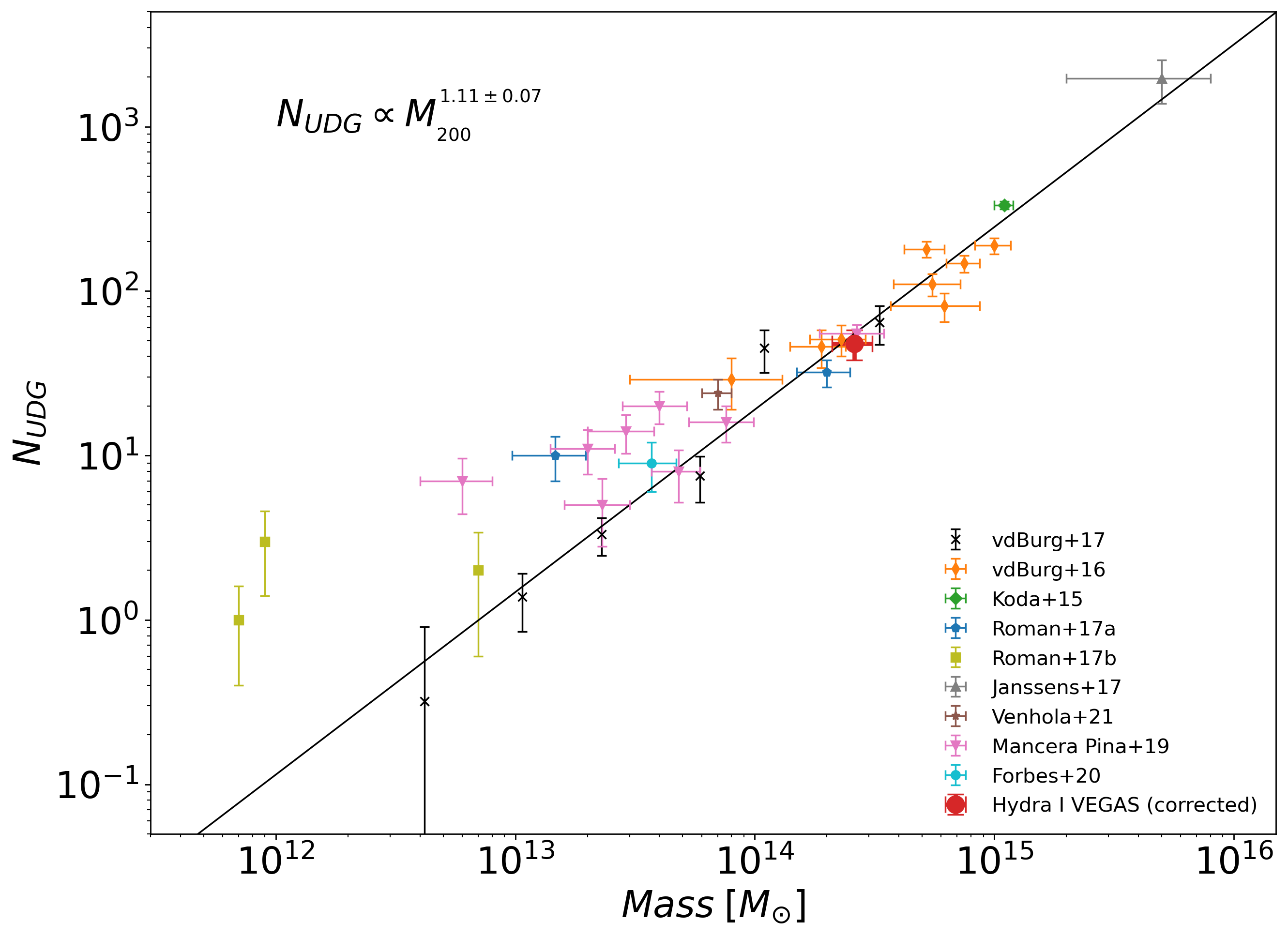}
	\includegraphics[width=9cm]{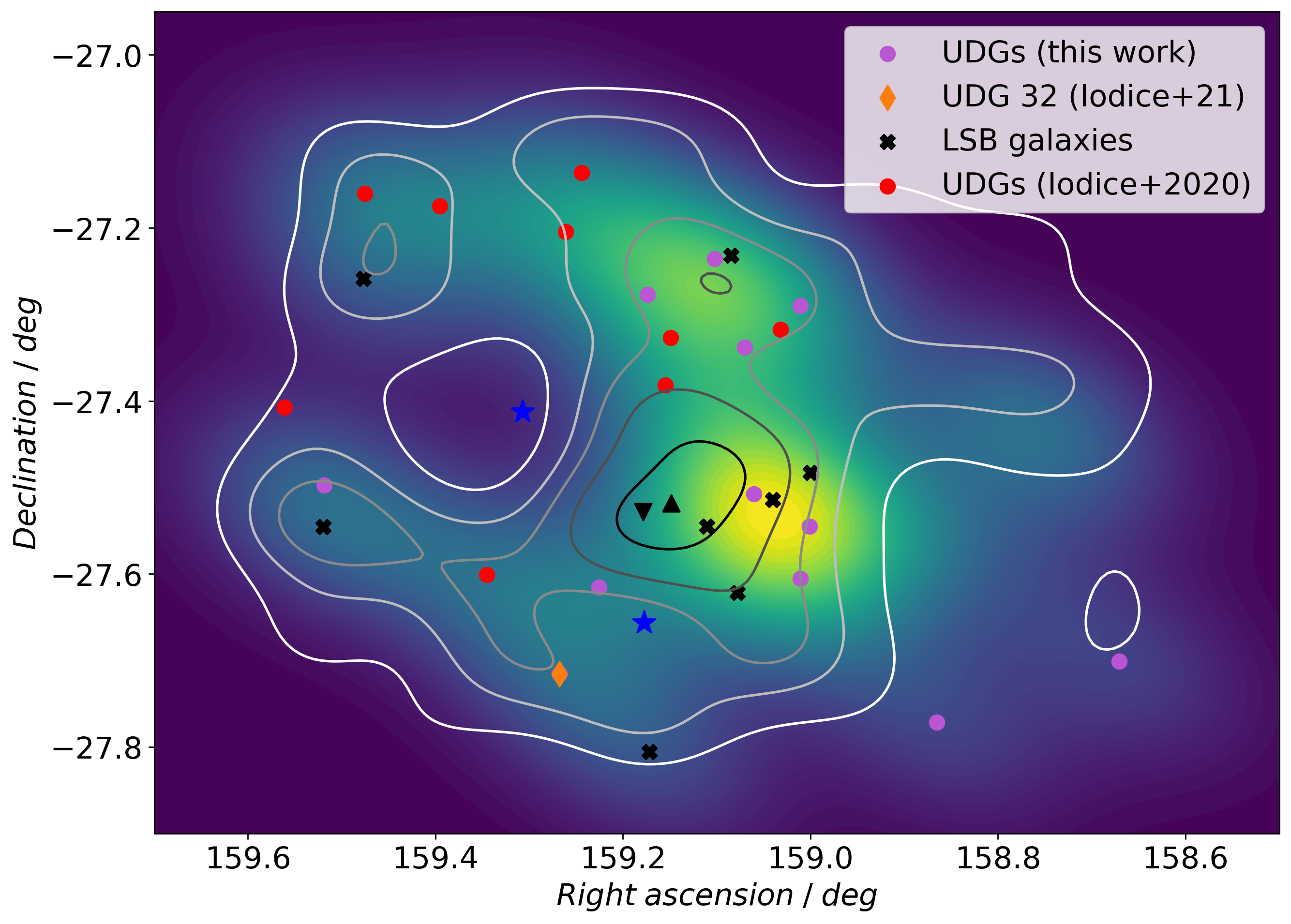}
	\caption{
	\emph{Left panel} - Abundance of UDGs as a function of the halo mass for available data \citep[][]{vanderBurg2017,VanderBurg2016,Koda2015,Roman2017a,Roman2017b,Janssens2017,Mancera-pina2019udg,Forbes2020b,Venhola2021}. 
	For the \object{Hydra\,I} cluster the number of UDGs within $\approx 0.4R_{vir}$ is scaled to $N_{UDG}(\leq R_{vir})$ on an assumed radial number density distribution (see Sect. \ref{sec:2D_distr}). 
	The black line represents the relation fitted by \citet[][]{vanderBurg2017}, which equation is reported in the plot.
	\emph{Right panel} - 2D projected distribution of all the UDGs detected in the Hydra I cluster, and of the 8 LSB galaxies presented in this work (sequential colormap).
	Contours represent the density distribution of dwarf galaxies (see Paper I), with increasing density from white to black. 
	The two brightest cluster members NGC 3311 and NGC 3309 are marked as black triangles. 
	The locations of the two brightest stars in the field are also indicated as blue stars.}
	\label{fig:N_density}
\end{figure*}

Considering rigorously the van Dokkum UDG definition \citep[][]{vanDokkum2015}, 
we find that the total number of UDGs within $\approx 0.4R_{vir}$ of the 
Hydra I cluster is 21:
11 are presented here, 9 by \citet[][]{Iodice2020c}, and UDG 32 discovered by \citet[][]{Iodice2021} in the stellar filaments of NGC 3314A.
In order to compare these counts with the UDG-abundance-halo mass relation \citep[][]{vanderBurg2017,Janssens2019}, we need to correct $N_{UDG}$.
To scale it up to 1 $R_{vir}$, we assume as radial number density distribution the \citet[][]{Einasto1965} profile fitted by \citet[][]{VanderBurg2016}.
As they point out, the Einasto profile provides a better description of the UDG radial density distribution than the Navarro-Frenk-White profile \citep[][]{Navarro1997}.
Therefore, we calculate the scaling number density profile factor for the Hydra I cluster based on our counts (21 UDGs within a circular area of $r=0.4R_{vir}$), and integrate the assumed number density profile on a disk of radius 1 $R_{vir}$.
We obtain $48\pm 10$ UDGs for the Hydra I cluster, where the error is calculated by propagating the Poissonian error on the UDG counts. 
A comparison with the literature UDG abundance-halo mass relation is presented in Fig. \ref{fig:N_density} (left panel).

Fig.~\ref{fig:N_density} (right panel) shows the 2D spatial distribution of all 21 Hydra I UDGs detected so far, and of the 8 LSB galaxies presented here. 
This is obtained by convolving the galaxy distribution with a Gaussian kernel with a standard deviation of $\sigma = 5$ arcmin.
The spatial distribution appears strongly asymmetric with respect to the cluster core.
Most galaxies seem to be concentrated towards the cluster core and around a sub-group of galaxies North of it, which are the two densest regions of the cluster, following a similar distribution found for the average dwarf galaxy population within the cluster central regions.

Like the dwarf density distribution, the 2D spatial distribution of the galaxies analyzed here presents a low density around the two brightest stars in the field.
Therefore, some of the apparent asymmetries in the distribution of dwarfs (both, normal and LSB/UDGs) could be due to the presence of these 
two bright stars and their residual light, rather than reflecting an intrinsic physical property.
To have a more quantitative idea of the impact of the two stars see also the right panel of Fig. 9 in Paper I.

\subsection{Cluster-centric properties}\label{clust-centric}

\begin{figure*}
    \centering
    \includegraphics[width=0.8\textwidth]{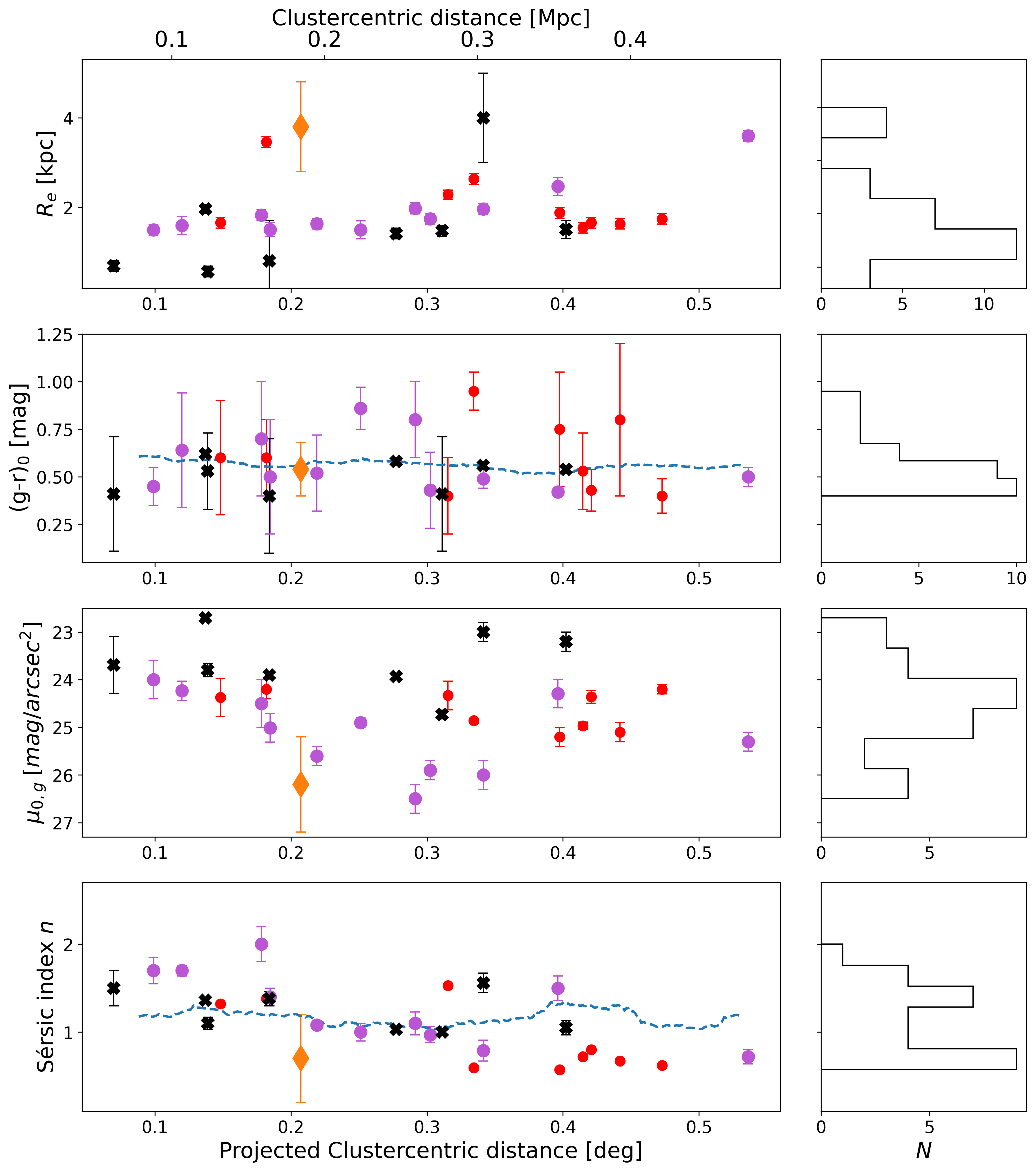}
    \caption{
    \textit{Left panel} - Structural parameters for UDGs (purple circles are UDGs in this work, red circles are UDGs from \citet[][]{Iodice2020c}, and the orange diamond marks UDG 32 \citet[][]{Iodice2021}) and LSB galaxies (black crosses) in the \object{Hydra\,I} cluster as function of the cluster-centric projected distance.
    From the top to the bottom: effective radius $R_e$, $g-r$ color, central surface brightness $\mu_0$ in the $g$ band and the Sérsic index $n$. 
    The blue dashed line is the mean locus derived for the dwarf galaxy sample in \object{Hydra\,I}, published in Paper~I. 
    The running mean shown is calculated considering a moving window of a fixed number of objects (50).
    \textit{Right panel} - Distributions of the parameters shown on the left panels.}
    \label{fig:trends}
\end{figure*}

In Fig.~\ref{fig:trends} we show the structural parameters ($R_e$, $\mu_0$, Sérsic index $n$) and the integrated $g-r$ color, derived for each UDG, as function of the projected cluster-centric distance.
For the three UDGs where we added a second component, the parameters of the outer component are taken as reference.

The effective radius peaks around 1.5 kpc, which is reasonably due to a selection effect, and does not show any trend with cluster-centric distance. 
However, a few UDGs have been found with larger values ($R_e\sim 2.5$~kpc).
The central surface brightness spans a wide range of values at all distances from the cluster center, being $24 \leq \mu_0 \leq 27$~mag/arcsec$^2$, and also in this case, a clear trend as function of the projected cluster-centric distance is not observed. 
It is worth noting that there are no UDGs fainter than $\mu_0\geq 25$~mag/arcsec$^2$ inside the angular distance of $\sim0.18$ degree from the cluster core.
The Sérsic index $n$ slightly increases moving inward in the cluster. 
Indeed, the UDGs with the highest $n$ values are located in the innermost region of Hydra\,I ($\leq0.2\;deg$), while no UDG with $n>1.0$ is farther than $\sim0.4\;deg$.
As a comparison, the Sérsic index $n$ derived for the dwarf galaxies in the cluster (see Paper I) remains nearly constant at all radii, and {  the mean value is slightly larger ($\sim$1) than that for UDGs ($\sim0.8$)}.

The $g-r$ color of UDGs is uniformly {  scattered} in the range $\sim0.4-0.9$~mag at all cluster-centric distances.
These values are consistent with colors derived for the faintest red dwarf galaxies, which also do not show
any trend with the distance from the cluster center (see Fig.~11 in Paper I).

\subsection{Specific frequency of GCs versus physical properties} \label{sec:Sn}

\begin{figure*}
    \centering
    \includegraphics[width=0.65\textwidth]{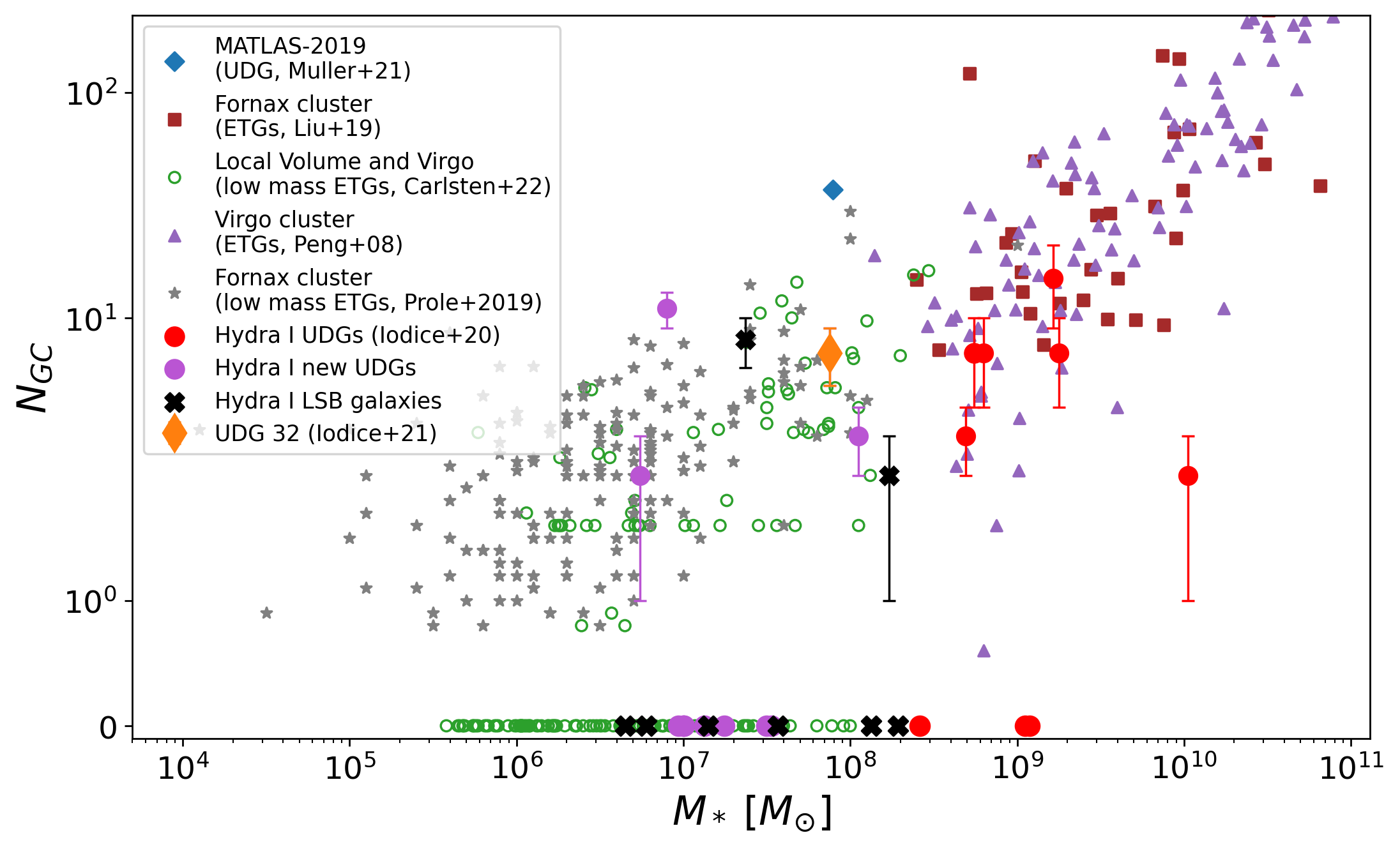}
    \includegraphics[width=0.65\textwidth]{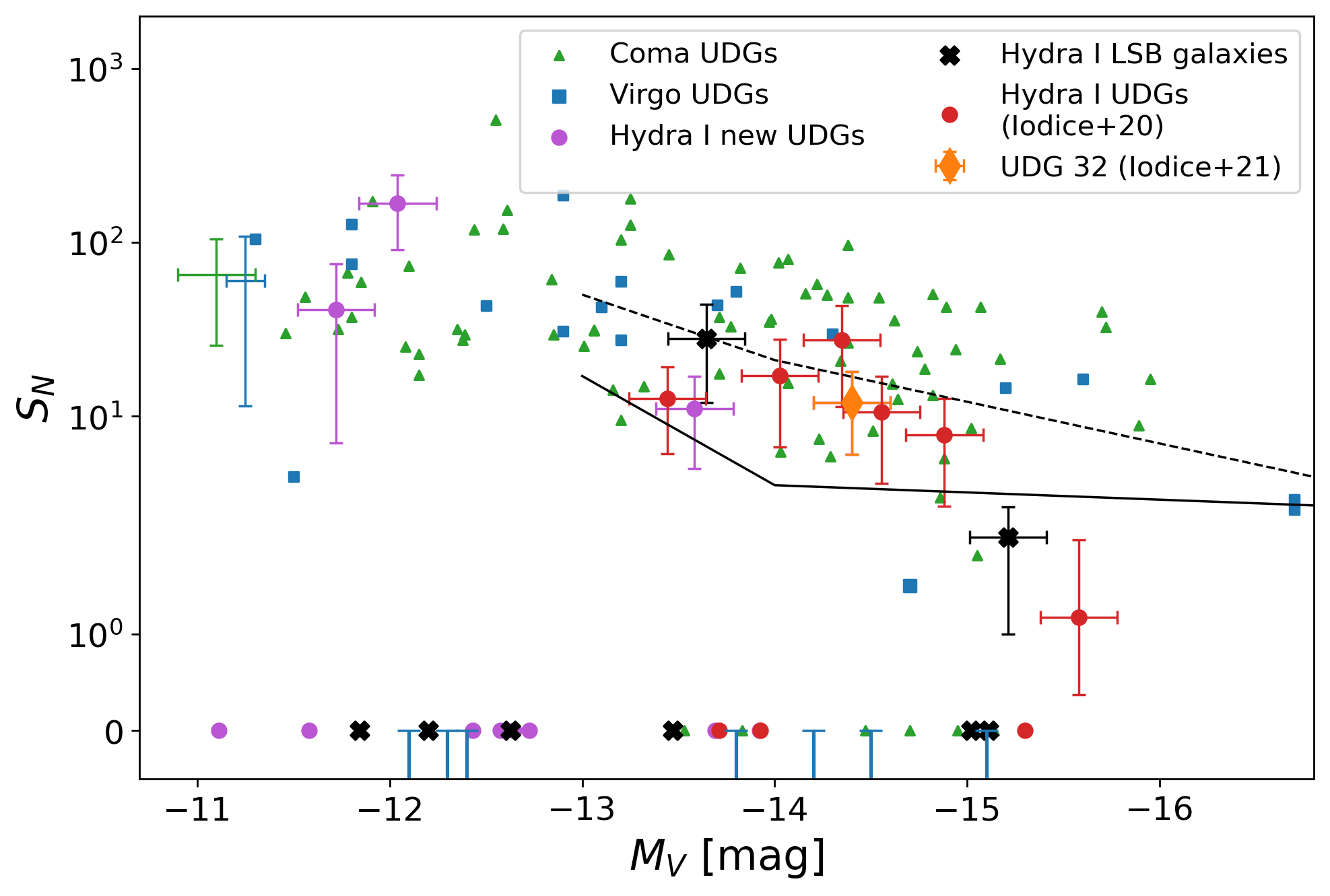}
    \caption{
    \textit{Top panel}. Number of GCs as a function of the host galaxy stellar mass.
    Filled purple circles are the new UDGs in our sample, while filled black crosses are Hydra I new LSB galaxies.
    Filled red circles are Hydra I UDGs from \citet[][]{Iodice2020c}, and the orange thin diamond is UDG32 \citep[][]{Iodice2021}.
    High surface brightness and LSB galaxies from other low- and high-density environments are plotted as comparison. 
    We show the Fornax and Virgo early-type galaxies \citep[ETGs,][respectively]{Liu2019Fornax,Peng2008virgo}, the Fornax low mass ETGs \citep[][]{Prole2019b}, the Local Volume and Virgo low mass ETGs \citep[][]{Carlsten2022}, and the MATLAS-2019 UDG \citep[][]{Muller2021} (markers are indicated in the legend).
    \textit{Bottom panel}. GCs specific frequency $S_N$ versus $V$-band absolute magnitude for UDGs and new LSB 
    galaxies in the Hydra I cluster \citep[symbols as top panel,][ and this work]{Iodice2020c}, compared to the UDGs in the Coma cluster \citep[green triangles,][]{Forbes2020a}, and to the UDGs in the Virgo cluster \citep[blue squares,][]{Lim2020}. 
    Virgo UDGs with a negative specific frequency are reported at $S_N=0$.
    In all three cases GCs are selected within 1.5$R_e$ and are locally corrected for background contamination. 
    The solid line shows the mean locus of dwarf galaxies, the dashed line represents the upper 2$\sigma$ bound \citep[see also Fig.4 in][]{Lim2018}. 
	The average uncertainties on Coma and Virgo are shown on the left side. 
	}
    \label{fig:SN_GC}
\end{figure*}

In Figure \ref{fig:SN_GC} (top panel) we show the total number of GCs, $N_{GC}$, as a function of the host galaxy stellar mass, for the \object{Hydra\,I}  LSB and UDG combined sample  \citep[both this work and][]{Iodice2020c}.
These are compared with galaxies in the Fornax and Virgo clusters, as well as Local Volume satellites \citep[][]{Liu2019Fornax,Prole2019b,Peng2008virgo,Carlsten2022}, and the UDG MATLAS-2019 \citep[][]{Muller2021}, covering a wide range of $M_*$ {  ($\sim10^6M_{\odot}<M_*<\sim 10^{11}M_{\odot}$)}.
The stellar masses in all these works are estimated adopting models based on a Chabrier initial mass function \citep[IMF,][]{Chabrier2003}.
The \citet[][]{Into2013} model we adopted, uses a Kroupa IMF \citep[][]{Kroupa2001}.
However, the properties obtained using a Chabrier IMF are very similar to those obtained using a Kroupa IMF \citep[see Fig. 4 in][]{Bruzual2003}. Therefore the assumption on IMF does not affect the results shown in Figure \ref{fig:SN_GC}.

The GCs have been selected in apertures comparable to our choice of 1.5$R_e$ by \citet[][]{Carlsten2022} ($\leq$ 1.5$R_e$), \citet[][]{Muller2021} ($\leq$ 1.75$R_e$). 
\citet[][]{Prole2019b} do not put any prior on the aperture size, but they find that the typical extension of GC systems is $1.73\pm0.27\;R_e$, comparable to 1.5$R_e$.
\citet[][]{Peng2008virgo,Liu2019Fornax} adopt larger apertures, but their analysis is focused on much brighter galaxies.

Several low-mass galaxies in literature have $N_{GC}$ consistent with zero.
Generally, the population of GCs increases with galaxy mass, with a larger scatter of the $N_{GC}\;vs\;M_*$ relation in the low mass region.
At visual inspection, \object{Hydra\,I} UDGs and LSB galaxies with $N_{GC}>0$ do not reveal substantial differences with respect to the galaxies in other environments.

For each galaxy of our new sample, $N_{GC}(\leq1.5R_e$, local corr.) is used to derive the GC specific frequency, $S_N = N_{GC} 10^{0.4 [M_v+15]}$ \citep[][]{Harris1981Sn}. 
Results are listed in Tab. \ref{tab:GC_1.5}.
We study the $S_N$ as a function of the general properties of the UDGs.
As shown in the bottom panel of Fig.~\ref{fig:SN_GC}, where $S_N$ is plotted vs. $M_V$, the values of $S_N$ are consistent with those obtained for the previous sample of UDGs in Hydra I \citep[$S_N \sim 5-20$, see][]{Iodice2020c} and with the average values obtained for dwarf galaxies \citep[][]{Lim2018}. 
Most of the UDGs and LSB galaxies in Hydra I have a $S_N$ consistent with that
typical for dwarf galaxies of similar magnitude.

Almost half of the Hydra I UDGs (10 out of 21) have specific frequency larger than 2;
in particular there are four individuals with $S_N\geq20$.
However, in all cases, the 1$\sigma$ errors have the same order of magnitude of the $S_N$ measurements, which {\bf do not} 
allow us to draw any strong conclusion.

Hydra I UDGs and LSB galaxies seem to have $S_N$ values comparable with Coma and Virgo UDGs \citep[][both works select GCs within an aperture of 1.5$R_e$, using a local background correction]{Forbes2020a,Lim2020}.
However, Hydra I might have a much larger fraction of UDGs with no GC over-density.
Nevertheless, this result needs new observational data, either spectroscopy or IR photometry, to be confirmed.

To check whether the $S_N$ distribution of UDGs for the three clusters are similar, we perform a Kolmogorov-Smirnov (KS) test \citep[][]{hodges1958significance}, including the zero values. 
The \textit{p-value} of the KS test gives the probability that two distributions are drawn from the same parent distribution. 
The three \textit{p-values} are:
\begin{itemize}
    \item \textit{Hydra I vs. Coma}: $p=1.1\times10^{-8}$;
    \item \textit{Hydra I vs. Virgo}: $p=0.008$;
    \item \textit{Coma vs. Virgo}: $p=0.03$.
\end{itemize}
Hence, the KS-tests indicate that the three $S_N$ distributions are statistically different from each other. 
This result is substantially biased by the large number of UDGs with zero $N_{GC}$ in our sample. 
Such large fraction is likely due to the difficulties in detecting GC over-densities using $g$ and $r$ data, 
for UDGs hosting less than a handful of GCs.
By rejecting the UDG candidates with non-positive $N_{GC}$ from both our, Coma and Virgo cluster samples, the KS test shows that the \textit{p-values} increase:
\begin{itemize}
    \item \textit{Hydra I vs. Coma}: $p=0.09$;
    \item \textit{Hydra I vs. Virgo}: $p=0.07$;
    \item \textit{Coma vs. Virgo}: $p=0.63$.
\end{itemize}
The \emph{p-values} increase in all cases.
Coma and Virgo $S_N$ distributions are statistically similar with a high degree of confidence, while for the Hydra I distribution the situation is less clear.
With the obtained \emph{p-values}, we cannot rule out the null hypothesis for which Hydra I $S_N$ distribution is drawn from the same parent distribution of Virgo and Coma.
However, these are still low to assert firmly that Hydra I $S_N$ distribution is statistically similar to Virgo and Coma ones.

\begin{figure*}
    \centering
    \includegraphics[width=0.9\textwidth]{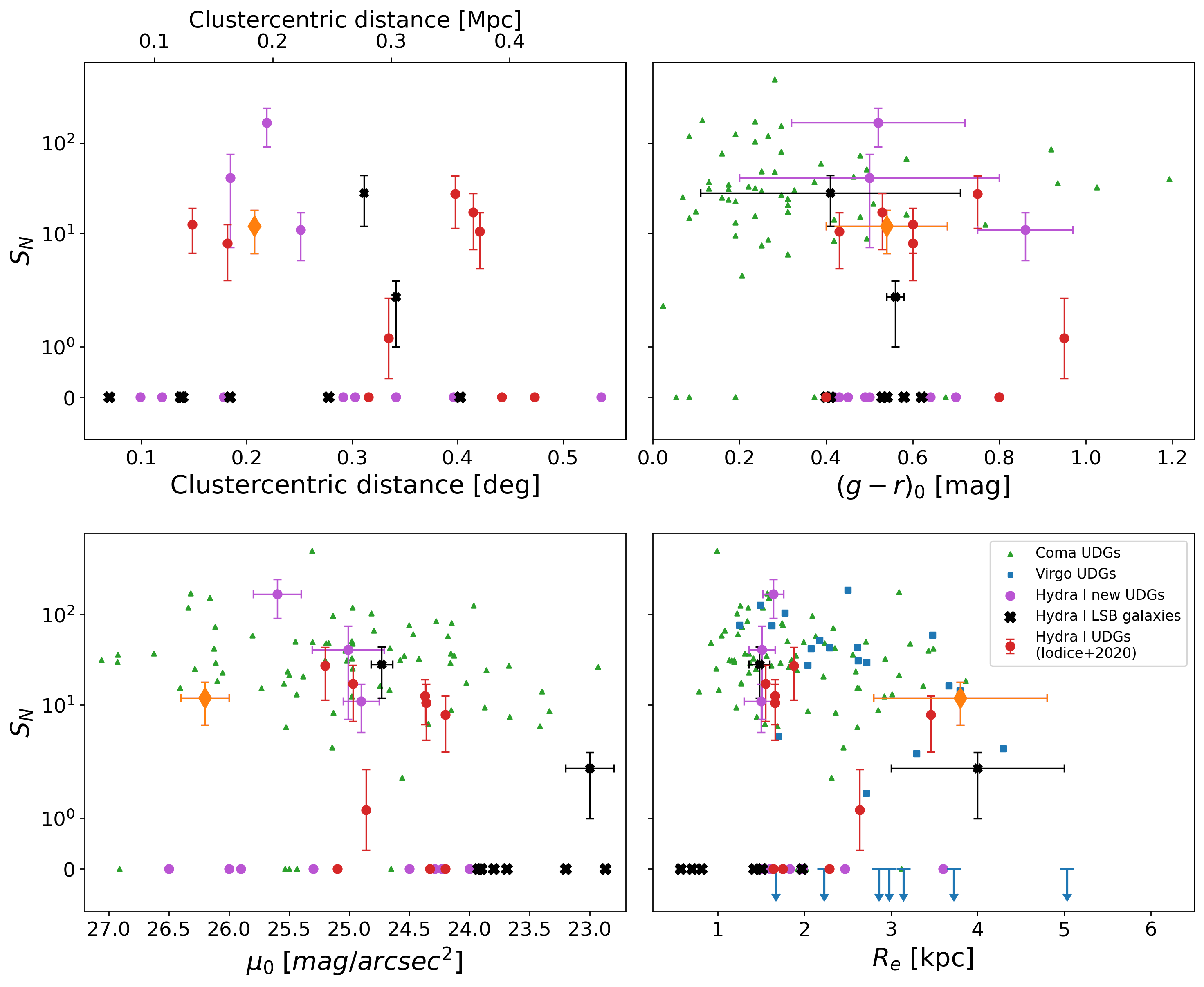}
    \caption{
    The GC specific frequency $S_N$ as a function of galaxies properties. Respectively, from left to right, and from top to bottom, $S_N$ as a function of the cluster-centric distance, $g{-}r$, $\mu_0$ and $R_e$.
    Symbols are the same as for bottom panel of Figure \ref{fig:SN_GC}.
    }
    \label{fig:SN_vs_param}
\end{figure*}

In Figure \ref{fig:SN_vs_param} we plot the $S_N$ versus the cluster-centric distance (upper left), the $g{-}r$ color (upper right), the central surface brightness and the effective radius (lower left and lower right panels, respectively). 
Where possible, we add the  Virgo and Coma cluster UDGs.

Considering the whole sample (including the galaxies with $S_N=0$), we cannot draw any obvious conclusion. Inspecting the galaxies with $S_N>0$, a trend seems to be present as a function of the effective radius and of the $\mu_0$. In particular, larger $S_N$ values are found at smaller $R_e$ and fainter central surface brightness. 
Conversely, there is no clear trend of $S_N$ as a function of the galaxy color and of the projected cluster-centric distance.
The latter finding is an indirect evidence of the reliability of GCs background subtraction: possible trends of $S_N$ against the cluster-centric distance may be evidence of spurious residual contamination from the ICGCs, generating false radial trends.
Finally, we observe that all LSB galaxies of the combined sample with $S_N>0$ (i.e. $N_{GC}>0$) lie at projected cluster-centric radii larger than 0.13 deg ($\sim 100 kpc$).

\section{Summary and Discussion: on the nature of the UDGs in the Hydra I cluster}\label{sec:disc}

In this paper we have presented a new catalog of 19 LSB galaxies detected in the Hydra~I cluster, based on the VST mosaic in the $g$- and $r$- bands.
Taking advantage of the large covered area and long integration times, these data allowed us to search for the LSB galaxies out to $\approx 0.4 R_{vir}$, therefore mapping the core of the cluster and the less dense regions at larger cluster-centric distances, as well.
The detection of LSB galaxies in this cluster has been carried out with the automatic search using {\sc SExtractor}, complemented by visual inspection, which contributed about half of the new detections. 

We found 11 new UDG candidates in this sample which, adding the 9 UDGs detected by \citet[][]{Iodice2020c} and the UDG reported by \citet[][]{Iodice2021}, 
led to a total number of 21 UDGs in the cluster, accordingly to the adopted UDG definition.
In addition, the remaining eight galaxies of the new sample are identified as {\it LSB dwarfs}.
Six out of eight are larger than the average dwarfs of similar luminosity inhabiting Hydra I, having $R_e\geq1.2$ kpc.

The 21 UDGs in \object{Hydra\,I} reach very faint levels of central surface  brightness (down to $\mu_{0,g} \sim 26.5$~mag/arcsec$^2$), and half of them are very diffuse objects, with effective radius from $\sim2$ to $\sim 4$~kpc. 
Their average $g-r$ color is in the range $0.4 \leq g-r \leq 0.9$~mag (see also Fig.~\ref{fig:corrRe}). 
The stellar mass estimates we derived vary from  $\sim 5\times  10^6$ to $8
1.2\times 10^8$~M$_{\odot}$. 
For all the newly detected Hydra\,I UDGs and LSB galaxies, we have searched for compact sources around them and estimated a total GCs population statistically larger than zero for five out of the 19 newly detected galaxies (after completeness and local contamination corrections).
Considering the whole Hydra I UDG population, 10 out of 21 UDGs have $N_{GC}>0$.

The available data allowed us to derive the projected distribution of UDGs inside the covered area of the cluster and to explore the structural properties as a function of the cluster-centric distances and how they compare with the population of dwarf and the newly detected  LSB galaxies in the cluster.
The main results are summarised below.

\begin{enumerate}

    \item The 2D projected distribution of UDGs and LSB galaxies seems to follow the distribution of dwarfs and giant galaxies. 
    It appears strongly asymmetric with respect to the cluster centre, even though the presence of two bright stars may alter this result.
    Half of the UDGs (11 out of 21) are concentrated close to the cluster core and around a sub-group of galaxies in the North, which are the two densest regions of the cluster.
    The remaining half of the UDGs (10 objects) are found roughly uniformly distributed at larger cluster-centric distances.
    The LSB galaxies follow qualitatively the same projected distribution: 5 out of 8 are found in the two densest regions, while 3 are found at larger distances. 
    
    \item The average $g-r$ color of all UDGs in the sample peaks around 0.4-0.6 mag, comparable with the average colors of dwarf and LSB dwarf galaxies in the cluster.
    The color distribution is constant with the cluster-centric distance (see Fig.~\ref{fig:trends}).
    
    \item The $R_e$ is uniformly distributed with respect to the projected cluster-centric distance. 
    Conversely, there is a lack of very faint UDGs ($\mu_0\geq25$~mag/arcsec$^2$) toward smaller cluster-centric distances (i.e. $\leq0.2$ deg, or $\le180$ kpc), where the Sérsic index $n$ seems to slightly increase.
    This trend is not observed in dwarf galaxies, where on average $n\sim 1$ at all distances (see lower panel in Fig.~\ref{fig:trends}).
    
    \item A handful of our candidates  (\object{UDG\,15}, \object{UDG\,17}, \object{UDG\,18}, \object{UDG\,32}, \object{LSB\,4}, and \object{LSB\,6}) show a total number of GCs within 1.5 $R_e$ larger than zero.
    These are consistent with the $N_{GC}$ found in the other cluster members \citep[][]{Iodice2020c}.
    Considering only the Hydra\,I UDG candidates with $N_{GC}>0$, the $N_{GC}$ vs. $M_*$ relation agrees with what found in other galaxies in the local Universe.
    Moreover, for the same UDGs, the GC specific frequency is comparable with the values derived for UDGs in the Coma and Virgo clusters (Fig. \ref{fig:SN_GC}).

\end{enumerate}

We now discuss the properties and results of the UDGs in \object{Hydra\,I} cluster in the general framework of the nature and formation for this class of LSB galaxies. 
We aim to address {\it i)} how the non-uniform 2D projected spatial distribution of UDGs in the cluster links to different 
formation scenarios; {\it ii)} how UDGs properties compare with those of higher surface brightness dwarf galaxies and the newly detected  LSB galaxies in \object{Hydra\,I} cluster and with UDGs in other environments 
(groups and clusters), and, therefore, {\it iii)} whether different populations of UDGs  exist in this cluster. 

\subsection{UDGs distribution and their assembly history in Hydra~I}\label{formation}

We have found that most of the UDGs are concentrated in two subgroups, one close to the cluster core and the other in the North. 
A similar 2D projected distribution has been found for the dwarf galaxies in \object{Hydra\,I} (see Fig.~5 in Paper I), which peaks close to the cluster core, off-centre on the NW side to the region where the BCGs are located, and on the N-NW, respectively. 
As pointed out in Sect. \ref{sec:2D_distr}, the asymmetries may be affected also by the two bright stars in the field and, therefore, may not reflect completely the true bi-dimensional galaxy distribution. 

All these regions are dominated by the light from the brightest cluster members. 
These are the two densest regions of the cluster, where there are evident signs of galaxy interactions and presence of intra-cluster diffuse light.
In a forthcoming paper, we will show clear signs of interactions and presence of intra-cluster LSB features inside the different sub-groups of the cluster (Iodice et al. in preparation). 
As pointed out in Paper I, the presence of several substructures in the \object{Hydra\,I} cluster that emerged from the 2D distribution of dwarfs and UDGs is consistent with the previous findings by \cite{Lima-dias2021}, suggesting that this environment is still in an active assembly phase, where several sub-groups of galaxies are merging into the cluster potential. In the core of the cluster several studies reported ongoing interactions and mass assembly around NGC 3311 \citep[see][and references therein]{Arnaboldi2012,Hilker2018,Barbosa2018}. 
The spatial association of UDGs with the location of dwarf and giant galaxies in the cluster is consistent with findings for other galaxy clusters, where over-densities of UDGs are found close to sub-groups of galaxies \citep[see e.g.][]{Janssens2019}. 
This would suggest that UDGs may be associated with groups infalling into the cluster, being a further evidence that UDGs might join the cluster environment also via accretion of sub-groups, differently from the native cluster UDGs, with large DM halos, as proposed by \citet[][]{Sales2020}.

\subsection{On the population of UDGs in Hydra~I compared to other galaxy environments}\label{pop}

\begin{figure*}
    \centering
    \includegraphics[width=0.47\textwidth]{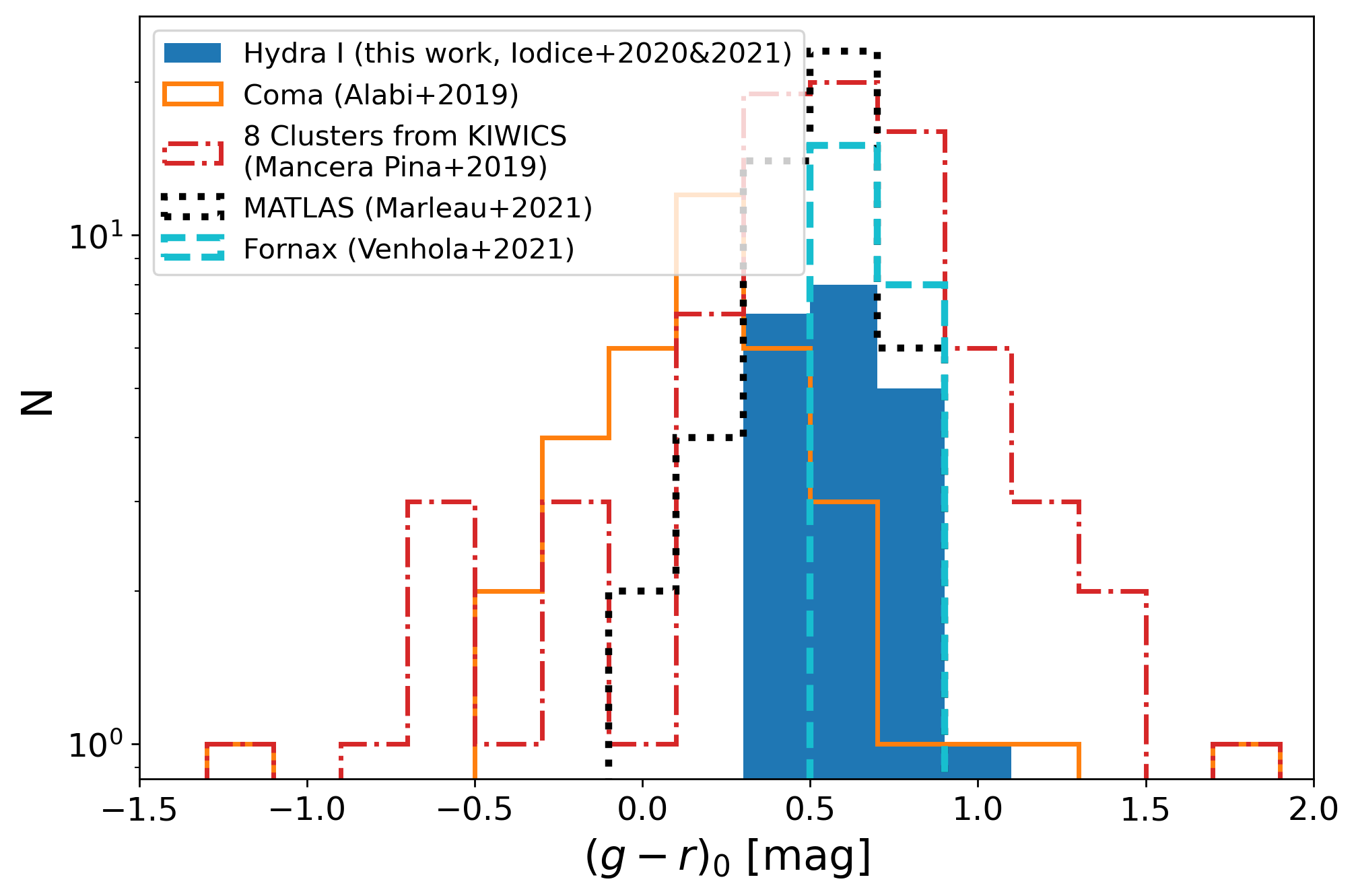}
    \includegraphics[width=0.47\textwidth]{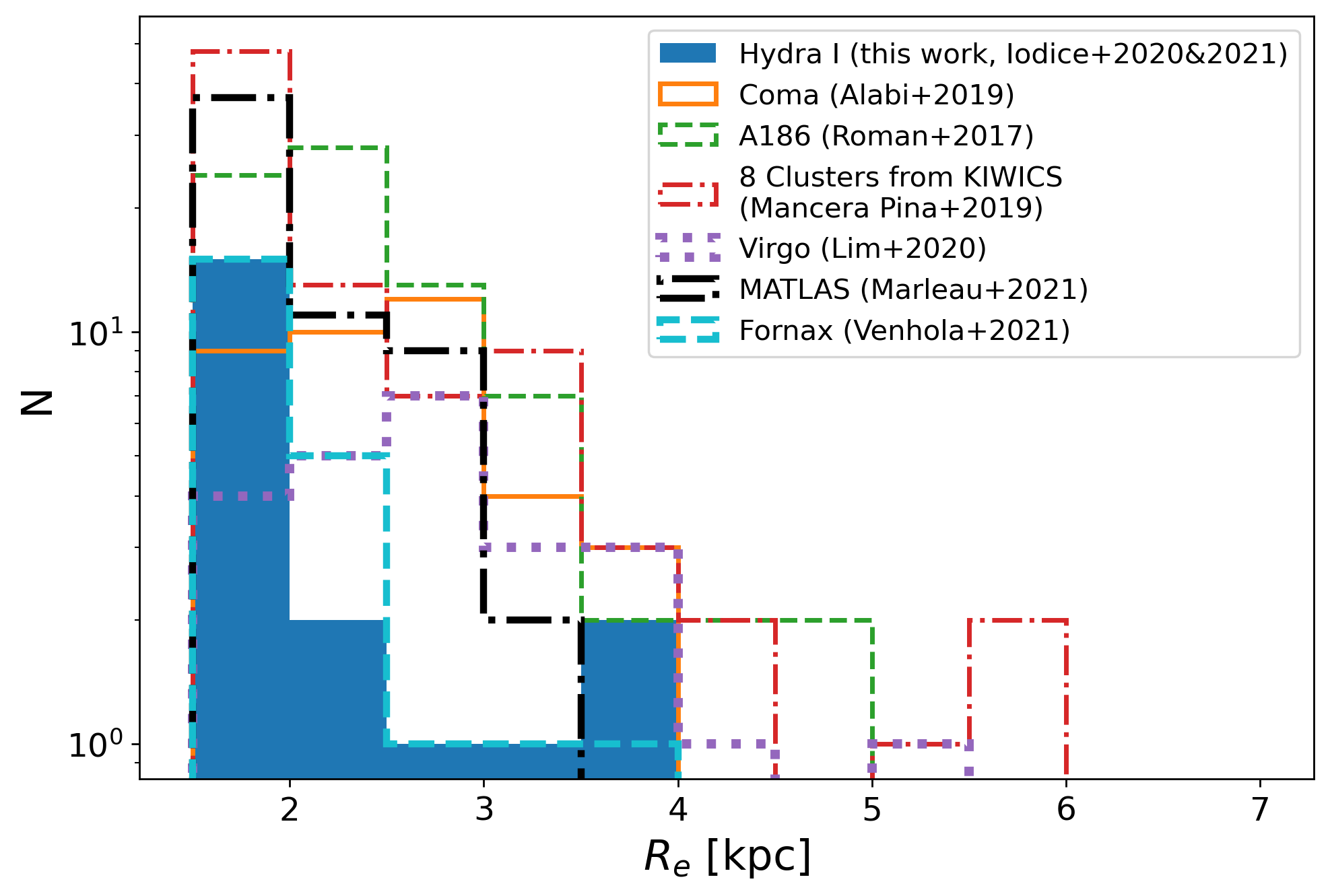}
    \includegraphics[width=0.47\textwidth]{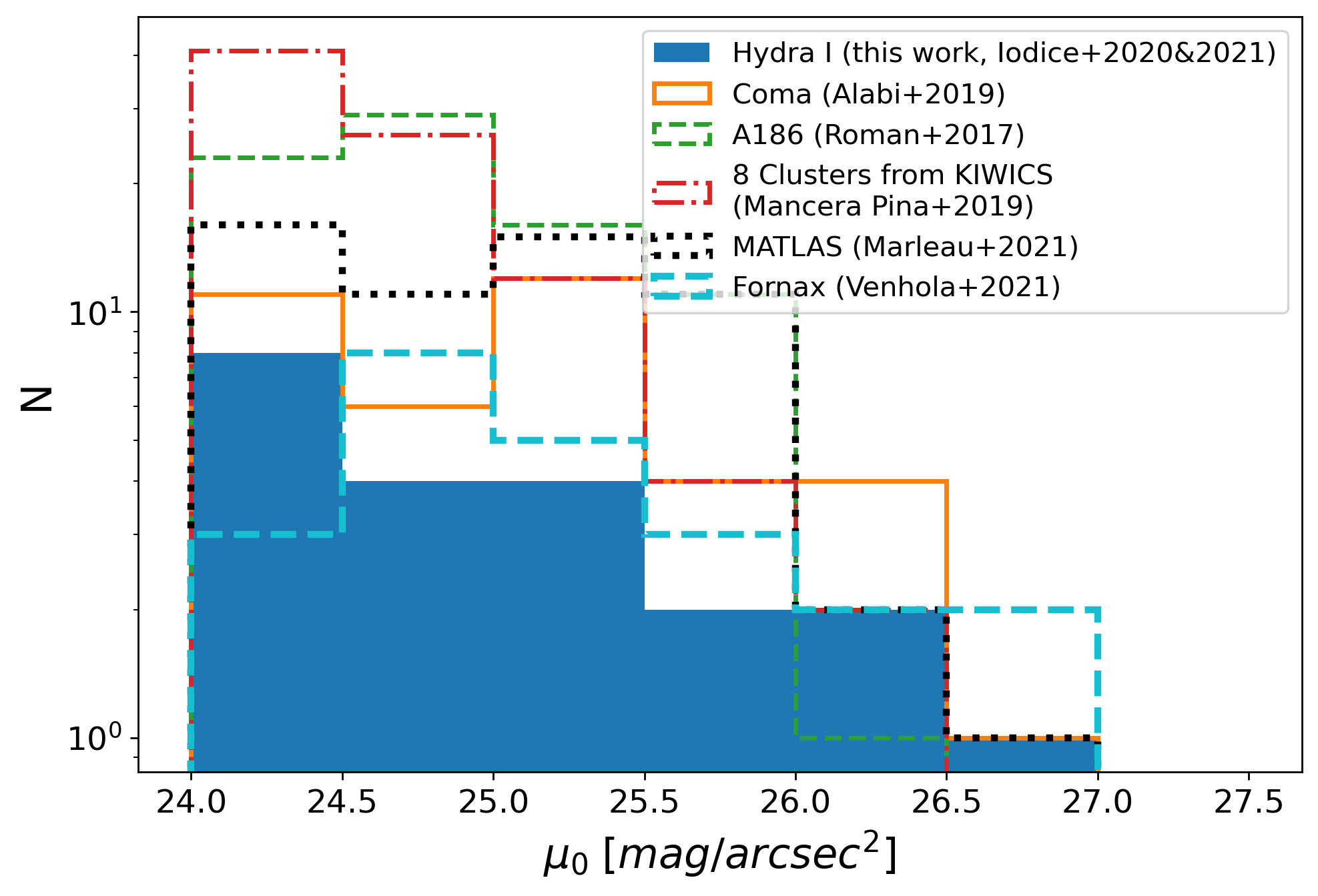}
    \includegraphics[width=0.47\textwidth]{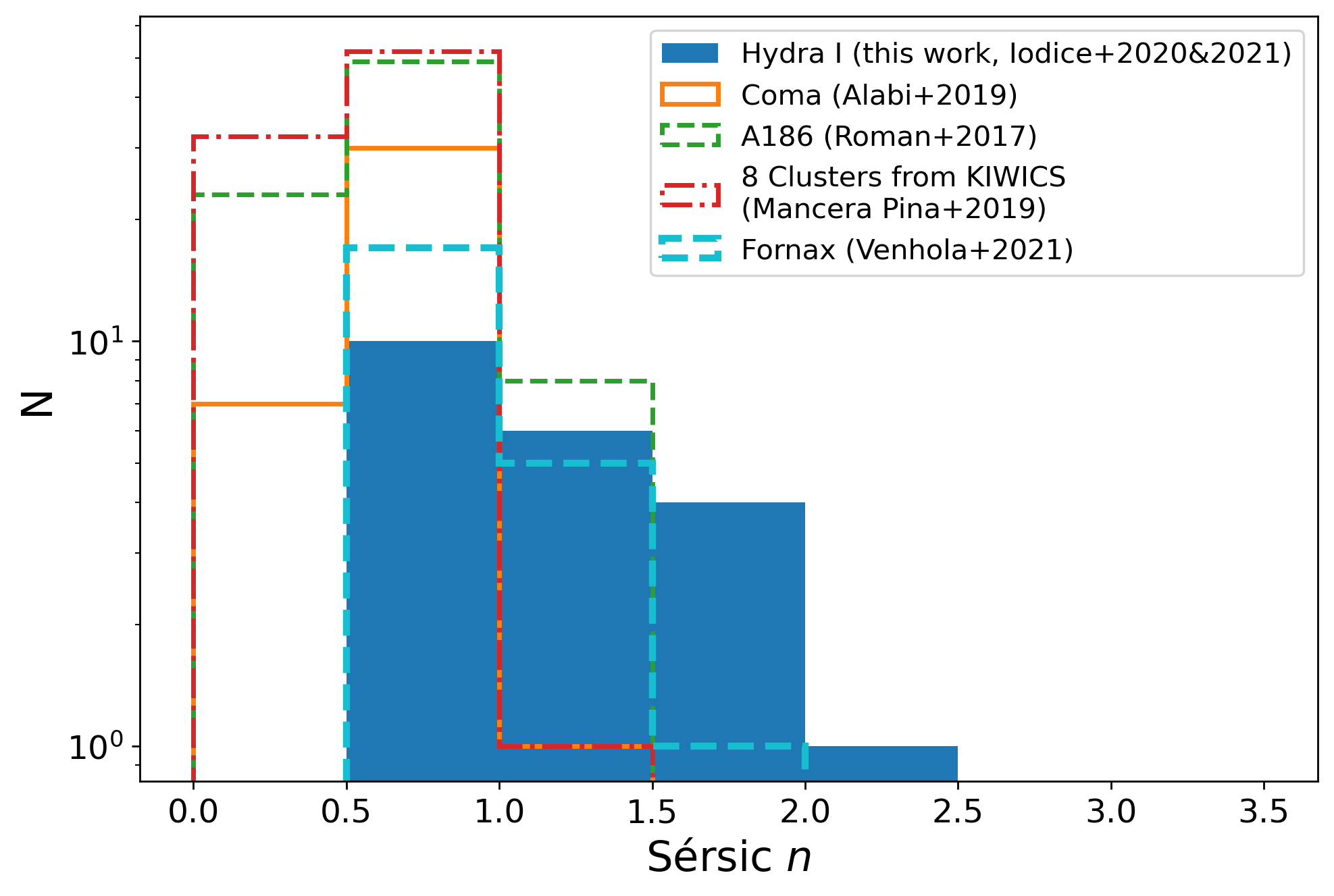}
    \caption{
    Distribution of the structural parameters of UDGs in the \object{Hydra\,I} cluster (filled blue histogram), compared to UDGs in other environments, as listed in the legend of each panel.
    For each sample we restricted the analysis only to those UDGs which strictly respect the \citet[][]{vanDokkum2015} definition.
    Optical $g-r$ color (upper-left panel), effective radius $R_e$ (upper-right), central surface brightness $\mu_0$ (bottom-right), and Sérsic index $n$ (bottom-right) are compared.
    References are given in the legend of each panel.
    }
    \label{fig:param_distr}
\end{figure*}

Based on their observed color distribution, it seems that two populations of UDGs exist: the red and quenched UDGs found primarily in clusters of galaxies, but also recently in low density environments \citep{Marleau2021matlas}, and a blue population of UDGs, which are mostly found in low density regions \citep[e.g.][]{Leisman2017,Roman2017a,Marleau2021matlas}.
In Fig. \ref{fig:param_distr} we compare the \object{Hydra\,I} UDG properties distributions with the UDGs from Fornax, Coma, Virgo, from the MATLAS survey, and from other 8 nearby galaxy clusters \citep[][]{Venhola2021,Alabi2020,Lim2020,Marleau2021matlas,Mancera-pina2019udg}.
For each sample, we restrict the comparison only to those UDGs that rigorously respect the definition limits $R_e\ge 1.5$ kpc and $\mu_0\ge 24$ mag/arcsec$^2$.
The upper-left panel shows a comparison with the color distributions. 
Inside $\approx 0.4 R_{vir}$ of the \object{Hydra\,I} cluster the UDGs are uniformly distributed in colors, having colors consistent with the red and quenched UDGs.
It is worth noting that this result is not a selection effect, since we did not detect blue UDG candidates that were excluded from the sample based on CMR.

\object{Hydra\,I} UDGs color suggest that, at least in this studied portion of the cluster, a single population of UDGs is found.
Based on the available data, we cannot exclude that bluer UDGs might exist in the outskirts of the cluster.
Since this population of UDGs has colors similar to the red dwarf galaxies in Hydra\,I, comparable stellar masses and share the same 2D projected distribution inside the cluster, we might consider them as the extreme LSB tail of the size-luminosity distribution of dwarfs in this environment. 
This result might be further supported once the colors, stellar masses and structural parameters are compared with those observed for the LSB galaxies also found in this work. 
These objects, which could reasonably be the largest and diffuse dwarf galaxies in the cluster, have colors and stellar masses similar to the UDGs in the sample, representing the physical link between dwarfs and UDGs.

We also report that the $g{-}r$ color stays constant as a function of the projected cluster-centric distance. 
By contrast, \citet[][]{Alabi2020} show that Coma LSB galaxies (and UDGs) have a color dependence on the cluster-centric distance. 
Coma LSB galaxies have redder average color within the cluster core, relative to the red sequence, suggesting that LSB galaxies may be most affected by the star formation quenching effects of the cluster-core environment. 
A similar color vs. cluster-centric trend is observed also by \citet[][]{Kadowaki2021}.
However, we point out that our data only cover the innermost portion of the Hydra\,I cluster ($\approx0.4R_{vir}$), so we cannot exclude that such a trend would appear if the population would be followed further. 

In the other panels of Fig. \ref{fig:param_distr} structural parameters of UDGs ($R_e$, $\mu_0$, $n$) from different clusters and less dense environments are compared \citep[][]{Venhola2021,Alabi2020,Mancera-pina2019udg,Roman2017a,Lim2020,Marleau2021matlas}.
A general agreement is observed for the effective radius and the central surface brightness distributions. 
Nevertheless, a relatively lower number of UDGs with $2.5\leq R_e \leq 3.5$ kpc is found in the studied portion of Hydra~I.
Sérsic index distributions have all similar shapes and average about $n=1$, suggesting that UDGs in clusters are better described by exponential profiles.
However, \object{Hydra\,I} $n$ distribution shows a larger number of UDGs with Sérsic index higher than 1.5.

Inside the whole sample of UDGs and LSB galaxies, the analysis of GCs suggests that a small yet non-negligible GC population is found in these galaxies, with $N_{GC}\geq 2$ within 1.5 $R_e$ for a bunch of galaxies (see Tab.~\ref{tab:UDG_GC}).
Given the poor statistics, new observational data (either spectroscopy or near-IR photometry) is needed to constrain the bulk of the GCs population and, therefore, to address the existence of a different class of UDGs in Hydra~I, where large $S_N$ might indicate a large halo mass ($>10^{10}$~M$_{\odot}$), as found in the Coma cluster \citep[][]{Burkert2020}.

\section{Conclusions}\label{sec:concl}

In this work, we studied the main photometric properties for 19 newly detected LSB galaxies in the \object{Hydra\,I} cluster.
Adopting the UDG definition criteria given by \citet[][]{vanDokkum2015}, 11 of the newly detected galaxies are UDG candidates, as well as other 9 UDGs reported in \citet[][]{Iodice2020c} and the UDG in the tail of NGC3314 discovered by \citet[][]{Iodice2021}.
Therefore, the total number of UDG candidates in the central region of the Hydra I cluster ($\lesssim0.4\;R_{vir}$) is 21.
The comparison with the high-surface brightness and LSB dwarf galaxies in this region of the cluster 
suggests that UDGs in this environment have similar properties.
Moreover, by comparing the sample with the analogous galaxies in other environments, we observed a general agreement for $R_e$, $\mu_0$ and $n$, while a lack of blue UDGs ($g{-}r<0.3$) is reported within the \object{Hydra\,I} cluster, out to $\approx0.4\;R_{vir}$, with respect to the other UDG populations (see Fig. \ref{fig:param_distr}). 

We analysed the GCs in each UDG.
The analysis of the GC systems is hampered by the availability of only two optical passbands close in terms of central wavelength, $g$ and $r$.
Nevertheless, the advantage of the large survey area, allowed us to highlight the presence of local GC over-densities by means of statistical background decontamination.
A handful of our candidates show a total number of GCs, $N_{GC}$, statistically larger than zero. 
However, to have more robust constraints on the GC systems in our UDG candidates, new spectroscopic or near-IR photometric data are required to further reduce GCs contamination and possibly identify more UDGs with a non-zero GC population.

Thanks to the deep and wide VEGAS coverage, the \object{Hydra\,I} cluster offers a rare opportunity to build a complete sample of UDGs and analyse this class of LSB galaxies across the cluster density.
This allows us to put constraints on the nature and formation of UDGs, also considering the wide range of parameters covered by the sample, and the existence of UDGs with GC specific frequency very different from each other. 
To further investigate this particular class of LSB galaxies, spectroscopic follow-up observations for the whole sample of UDGs in \object{Hydra\,I} will be soon available from the large programme recently approved in ESO period 108 (Run ID:108.222P), entitled {\it Looking into the faintEst WIth muSe (LEWIS): on the nature of ultra-diffuse galaxies in the Hydra-I cluster}.

\begin{acknowledgements}
We thank the anonymous referee for his/her comments and constructive suggestions.
This work is based on visitor mode observations collected at the European Southern Observatory (ESO) La Silla Paranal Observatory within the VST Guaranteed Time Observations, Programme ID: 099.B-0560(A).
ALM acknowledges financial support from the INAF-OAC funds. Authors acknowledge financial support from the VST INAF funds.
MC acknowledges financial support from MIUR (PRIN 2017 grant 20179ZF5KS).
CS is supported by an `Hintze Fellow' at the Oxford Centre for Astrophysical Surveys, which is funded through generous support from the Hintze Family Charitable Foundation. 
GD acknowledges support from FONDECYT REGULAR 1200495, and ANID project Basal FB-210003.
ALM wishes to thank Pavel E. Mancera Piña for providing the galaxy catalogs from his works.
This research made use of Astropy,\footnote{\url{http://www.astropy.org}} a community-developed core Python package for Astronomy \citep{astropy:2013, astropy:2018}.

\end{acknowledgements}

 \bibliographystyle{aa.bst}
  \bibliography{Hydra}

\begin{thebibliography}{115}
\expandafter\ifx\csname natexlab\endcsname\relax\def\natexlab#1{#1}\fi

\bibitem[{{Alabi} {et~al.}(2020){Alabi}, {Romanowsky}, {Forbes}, {Brodie}, \&
  {Okabe}}]{Alabi2020}
{Alabi}, A.~B., {Romanowsky}, A.~J., {Forbes}, D.~A., {Brodie}, J.~P., \&
  {Okabe}, N. 2020, \mnras, 496, 3182

\bibitem[{{Amorisco} \& {Loeb}(2016)}]{Amorisco2016}
{Amorisco}, N.~C. \& {Loeb}, A. 2016, \mnras, 459, L51

\bibitem[{{Arnaboldi} {et~al.}(2012){Arnaboldi}, {Ventimiglia}, {Iodice},
  {Gerhard}, \& {Coccato}}]{Arnaboldi2012}
{Arnaboldi}, M., {Ventimiglia}, G., {Iodice}, E., {Gerhard}, O., \& {Coccato},
  L. 2012, \aap, 545, A37

\bibitem[{{Astropy Collaboration} {et~al.}(2018){Astropy Collaboration},
  {Price-Whelan}, {Sip{\H{o}}cz}, {G{\"u}nther}, {Lim}, {Crawford}, {Conseil},
  {Shupe}, {Craig}, {Dencheva}, {Ginsburg}, {Vand erPlas}, {Bradley},
  {P{\'e}rez-Su{\'a}rez}, {de Val-Borro}, {Aldcroft}, {Cruz}, {Robitaille},
  {Tollerud}, {Ardelean}, {Babej}, {Bach}, {Bachetti}, {Bakanov}, {Bamford},
  {Barentsen}, {Barmby}, {Baumbach}, {Berry}, {Biscani}, {Boquien}, {Bostroem},
  {Bouma}, {Brammer}, {Bray}, {Breytenbach}, {Buddelmeijer}, {Burke},
  {Calderone}, {Cano Rodr{\'\i}guez}, {Cara}, {Cardoso}, {Cheedella}, {Copin},
  {Corrales}, {Crichton}, {D'Avella}, {Deil}, {Depagne}, {Dietrich}, {Donath},
  {Droettboom}, {Earl}, {Erben}, {Fabbro}, {Ferreira}, {Finethy}, {Fox},
  {Garrison}, {Gibbons}, {Goldstein}, {Gommers}, {Greco}, {Greenfield},
  {Groener}, {Grollier}, {Hagen}, {Hirst}, {Homeier}, {Horton}, {Hosseinzadeh},
  {Hu}, {Hunkeler}, {Ivezi{\'c}}, {Jain}, {Jenness}, {Kanarek}, {Kendrew},
  {Kern}, {Kerzendorf}, {Khvalko}, {King}, {Kirkby}, {Kulkarni}, {Kumar},
  {Lee}, {Lenz}, {Littlefair}, {Ma}, {Macleod}, {Mastropietro}, {McCully},
  {Montagnac}, {Morris}, {Mueller}, {Mumford}, {Muna}, {Murphy}, {Nelson},
  {Nguyen}, {Ninan}, {N{\"o}the}, {Ogaz}, {Oh}, {Parejko}, {Parley}, {Pascual},
  {Patil}, {Patil}, {Plunkett}, {Prochaska}, {Rastogi}, {Reddy Janga},
  {Sabater}, {Sakurikar}, {Seifert}, {Sherbert}, {Sherwood-Taylor}, {Shih},
  {Sick}, {Silbiger}, {Singanamalla}, {Singer}, {Sladen}, {Sooley},
  {Sornarajah}, {Streicher}, {Teuben}, {Thomas}, {Tremblay}, {Turner},
  {Terr{\'o}n}, {van Kerkwijk}, {de la Vega}, {Watkins}, {Weaver}, {Whitmore},
  {Woillez}, {Zabalza}, \& {Astropy Contributors}}]{astropy:2018}
{Astropy Collaboration}, {Price-Whelan}, A.~M., {Sip{\H{o}}cz}, B.~M., {et~al.}
  2018, \aj, 156, 123

\bibitem[{{Astropy Collaboration} {et~al.}(2013){Astropy Collaboration},
  {Robitaille}, {Tollerud}, {Greenfield}, {Droettboom}, {Bray}, {Aldcroft},
  {Davis}, {Ginsburg}, {Price-Whelan}, {Kerzendorf}, {Conley}, {Crighton},
  {Barbary}, {Muna}, {Ferguson}, {Grollier}, {Parikh}, {Nair}, {Unther},
  {Deil}, {Woillez}, {Conseil}, {Kramer}, {Turner}, {Singer}, {Fox}, {Weaver},
  {Zabalza}, {Edwards}, {Azalee Bostroem}, {Burke}, {Casey}, {Crawford},
  {Dencheva}, {Ely}, {Jenness}, {Labrie}, {Lim}, {Pierfederici}, {Pontzen},
  {Ptak}, {Refsdal}, {Servillat}, \& {Streicher}}]{astropy:2013}
{Astropy Collaboration}, {Robitaille}, T.~P., {Tollerud}, E.~J., {et~al.} 2013,
  \aap, 558, A33

\bibitem[{{Barbosa} {et~al.}(2018){Barbosa}, {Arnaboldi}, {Coccato}, {Gerhard},
  {Mendes de Oliveira}, {Hilker}, \& {Richtler}}]{Barbosa2018}
{Barbosa}, C.~E., {Arnaboldi}, M., {Coccato}, L., {et~al.} 2018, \aap, 609, A78

\bibitem[{{Benavides} {et~al.}(2021){Benavides}, {Sales}, {Abadi}, {Pillepich},
  {Nelson}, {Marinacci}, {Cooper}, {Pakmor}, {Torrey}, {Vogelsberger}, \&
  {Hernquist}}]{Benavides2021}
{Benavides}, J.~A., {Sales}, L.~V., {Abadi}, M.~G., {et~al.} 2021, arXiv
  e-prints, arXiv:2109.01677

\bibitem[{{Bennet} {et~al.}(2018){Bennet}, {Sand}, {Zaritsky}, {Crnojevi{\'c}},
  {Spekkens}, \& {Karunakaran}}]{Bennet2018}
{Bennet}, P., {Sand}, D.~J., {Zaritsky}, D., {et~al.} 2018, \apjl, 866, L11

\bibitem[{{Bertin} \& {Arnouts}(1996)}]{Bertin1996}
{Bertin}, E. \& {Arnouts}, S. 1996, \aaps, 117, 393

\bibitem[{{Binggeli} \& {Cameron}(1991)}]{binggeli1991}
{Binggeli}, B. \& {Cameron}, L.~M. 1991, \aap, 252, 27

\bibitem[{Bradley {et~al.}(2020)Bradley, Sip{\H o}cz, Robitaille, Tollerud,
  Vin{\'{\i}}cius, Deil, Barbary, Wilson, Busko, G{\"u}nther, Cara, Conseil,
  Bostroem, Droettboom, Bray, Bratholm, Lim, Barentsen, Craig, Pascual, Perren,
  Greco, Donath, de~Val-Borro, Kerzendorf, Bach, Weaver, D'Eugenio, Souchereau,
  \& Ferreira}]{larry_bradley_2020_4044744}
Bradley, L., Sip{\H o}cz, B., Robitaille, T., {et~al.} 2020, astropy/photutils:
  1.0.0

\bibitem[{{Bruzual} \& {Charlot}(2003)}]{Bruzual2003}
{Bruzual}, G. \& {Charlot}, S. 2003, \mnras, 344, 1000

\bibitem[{{Burkert} \& {Forbes}(2020)}]{Burkert2020}
{Burkert}, A. \& {Forbes}, D.~A. 2020, \aj, 159, 56

\bibitem[{{Cantiello} {et~al.}(2018{\natexlab{a}}){Cantiello}, {D'Abrusco},
  {Spavone}, {Paolillo}, {Capaccioli}, {Limatola}, {Grado}, {Iodice},
  {Raimondo}, {Napolitano}, {Blakeslee}, {Brocato}, {Forbes}, {Hilker},
  {Mieske}, {Peletier}, {van de Ven}, \& {Schipani}}]{cantiello2018}
{Cantiello}, M., {D'Abrusco}, R., {Spavone}, M., {et~al.} 2018{\natexlab{a}},
  \aap, 611, A93

\bibitem[{{Cantiello} {et~al.}(2018{\natexlab{b}}){Cantiello}, {Grado},
  {Rejkuba}, {Arnaboldi}, {Capaccioli}, {Greggio}, {Iodice}, \&
  {Limatola}}]{Cantiello2018a}
{Cantiello}, M., {Grado}, A., {Rejkuba}, M., {et~al.} 2018{\natexlab{b}}, \aap,
  611, A21

\bibitem[{{Cantiello} {et~al.}(2020){Cantiello}, {Venhola}, {Grado},
  {Paolillo}, {D'Abrusco}, {Raimondo}, {Quintini}, {Hilker}, {Mieske},
  {Tortora}, {Spavone}, {Capaccioli}, {Iodice}, {Peletier}, {Falcon Barroso},
  {Limatola}, {Napolitano}, {Schipani}, {van de Ven}, {Gentile}, \&
  {Covone}}]{Cantiello2020}
{Cantiello}, M., {Venhola}, A., {Grado}, A., {et~al.} 2020, arXiv e-prints,
  arXiv:2005.12085

\bibitem[{{Carleton} {et~al.}(2019){Carleton}, {Errani}, {Cooper},
  {Kaplinghat}, {Pe{\~n}arrubia}, \& {Guo}}]{Carleton2019}
{Carleton}, T., {Errani}, R., {Cooper}, M., {et~al.} 2019, \mnras, 485, 382

\bibitem[{{Carleton} {et~al.}(2021){Carleton}, {Guo}, {Munshi}, {Tremmel}, \&
  {Wright}}]{Carleton2021}
{Carleton}, T., {Guo}, Y., {Munshi}, F., {Tremmel}, M., \& {Wright}, A. 2021,
  \mnras, 502, 398

\bibitem[{{Carlsten} {et~al.}(2022){Carlsten}, {Greene}, {Beaton}, \&
  {Greco}}]{Carlsten2022}
{Carlsten}, S.~G., {Greene}, J.~E., {Beaton}, R.~L., \& {Greco}, J.~P. 2022,
  \apj, 927, 44

\bibitem[{{Caso} {et~al.}(2019){Caso}, {De B{\'o}rtoli}, {Ennis}, \&
  {Bassino}}]{Caso2019}
{Caso}, J.~P., {De B{\'o}rtoli}, B.~J., {Ennis}, A.~I., \& {Bassino}, L.~P.
  2019, \mnras, 488, 4504

\bibitem[{{Chabrier}(2003)}]{Chabrier2003}
{Chabrier}, G. 2003, \pasp, 115, 763

\bibitem[{{Christlein} \& {Zabludoff}(2003)}]{Christlein2003}
{Christlein}, D. \& {Zabludoff}, A.~I. 2003, \apj, 591, 764

\bibitem[{{Chung} {et~al.}(2020){Chung}, {Yoon}, {Cho}, {Lee}, \&
  {Lee}}]{chung2020}
{Chung}, C., {Yoon}, S.-J., {Cho}, H., {Lee}, S.-Y., \& {Lee}, Y.-W. 2020,
  \apjs, 250, 33

\bibitem[{{Collins} {et~al.}(2021){Collins}, {Read}, {Ibata}, {Rich}, {Martin},
  {Pe{\~n}arrubia}, {Chapman}, {Tollerud}, \& {Weisz}}]{Collins2021}
{Collins}, M. L.~M., {Read}, J.~I., {Ibata}, R.~A., {et~al.} 2021, arXiv
  e-prints, arXiv:2102.11890

\bibitem[{{Conselice}(2018)}]{Conselice2018}
{Conselice}, C.~J. 2018, Research Notes of the American Astronomical Society,
  2, 43

\bibitem[{{Di Cintio} {et~al.}(2017){Di Cintio}, {Brook}, {Dutton},
  {Macci{\`o}}, {Obreja}, \& {Dekel}}]{diCintio2017}
{Di Cintio}, A., {Brook}, C.~B., {Dutton}, A.~A., {et~al.} 2017, \mnras, 466,
  L1

\bibitem[{{Duc} {et~al.}(2014){Duc}, {Paudel}, {McDermid}, {Cuillandre},
  {Serra}, {Bournaud}, {Cappellari}, \& {Emsellem}}]{Duc2014}
{Duc}, P.-A., {Paudel}, S., {McDermid}, R.~M., {et~al.} 2014, \mnras, 440, 1458

\bibitem[{{Einasto}(1965)}]{Einasto1965}
{Einasto}, J. 1965, Trudy Astrofizicheskogo Instituta Alma-Ata, 5, 87

\bibitem[{{Emsellem} {et~al.}(2019){Emsellem}, {van der Burg}, {Fensch},
  {Je{\v{r}}{\'a}bkov{\'a}}, {Zanella}, {Agnello}, {Hilker}, {M{\"u}ller},
  {Rejkuba}, {Duc}, {Durrell}, {Habas}, {Lelli}, {Lim}, {Marleau}, {Peng}, \&
  {S{\'a}nchez-Janssen}}]{Emsellem2019}
{Emsellem}, E., {van der Burg}, R. F.~J., {Fensch}, J., {et~al.} 2019, \aap,
  625, A76

\bibitem[{{Fensch} {et~al.}(2019){Fensch}, {van der Burg},
  {Je{\v{r}}{\'a}bkov{\'a}}, {Emsellem}, {Zanella}, {Agnello}, {Hilker},
  {M{\"u}ller}, {Rejkuba}, {Duc}, {Durrell}, {Habas}, {Lim}, {Marleau}, {Peng},
  \& {S{\'a}nchez Janssen}}]{Fensch2019}
{Fensch}, J., {van der Burg}, R. F.~J., {Je{\v{r}}{\'a}bkov{\'a}}, T., {et~al.}
  2019, \aap, 625, A77

\bibitem[{{Ferr{\'e}-Mateu} {et~al.}(2018){Ferr{\'e}-Mateu}, {Alabi}, {Forbes},
  {Romanowsky}, {Brodie}, {Pandya}, {Mart{\'\i}n-Navarro}, {Bellstedt},
  {Wasserman}, {Stone}, \& {Okabe}}]{Ferre-Mateu2018}
{Ferr{\'e}-Mateu}, A., {Alabi}, A., {Forbes}, D.~A., {et~al.} 2018, \mnras,
  479, 4891

\bibitem[{{Forbes}(2017)}]{Forbes2017}
{Forbes}, D.~A. 2017, \mnras, 472, L104

\bibitem[{{Forbes} {et~al.}(2020{\natexlab{a}}){Forbes}, {Alabi}, {Romanowsky},
  {Brodie}, \& {Arimoto}}]{Forbes2020a}
{Forbes}, D.~A., {Alabi}, A., {Romanowsky}, A.~J., {Brodie}, J.~P., \&
  {Arimoto}, N. 2020{\natexlab{a}}, \mnras, 492, 4874

\bibitem[{{Forbes} {et~al.}(2020{\natexlab{b}}){Forbes}, {Dullo}, {Gannon},
  {Couch}, {Iodice}, {Spavone}, {Cantiello}, \& {Schipani}}]{Forbes2020b}
{Forbes}, D.~A., {Dullo}, B.~T., {Gannon}, J., {et~al.} 2020{\natexlab{b}},
  \mnras [\eprint[arXiv]{2004.10855}]

\bibitem[{{Forbes} {et~al.}(2019){Forbes}, {Gannon}, {Couch}, {Iodice},
  {Spavone}, {Cantiello}, {Napolitano}, \& {Schipani}}]{Forbes2019}
{Forbes}, D.~A., {Gannon}, J., {Couch}, W.~J., {et~al.} 2019, \aap, 626, A66

\bibitem[{{Forbes} {et~al.}(2021){Forbes}, {Gannon}, {Romanowsky}, {Alabi},
  {Brodie}, {Couch}, \& {Ferr{\'e}-Mateu}}]{Forbes2021}
{Forbes}, D.~A., {Gannon}, J.~S., {Romanowsky}, A.~J., {et~al.} 2021, \mnras,
  500, 1279

\bibitem[{{Gannon} {et~al.}(2021){Gannon}, {Dullo}, {Forbes}, {Rich},
  {Rom{\'a}n}, {Couch}, {Brodie}, {Ferr{\'e}-Mateu}, {Alabi}, \&
  {Mould}}]{Gannon2021}
{Gannon}, J.~S., {Dullo}, B.~T., {Forbes}, D.~A., {et~al.} 2021, \mnras, 502,
  3144

\bibitem[{{Georgiev} {et~al.}(2010){Georgiev}, {Puzia}, {Goudfrooij}, \&
  {Hilker}}]{Georgiev2010}
{Georgiev}, I.~Y., {Puzia}, T.~H., {Goudfrooij}, P., \& {Hilker}, M. 2010,
  \mnras, 406, 1967

\bibitem[{{Habas} {et~al.}(2020){Habas}, {Marleau}, {Duc}, {Durrell}, {Paudel},
  {Poulain}, {S{\'a}nchez-Janssen}, {Sreejith}, {Ramasawmy}, {Stemock},
  {Leach}, {Cuillandre}, {Gwyn}, {Agnello}, {B{\'\i}lek}, {Fensch},
  {M{\"u}ller}, {Peng}, \& {van der Burg}}]{Habas2020}
{Habas}, R., {Marleau}, F.~R., {Duc}, P.-A., {et~al.} 2020, \mnras, 491, 1901

\bibitem[{{Harris} \& {van den Bergh}(1981)}]{Harris1981Sn}
{Harris}, W.~E. \& {van den Bergh}, S. 1981, \aj, 86, 1627

\bibitem[{{Hilker} {et~al.}(2018){Hilker}, {Richtler}, {Barbosa}, {Arnaboldi},
  {Coccato}, \& {Mendes de Oliveira}}]{Hilker2018}
{Hilker}, M., {Richtler}, T., {Barbosa}, C.~E., {et~al.} 2018, \aap, 619, A70

\bibitem[{Hodges(1958)}]{hodges1958significance}
Hodges, J.~L. 1958, Arkiv f{\"o}r Matematik, 3, 469

\bibitem[{{Hudson} {et~al.}(2014){Hudson}, {Harris}, \& {Harris}}]{Hudson2014}
{Hudson}, M.~J., {Harris}, G.~L., \& {Harris}, W.~E. 2014, \apjl, 787, L5

\bibitem[{{Into} \& {Portinari}(2013)}]{Into2013}
{Into}, T. \& {Portinari}, L. 2013, \mnras, 430, 2715

\bibitem[{{Iodice} {et~al.}(2020){Iodice}, {Cantiello}, {Hilker}, {Rejkuba},
  {Arnaboldi}, {Spavone}, {Greggio}, {Forbes}, {D'Ago}, {Mieske}, {Spiniello},
  {La Marca}, {Rampazzo}, {Paolillo}, {Capaccioli}, \&
  {Schipani}}]{Iodice2020c}
{Iodice}, E., {Cantiello}, M., {Hilker}, M., {et~al.} 2020, \aap, 642, A48

\bibitem[{{Iodice} {et~al.}(2021){Iodice}, {La Marca}, {Hilker}, {Cantiello},
  {D'Ago}, {Gullieuszik}, {Rejkuba}, {Arnaboldi}, {Spavone}, {Spiniello},
  {Forbes}, {Greggio}, {Rampazzo}, {Mieske}, {Paolillo}, \&
  {Schipani}}]{Iodice2021}
{Iodice}, E., {La Marca}, A., {Hilker}, M., {et~al.} 2021, \aap, 652, L11

\bibitem[{{Janssens} {et~al.}(2017){Janssens}, {Abraham}, {Brodie}, {Forbes},
  {Romanowsky}, \& {van Dokkum}}]{Janssens2017}
{Janssens}, S., {Abraham}, R., {Brodie}, J., {et~al.} 2017, \apjl, 839, L17

\bibitem[{{Janssens} {et~al.}(2019){Janssens}, {Abraham}, {Brodie}, {Forbes},
  \& {Romanowsky}}]{Janssens2019}
{Janssens}, S.~R., {Abraham}, R., {Brodie}, J., {Forbes}, D.~A., \&
  {Romanowsky}, A.~J. 2019, \apj, 887, 92

\bibitem[{{Jones} {et~al.}(2021){Jones}, {Bennet}, {Mutlu-Pakdil}, {Sand},
  {Spekkens}, {Crnojevic}, {Karunakaran}, \& {Zaritsky}}]{Jones2021}
{Jones}, M.~G., {Bennet}, P., {Mutlu-Pakdil}, B., {et~al.} 2021, arXiv
  e-prints, arXiv:2104.12805

\bibitem[{{Kadowaki} {et~al.}(2021){Kadowaki}, {Zaritsky}, {Donnerstein}, {RS},
  {Karunakaran}, \& {Spekkens}}]{Kadowaki2021}
{Kadowaki}, J., {Zaritsky}, D., {Donnerstein}, R.~L., {et~al.} 2021, arXiv
  e-prints, arXiv:2110.00015

\bibitem[{{Kartha} {et~al.}(2014){Kartha}, {Forbes}, {Spitler}, {Romanowsky},
  {Arnold}, \& {Brodie}}]{Kartha2014}
{Kartha}, S.~S., {Forbes}, D.~A., {Spitler}, L.~R., {et~al.} 2014, \mnras, 437,
  273

\bibitem[{{Koch} {et~al.}(2012){Koch}, {Burkert}, {Rich}, {Collins}, {Black},
  {Hilker}, \& {Benson}}]{Koch2012}
{Koch}, A., {Burkert}, A., {Rich}, R.~M., {et~al.} 2012, \apjl, 755, L13

\bibitem[{{Koda} {et~al.}(2015){Koda}, {Yagi}, {Yamanoi}, \&
  {Komiyama}}]{Koda2015}
{Koda}, J., {Yagi}, M., {Yamanoi}, H., \& {Komiyama}, Y. 2015, \apjl, 807, L2

\bibitem[{{Kostov} \& {Bonev}(2018)}]{Kostov2018}
{Kostov}, A. \& {Bonev}, T. 2018, Bulgarian Astronomical Journal, 28, 3

\bibitem[{{Kroupa}(2001)}]{Kroupa2001}
{Kroupa}, P. 2001, \mnras, 322, 231

\bibitem[{{Kuijken}(2011)}]{kujiken2011}
{Kuijken}, K. 2011, The Messenger, 146, 8

\bibitem[{{La Marca} {et~al.}(2021){La Marca}, {Peletier}, {Iodice},
  {Paolillo}, {Choque Challapa}, {Venhola}, {Forbes}, {Cantiello}, {Hilker},
  {Rejkuba}, {Arnaboldi}, {Spavone}, {D'Ago}, {Raj}, {Ragusa}, {Mirabile},
  {Rampazzo}, {Spiniello}, {Mieske}, \& {Schipani}}]{LaMarca2021}
{La Marca}, A., {Peletier}, R., {Iodice}, E., {et~al.} 2021, arXiv e-prints,
  arXiv:2112.00711

\bibitem[{{Lee} {et~al.}(2021){Lee}, {Shin}, \& {Kim}}]{Lee2021}
{Lee}, J., {Shin}, E.-j., \& {Kim}, J.-h. 2021, \apjl, 917, L15

\bibitem[{{Lee} {et~al.}(2020){Lee}, {Kang}, {Lee}, \& {Jang}}]{Lee2020}
{Lee}, J.~H., {Kang}, J., {Lee}, M.~G., \& {Jang}, I.~S. 2020, arXiv e-prints,
  arXiv:2004.01340

\bibitem[{{Leisman} {et~al.}(2017){Leisman}, {Haynes}, {Janowiecki},
  {Hallenbeck}, {J{\'o}zsa}, {Giovanelli}, {Adams}, {Bernal Neira}, {Cannon},
  {Janesh}, {Rhode}, \& {Salzer}}]{Leisman2017}
{Leisman}, L., {Haynes}, M.~P., {Janowiecki}, S., {et~al.} 2017, \apj, 842, 133

\bibitem[{{Lelli} {et~al.}(2015){Lelli}, {Duc}, {Brinks}, {Bournaud},
  {McGaugh}, {Lisenfeld}, {Weilbacher}, {Boquien}, {Revaz}, {Braine},
  {Koribalski}, \& {Belles}}]{Lelli2015}
{Lelli}, F., {Duc}, P.-A., {Brinks}, E., {et~al.} 2015, \aap, 584, A113

\bibitem[{{Lim} {et~al.}(2020){Lim}, {C{\^o}t{\'e}}, {Peng}, {Ferrarese},
  {Roediger}, {Durrell}, {Mihos}, {Wang}, {Gwyn}, {Cuillandre}, {Liu},
  {S{\'a}nchez-Janssen}, {Toloba}, {Sales}, {Guhathakurta}, {Lan{\c{c}}on}, \&
  {Puzia}}]{Lim2020}
{Lim}, S., {C{\^o}t{\'e}}, P., {Peng}, E.~W., {et~al.} 2020, \apj, 899, 69

\bibitem[{{Lim} {et~al.}(2018){Lim}, {Peng}, {C{\^o}t{\'e}}, {Sales}, {den
  Brok}, {Blakeslee}, \& {Guhathakurta}}]{Lim2018}
{Lim}, S., {Peng}, E.~W., {C{\^o}t{\'e}}, P., {et~al.} 2018, \apj, 862, 82

\bibitem[{{Lima-Dias} {et~al.}(2021){Lima-Dias}, {Monachesi}, {Torres-Flores},
  {Cortesi}, {Hern{\'a}ndez-Lang}, {Barbosa}, {Mendes de Oliveira},
  {Olave-Rojas}, {Pallero}, {Sampedro}, {Molino}, {Herpich}, {Jaff{\'e}},
  {Amor{\'\i}n}, {Chies-Santos}, {Dimauro}, {Telles}, {Lopes},
  {Alvarez-Candal}, {Ferrari}, {Kanaan}, {Ribeiro}, \&
  {Schoenell}}]{Lima-dias2021}
{Lima-Dias}, C., {Monachesi}, A., {Torres-Flores}, S., {et~al.} 2021, \mnras,
  500, 1323

\bibitem[{{Liu} {et~al.}(2019){Liu}, {Peng}, {Jord{\'a}n}, {Blakeslee},
  {C{\^o}t{\'e}}, {Ferrarese}, \& {Puzia}}]{Liu2019Fornax}
{Liu}, Y., {Peng}, E.~W., {Jord{\'a}n}, A., {et~al.} 2019, \apj, 875, 156

\bibitem[{{Mancera Pi{\~n}a} {et~al.}(2019){Mancera Pi{\~n}a}, {Aguerri},
  {Peletier}, {Venhola}, {Trager}, \& {Choque Challapa}}]{Mancera-pina2019udg}
{Mancera Pi{\~n}a}, P.~E., {Aguerri}, J.~A.~L., {Peletier}, R.~F., {et~al.}
  2019, \mnras, 485, 1036

\bibitem[{{Mancera Pi{\~n}a} {et~al.}(2018){Mancera Pi{\~n}a}, {Peletier},
  {Aguerri}, {Venhola}, {Trager}, \& {Choque Challapa}}]{Mancera-pina2018udg}
{Mancera Pi{\~n}a}, P.~E., {Peletier}, R.~F., {Aguerri}, J.~A.~L., {et~al.}
  2018, \mnras, 481, 4381

\bibitem[{{Marleau} {et~al.}(2021){Marleau}, {Habas}, {Poulain}, {Duc},
  {Mueller}, {Lim}, {Durrell}, {Sanchez-Janssen}, {Paudel}, {Lammim Ahad},
  {Chougule}, {Bilek}, \& {Fensch}}]{Marleau2021matlas}
{Marleau}, F.~R., {Habas}, R., {Poulain}, M., {et~al.} 2021, arXiv e-prints,
  arXiv:2109.13173

\bibitem[{{Mart{\'\i}n-Navarro} {et~al.}(2019){Mart{\'\i}n-Navarro},
  {Romanowsky}, {Brodie}, {Ferr{\'e}-Mateu}, {Alabi}, {Forbes}, {Sharina},
  {Villaume}, {Pandya}, \& {Martinez-Delgado}}]{Martin-Navarro2019}
{Mart{\'\i}n-Navarro}, I., {Romanowsky}, A.~J., {Brodie}, J.~P., {et~al.} 2019,
  \mnras, 484, 3425

\bibitem[{{Merritt} {et~al.}(2016){Merritt}, {van Dokkum}, {Danieli},
  {Abraham}, {Zhang}, {Karachentsev}, \& {Makarova}}]{Merritt2016a}
{Merritt}, A., {van Dokkum}, P., {Danieli}, S., {et~al.} 2016, \apj, 833, 168

\bibitem[{{Mihos} {et~al.}(2005){Mihos}, {Harding}, {Feldmeier}, \&
  {Morrison}}]{Mihos2005}
{Mihos}, J.~C., {Harding}, P., {Feldmeier}, J., \& {Morrison}, H. 2005, \apjl,
  631, L41

\bibitem[{{Misgeld} {et~al.}(2008){Misgeld}, {Mieske}, \&
  {Hilker}}]{Misgeld2008}
{Misgeld}, I., {Mieske}, S., \& {Hilker}, M. 2008, \aap, 486, 697

\bibitem[{{Montes} {et~al.}(2020){Montes}, {Infante-Sainz}, {Madrigal-Aguado},
  {Rom{\'a}n}, {Monelli}, {Borlaff}, \& {Trujillo}}]{Montes2020}
{Montes}, M., {Infante-Sainz}, R., {Madrigal-Aguado}, A., {et~al.} 2020, \apj,
  904, 114

\bibitem[{{M{\"u}ller} {et~al.}(2021){M{\"u}ller}, {Durrell}, {Marleau}, {Duc},
  {Lim}, {Posti}, {Agnello}, {S{\'a}nchez-Janssen}, {Poulain}, {Habas},
  {Emsellem}, {Paudel}, {van der Burg}, \& {Fensch}}]{Muller2021}
{M{\"u}ller}, O., {Durrell}, P.~R., {Marleau}, F.~R., {et~al.} 2021, arXiv
  e-prints, arXiv:2101.10659

\bibitem[{{M{\"u}ller} {et~al.}(2020){M{\"u}ller}, {Marleau}, {Duc}, {Habas},
  {Fensch}, {Emsellem}, {Poulain}, {Lim}, {Agnello}, {Durrell}, {Paudel},
  {S{\'a}nchez-Janssen}, \& {van der Burg}}]{Muller2020}
{M{\"u}ller}, O., {Marleau}, F.~R., {Duc}, P.-A., {et~al.} 2020, \aap, 640,
  A106

\bibitem[{{M{\"u}ller} {et~al.}(2019){M{\"u}ller}, {Rich}, {Rom{\'a}n},
  {Y{\i}ld{\i}z}, {B{\'\i}lek}, {Duc}, {Fensch}, {Trujillo}, \&
  {Koch}}]{Muller2019}
{M{\"u}ller}, O., {Rich}, R.~M., {Rom{\'a}n}, J., {et~al.} 2019, \aap, 624, L6

\bibitem[{{Navarro} {et~al.}(1997){Navarro}, {Frenk}, \& {White}}]{Navarro1997}
{Navarro}, J.~F., {Frenk}, C.~S., \& {White}, S. D.~M. 1997, \apj, 490, 493

\bibitem[{{Pandya} {et~al.}(2018){Pandya}, {Romanowsky}, {Laine}, {Brodie},
  {Johnson}, {Glaccum}, {Villaume}, {Cuillandre}, {Gwyn}, {Krick}, {Lasker},
  {Mart{\'\i}n-Navarro}, {Martinez-Delgado}, \& {van Dokkum}}]{Pandya2018}
{Pandya}, V., {Romanowsky}, A.~J., {Laine}, S., {et~al.} 2018, \apj, 858, 29

\bibitem[{{Peng} {et~al.}(2011){Peng}, {Ferguson}, {Goudfrooij}, {Hammer},
  {Lucey}, {Marzke}, {Puzia}, {Carter}, {Balcells}, {Bridges}, {Chiboucas},
  {del Burgo}, {Graham}, {Guzm{\'a}n}, {Hudson}, {Matkovi{\'c}}, {Merritt},
  {Miller}, {Mouhcine}, {Phillipps}, {Sharples}, {Smith}, {Tully}, \& {Verdoes
  Kleijn}}]{peng11}
{Peng}, E.~W., {Ferguson}, H.~C., {Goudfrooij}, P., {et~al.} 2011, \apj, 730,
  23

\bibitem[{{Peng} {et~al.}(2008){Peng}, {Jord{\'a}n}, {C{\^o}t{\'e}},
  {Takamiya}, {West}, {Blakeslee}, {Chen}, {Ferrarese}, {Mei}, {Tonry}, \&
  {West}}]{Peng2008virgo}
{Peng}, E.~W., {Jord{\'a}n}, A., {C{\^o}t{\'e}}, P., {et~al.} 2008, \apj, 681,
  197

\bibitem[{{Ploeckinger} {et~al.}(2018){Ploeckinger}, {Sharma}, {Schaye},
  {Crain}, {Schaller}, \& {Barber}}]{Ploeckinger2018}
{Ploeckinger}, S., {Sharma}, K., {Schaye}, J., {et~al.} 2018, \mnras, 474, 580

\bibitem[{{Poggianti} {et~al.}(2019){Poggianti}, {Gullieuszik}, {Tonnesen},
  {Moretti}, {Vulcani}, {Radovich}, {Jaff{\'e}}, {Fritz}, {Bettoni},
  {Franchetto}, {Fasano}, {Bellhouse}, \& {Omizzolo}}]{Poggianti2019}
{Poggianti}, B.~M., {Gullieuszik}, M., {Tonnesen}, S., {et~al.} 2019, \mnras,
  482, 4466

\bibitem[{{Prole} {et~al.}(2019{\natexlab{a}}){Prole}, {Hilker}, {van der
  Burg}, {Cantiello}, {Venhola}, {Iodice}, {van de Ven}, {Wittmann},
  {Peletier}, {Mieske}, {Capaccioli}, {Napolitano}, {Paolillo}, {Spavone}, \&
  {Valentijn}}]{Prole2019b}
{Prole}, D.~J., {Hilker}, M., {van der Burg}, R.~F.~J., {et~al.}
  2019{\natexlab{a}}, \mnras, 484, 4865

\bibitem[{{Prole} {et~al.}(2019{\natexlab{b}}){Prole}, {van der Burg},
  {Hilker}, \& {Davies}}]{Prole2019a}
{Prole}, D.~J., {van der Burg}, R.~F.~J., {Hilker}, M., \& {Davies}, J.~I.
  2019{\natexlab{b}}, \mnras, 488, 2143

\bibitem[{{Rejkuba}(2012)}]{Rejkuba2012}
{Rejkuba}, M. 2012, \apss, 341, 195

\bibitem[{{Richter}(1987)}]{Richter1987}
{Richter}, O.~G. 1987, \aaps, 67, 237

\bibitem[{{Richter} {et~al.}(1982){Richter}, {Materne}, \&
  {Huchtmeier}}]{Richter1982}
{Richter}, O.~G., {Materne}, J., \& {Huchtmeier}, W.~K. 1982, \aap, 111, 193

\bibitem[{{Rom{\'a}n} {et~al.}(2019){Rom{\'a}n}, {Beasley}, {Ruiz-Lara}, \&
  {Valls-Gabaud}}]{Roman2019}
{Rom{\'a}n}, J., {Beasley}, M.~A., {Ruiz-Lara}, T., \& {Valls-Gabaud}, D. 2019,
  \mnras, 486, 823

\bibitem[{{Rom{\'a}n} \& {Trujillo}(2017{\natexlab{a}})}]{Roman2017a}
{Rom{\'a}n}, J. \& {Trujillo}, I. 2017{\natexlab{a}}, \mnras, 468, 703

\bibitem[{{Rom{\'a}n} \& {Trujillo}(2017{\natexlab{b}})}]{Roman2017b}
{Rom{\'a}n}, J. \& {Trujillo}, I. 2017{\natexlab{b}}, \mnras, 468, 4039

\bibitem[{{Rong} {et~al.}(2017){Rong}, {Guo}, {Gao}, {Liao}, {Xie}, {Puzia},
  {Sun}, \& {Pan}}]{Rong2017}
{Rong}, Y., {Guo}, Q., {Gao}, L., {et~al.} 2017, \mnras, 470, 4231

\bibitem[{{Saifollahi} {et~al.}(2021){Saifollahi}, {Trujillo}, {Beasley},
  {Peletier}, \& {Knapen}}]{Saifollahi2021}
{Saifollahi}, T., {Trujillo}, I., {Beasley}, M.~A., {Peletier}, R.~F., \&
  {Knapen}, J.~H. 2021, \mnras, 502, 5921

\bibitem[{{Saifollahi} {et~al.}(2022){Saifollahi}, {Zaritsky}, {Trujillo},
  {Peletier}, {Knapen}, {Amorisco}, {Beasley}, \&
  {Donnerstein}}]{Saifollahi2022}
{Saifollahi}, T., {Zaritsky}, D., {Trujillo}, I., {et~al.} 2022, \mnras, 511,
  4633

\bibitem[{{Sales} {et~al.}(2020){Sales}, {Navarro}, {Pe{\~n}afiel}, {Peng},
  {Lim}, \& {Hernquist}}]{Sales2020}
{Sales}, L.~V., {Navarro}, J.~F., {Pe{\~n}afiel}, L., {et~al.} 2020, \mnras,
  494, 1848

\bibitem[{{Schlafly} \& {Finkbeiner}(2011)}]{Schlafly2011}
{Schlafly}, E.~F. \& {Finkbeiner}, D.~P. 2011, \apj, 737, 103

\bibitem[{{S{\'e}rsic}(1963)}]{Sersic}
{S{\'e}rsic}, J.~L. 1963, Boletin de la Asociacion Argentina de Astronomia La
  Plata Argentina, 6, 41

\bibitem[{{Silk}(2019)}]{Silk2019}
{Silk}, J. 2019, \mnras, 488, L24

\bibitem[{{Springel} {et~al.}(2005){Springel}, {White}, {Jenkins}, {Frenk},
  {Yoshida}, {Gao}, {Navarro}, {Thacker}, {Croton}, {Helly}, {Peacock}, {Cole},
  {Thomas}, {Couchman}, {Evrard}, {Colberg}, \& {Pearce}}]{springel2005}
{Springel}, V., {White}, S. D.~M., {Jenkins}, A., {et~al.} 2005, \nat, 435, 629

\bibitem[{{Tamura} {et~al.}(2000){Tamura}, {Makishima}, {Fukazawa}, {Ikebe}, \&
  {Xu}}]{Tamura2000}
{Tamura}, T., {Makishima}, K., {Fukazawa}, Y., {Ikebe}, Y., \& {Xu}, H. 2000,
  \apj, 535, 602

\bibitem[{{Toloba} {et~al.}(2018){Toloba}, {Lim}, {Peng}, {Sales},
  {Guhathakurta}, {Mihos}, {C{\^o}t{\'e}}, {Boselli}, {Cuillandre},
  {Ferrarese}, {Gwyn}, {Lan{\c{c}}on}, {Mu{\~n}oz}, \& {Puzia}}]{Toloba2018}
{Toloba}, E., {Lim}, S., {Peng}, E., {et~al.} 2018, \apjl, 856, L31

\bibitem[{{Tremmel} {et~al.}(2020){Tremmel}, {Wright}, {Brooks}, {Munshi},
  {Nagai}, \& {Quinn}}]{Tremmel2020}
{Tremmel}, M., {Wright}, A.~C., {Brooks}, A.~M., {et~al.} 2020, \mnras, 497,
  2786

\bibitem[{{Tremmel} {et~al.}(2019){Tremmel}, {Wright}, {Munshi}, {Quinn},
  {Nagai}, {Brooks}, \& {Pontzen}}]{Tremmel2019}
{Tremmel}, M.~J., {Wright}, A., {Munshi}, F., {et~al.} 2019, in American
  Astronomical Society Meeting Abstracts, Vol. 233, American Astronomical
  Society Meeting Abstracts \#233, 416.04

\bibitem[{{Trujillo} {et~al.}(2017){Trujillo}, {Roman}, {Filho}, \&
  {S{\'a}nchez Almeida}}]{Trujillo2017}
{Trujillo}, I., {Roman}, J., {Filho}, M., \& {S{\'a}nchez Almeida}, J. 2017,
  \apj, 836, 191

\bibitem[{{van der Burg} {et~al.}(2017){van der Burg}, {Hoekstra}, {Muzzin},
  {Sif{\'o}n}, {Viola}, {Bremer}, {Brough}, {Driver}, {Erben}, {Heymans},
  {Hildebrandt}, {Holwerda}, {Klaes}, {Kuijken}, {McGee}, {Nakajima},
  {Napolitano}, {Norberg}, {Taylor}, \& {Valentijn}}]{vanderBurg2017}
{van der Burg}, R. F.~J., {Hoekstra}, H., {Muzzin}, A., {et~al.} 2017, \aap,
  607, A79

\bibitem[{{van der Burg} {et~al.}(2016){van der Burg}, {Muzzin}, \&
  {Hoekstra}}]{VanderBurg2016}
{van der Burg}, R. F.~J., {Muzzin}, A., \& {Hoekstra}, H. 2016, \aap, 590, A20

\bibitem[{{van Dokkum} {et~al.}(2016){van Dokkum}, {Abraham}, {Brodie},
  {Conroy}, {Danieli}, {Merritt}, {Mowla}, {Romanowsky}, \&
  {Zhang}}]{vanDokkum2016}
{van Dokkum}, P., {Abraham}, R., {Brodie}, J., {et~al.} 2016, \apjl, 828, L6

\bibitem[{{van Dokkum} {et~al.}(2018){van Dokkum}, {Danieli}, {Cohen},
  {Merritt}, {Romanowsky}, {Abraham}, {Brodie}, {Conroy}, {Lokhorst}, {Mowla},
  {O'Sullivan}, \& {Zhang}}]{vanDokkum2018}
{van Dokkum}, P., {Danieli}, S., {Cohen}, Y., {et~al.} 2018, \nat, 555, 629

\bibitem[{{van Dokkum} {et~al.}(2019){van Dokkum}, {Wasserman}, {Danieli},
  {Abraham}, {Brodie}, {Conroy}, {Forbes}, {Martin}, {Matuszewski},
  {Romanowsky}, \& {Villaume}}]{vanDokkum2019}
{van Dokkum}, P., {Wasserman}, A., {Danieli}, S., {et~al.} 2019, \apj, 880, 91

\bibitem[{{van Dokkum} {et~al.}(2015){van Dokkum}, {Romanowsky}, {Abraham},
  {Brodie}, {Conroy}, {Geha}, {Merritt}, {Villaume}, \&
  {Zhang}}]{vanDokkum2015}
{van Dokkum}, P.~G., {Romanowsky}, A.~J., {Abraham}, R., {et~al.} 2015, \apjl,
  804, L26

\bibitem[{{Venhola} {et~al.}(2017){Venhola}, {Peletier}, {Laurikainen}, {Salo},
  {Lisker}, {Iodice}, {Capaccioli}, {Kleijn}, {Valentijn}, {Mieske}, {Hilker},
  {Wittmann}, {van de Ven}, {Grado}, {Spavone}, {Cantiello}, {Napolitano},
  {Paolillo}, \& {Falc{\'o}n-Barroso}}]{Venhola2017}
{Venhola}, A., {Peletier}, R., {Laurikainen}, E., {et~al.} 2017, \aap, 608,
  A142

\bibitem[{{Venhola} {et~al.}(2021){Venhola}, {Peletier}, {Salo}, {Laurikainen},
  {Janz}, {Haigh}, {Wilkinson}, {Iodice}, {Hilker}, {Mieske}, {Cantiello}, \&
  {Spavone}}]{Venhola2021}
{Venhola}, A., {Peletier}, R.~F., {Salo}, H., {et~al.} 2021, arXiv e-prints,
  arXiv:2111.01855

\bibitem[{{Villegas} {et~al.}(2010){Villegas}, {Jord{\'a}n}, {Peng},
  {Blakeslee}, {C{\^o}t{\'e}}, {Ferrarese}, {Kissler-Patig}, {Mei}, {Infante},
  {Tonry}, \& {West}}]{villegas10}
{Villegas}, D., {Jord{\'a}n}, A., {Peng}, E.~W., {et~al.} 2010, \apj, 717, 603

\bibitem[{{Wright} {et~al.}(2021){Wright}, {Tremmel}, {Brooks}, {Munshi},
  {Nagai}, {Sharma}, \& {Quinn}}]{Wright2021}
{Wright}, A.~C., {Tremmel}, M., {Brooks}, A.~M., {et~al.} 2021, \mnras, 502,
  5370

\bibitem[{{Yagi} {et~al.}(2016){Yagi}, {Koda}, {Komiyama}, \&
  {Yamanoi}}]{Yagi2016}
{Yagi}, M., {Koda}, J., {Komiyama}, Y., \& {Yamanoi}, H. 2016, \apjs, 225, 11

\bibitem[{{Zaritsky} {et~al.}(2019){Zaritsky}, {Donnerstein}, {Dey},
  {Kadowaki}, {Zhang}, {Karunakaran}, {Mart{\'\i}nez-Delgado}, {Rahman}, \&
  {Spekkens}}]{Zaritsky2019}
{Zaritsky}, D., {Donnerstein}, R., {Dey}, A., {et~al.} 2019, \apjs, 240, 1

\end{thebibliography}
  
  \begin{appendix} 

\section{An example of globular clusters detection in UDGs}\label{sec:GCs_selection}
{ 
In section \ref{sec:GCs} we explained in details how GC candidates around each UDG are selected. 
Here we present the selection cuts for an example galaxy, namely UDG\,13. 
In Figure \ref{fig:GC_sel} we show the following $g$-band cuts: {\it i)} magnitude $23.5\ge m_g \ge 26.0$ mag; {\it ii)} color $0.25\leq g-r \leq 1.25$ mag; {\it iii)} {\sc SExtractor} star/galaxy classifier $CLASS\_STAR\ge 0.4$; {\it iv)} concentration index $CI_g$ within $\pm 0.1$ mag of the local sequence of point sources; {\it v)} location selection, i.e. within 1.5, 3, $5\;R_e$. 
For sake of clarity, we do not show the elongation selection, which is the least constraining criterion in our procedure. 

In Figure \ref{fig:GC_sel} we report all the detected objects and mark differently the likely MW stars, the GC candidates in the analyzed cutout, and those that are within 5$R_e$ from the galaxy center. 

\begin{figure*}[t]
    \centering
    \includegraphics[width=0.8\textwidth]{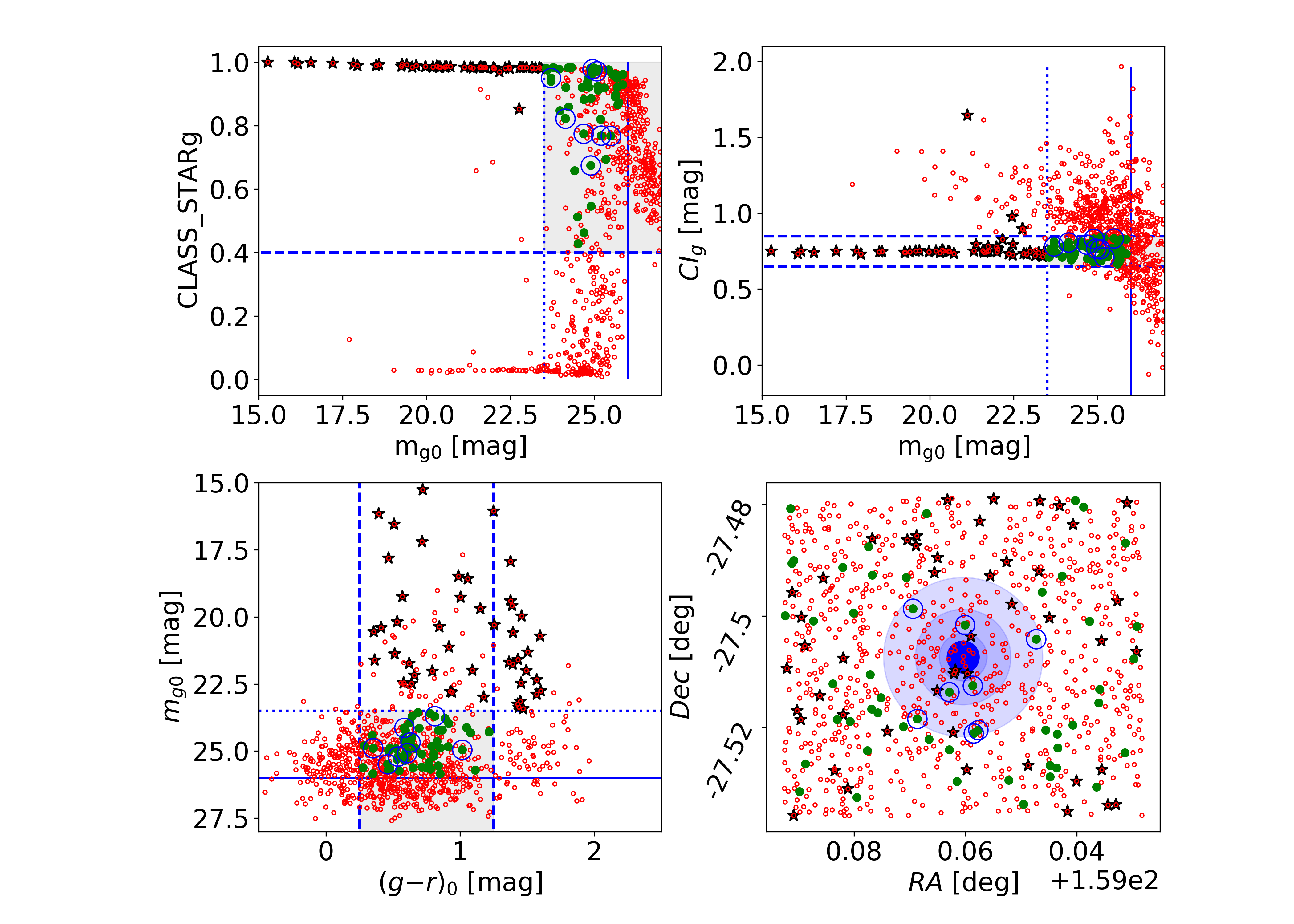}
    \caption{
    Globular clusters selection process and cuts for \object{UDG\,13}. 
    Red dots are all the sources detected in the $20R_e\times 20R_e$ cutout centred on the galaxy.
    The green filled dots are all the sources matching the GC selection criteria described in Sec. \ref{sec:GCs}. 
    The blue empty circles around the green filled dots identify GC candidates within $5R_e$ of the UDG. 
    Filled black five-pointed stars mark bright point sources, likely MW stars. 
    \textit{Upper left} - $g$-band CLASS\_STAR vs. magnitude. The thin solid blue line at 26 mag represents the TOM, the blue dotted line is the the $3\sigma_{GCLF}$ brighter selection cut. The horizontal blue dashed line shows the adopted star/galaxy cut. 
    \textit{Upper right} - $g$-band concentration index $CI$ vs. magnitude. Vertical lines are the same as the upper left panel, while horizontal blue dashed lines show the $\pm 0.1$ mag region around the local $CI_g$ sequence of point sources.
    \textit{Bottom left} - color magnitude diagram. The dashed blue lines show the $0.25\leq (g-r)_0 \leq 1.25$ mag color cut. The other lines are as in previous panels.
    \textit{Bottom right} - UDG\,13 cutout field. Concentric shaded blue circles represent the 1, 1.5, 3, and $5\;R_e$ regions.  
    }
    
    \label{fig:GC_sel}
\end{figure*}

\section{$N_{GC}$ and $S_N$ within the other apertures}

In this appendix we present the GC counts within 3 and 5 $R_e$, for both new LSB dwarfs and UDGs in our sample.
We show both the global and local corrected numbers, and the respective specific frequency.
Additionally, we list the globally corrected $N_{GC}$ within 1.5 $R_e$ aperture radius.
We include the GC counts for the UDG 32 presented by \citet[][]{Iodice2021}.
Values are reported in Table \ref{tab:UDG_GC}.

%------------- Appendix Table

\begin{table*}

\setlength{\tabcolsep}{2pt}
\begin{center}
\small 
%%\scriptsize
\caption{Globular clusters and associated specific frequencies of the new UDG candidates in the Hydra I cluster.} 
\label{tab:UDG_GC}
\begin{tabular}{lcccccc|cccc}

\hline \hline
Object & \multicolumn{6}{c|}{Global determination} & \multicolumn{4}{|c}{Local determination} \\
\hline
 & \multicolumn{3}{c|}{$N_{GC}$} & \multicolumn{3}{|c|}{$S_N$} & \multicolumn{2}{|c|}{$N_{GC}$} & \multicolumn{2}{|c}{$S_N$} \\
\hline

 & $1.5\;R_e$ & $3\; R_e$ & $5\;R_e$ & $1.5\;R_e$ & $3\; R_e$ & $5\;R_e$  & $3\; R_e$ & $5\;R_e$  & $3\; R_e$ & $5\;R_e$ \\% & $1.5\;R_e$ & $3\; R_e$ & $5\;R_e$ & $1.5\;R_e$ & $3\; R_e$ & $5\;R_e$ \\

\hline
UDG 13 & - & - & - & - & - & - & - &  1 $\pm$ 9  & - & 11 $\pm$ 100 \\
UDG 14 & - & 4 $\pm$ 3 & 1 $\pm$ 5 & - & 93 $\pm$ 93 & 23 $\pm$ 131 & 4 $\pm$ 6 & 3 $\pm$ 9 &  93 $\pm$ 170 & 70 $\pm$ 233 \\
UDG 15 &2 $\pm$ 1 & - & - & 41 $\pm$ 36 & - & - & - & -  & - & - \\
UDG 16 & - & - & - & - & - & - & - & - & - & - \\
UDG 17 & 3 $\pm$ 1 & - & - & 11 $\pm$ 7 & - & -  & - & - & - & - \\
UDG 18 & 11 $\pm$ 4 & 5 $\pm$ 2 & 7 $\pm$ 3 & 195 $\pm$ 108 & 76 $\pm$ 52 & 107 $\pm$ 78  & 7 $\pm$ 7 & 10 $\pm$ 10 & 107 $\pm$ 136 & 153 $\pm$ 96  \\
UDG 19 & - & - & - & - & - & - & - & - & -  & - \\
UDG 20 & - & - & - & - & - & - & - & - & - & - \\
UDG 21 & - & - & - & - & - & - & - & - & - & - \\
UDG 22 & - & 1.0 $\pm$ 0.4 & - & - & 3 $\pm$ 2 & - & - & -  & - & - \\
UDG 23 & - & - & - & - & - & - & - & - &  - & - \\
UDG 32 & 7 $\pm$ 3 & 1 $\pm$ 3 & - & 12 $\pm$ 8 & 2 $\pm$ 5 & - & 3 $\pm$ 5 & 4 $\pm$ 10 & 5 $\pm$ 10 & 7 $\pm$ 19  \\ 
\hline
LSB 1 & - & 2 $\pm$ 1 & 3 $\pm$ 3 & - & 26 $\pm$ 26 & 40 $\pm$ 54& 3 $\pm$ 5 & 5 $\pm$ 8 & 40 $\pm$ 79 & 66 $\pm$ 127  \\ 
LSB 2 & - & - & - & - & - & - & - & - & - & - \\
LSB 3 &- & 3 $\pm$ 3 & - & - & 55 $\pm$ 81 & - & 4 $\pm$ 7 & 2 $\pm$ 10 & 73 $\pm$ 151 & 36 $\pm$ 194 \\
LSB 4 & 7 $\pm$ 3 & 3 $\pm$ 1 & 3 $\pm$ 1 & 24 $\pm$ 17 & 10 $\pm$ 7 & 10 $\pm$ 8 & 3 $\pm$ 5 & 3 $\pm$ 6 & 10 $\pm$ 21 & 10 $\pm$ 24 \\
LSB 5 & - & - & - & - & - & - & - & - & - & - \\
LSB 6 & 2 $\pm$ 1 & 5 $\pm$ 2 & 4 $\pm$ 3 & 2 $\pm$ 1 & 4 $\pm$ 2 & 3 $\pm$ 3 & 4 $\pm$ 6 & 2 $\pm$ 7 & 3 $\pm$ 6 & 2 $\pm$ 6 \\
LSB 7 & - & - & - & - & - & - & - & - & - & - \\
LSB 8 & - & - & - & - & - & - & - & - & - & - \\
\hline

\end{tabular}
\tablefoot{$N_{GCs}$ listed are all corrected for background contamination. Values corrected for the global and local contamination are reported on the left and right columns, respectively. 
}
\end{center}
\end{table*}
% \end{sidewaystable*}
}

% \clearpage

\section{Images and surface brightness distribution of twelve new UDGs}\label{sec:UDGimages}

\begin{figure*}
\centering
\begin{tabular}{cc}

    \includegraphics[width=7cm]{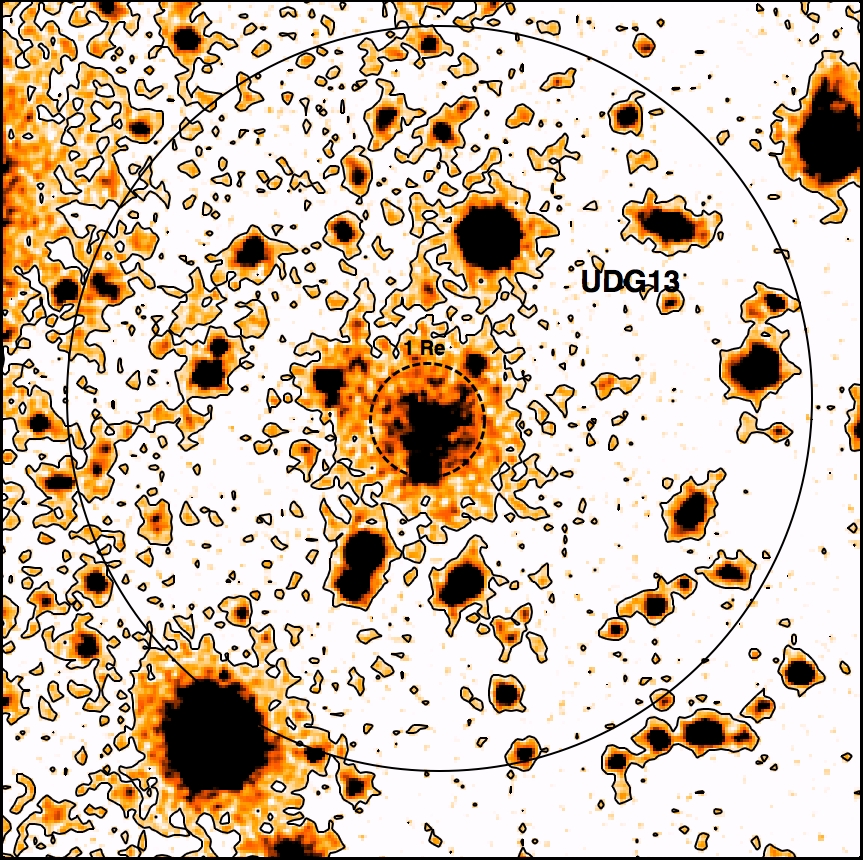} &
    \includegraphics[width=7cm]{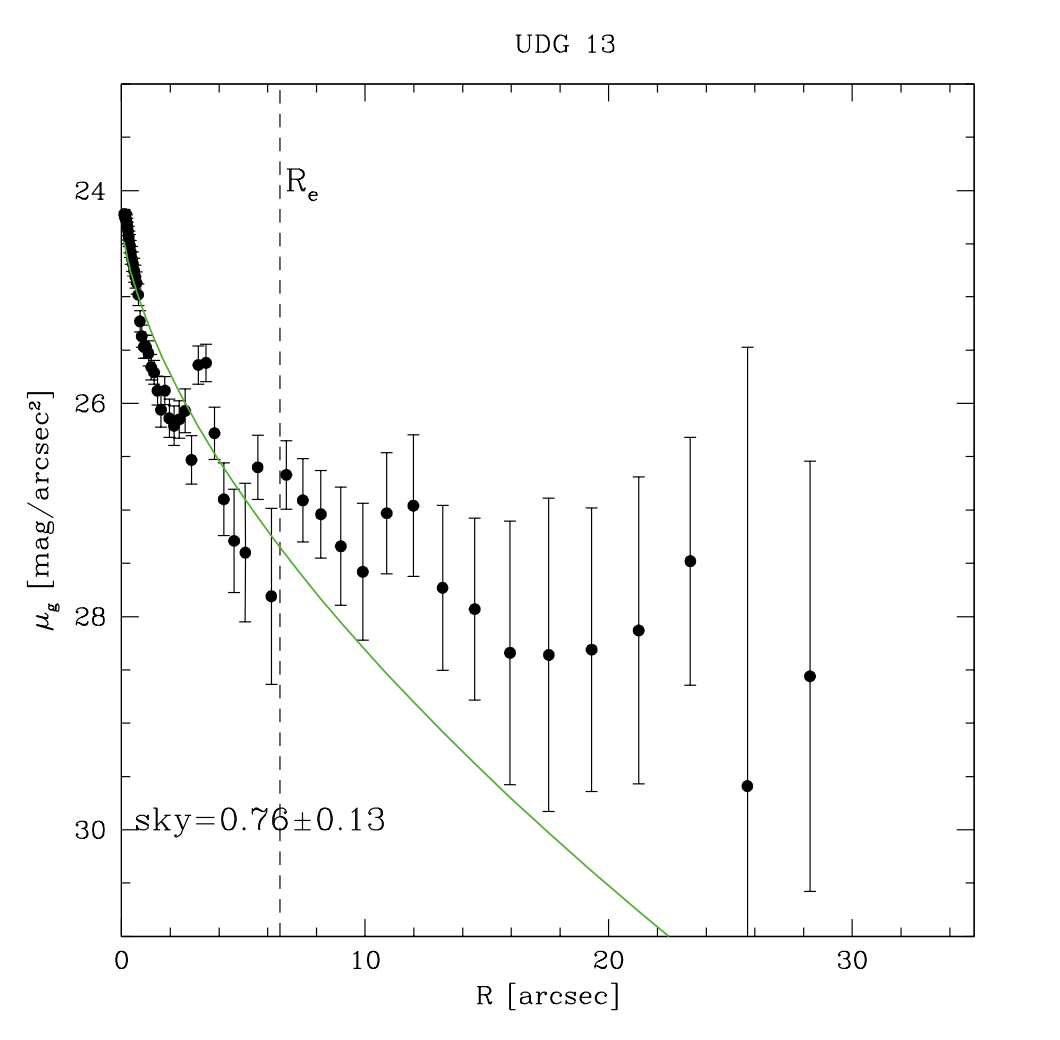} \\
    \includegraphics[width=7cm]{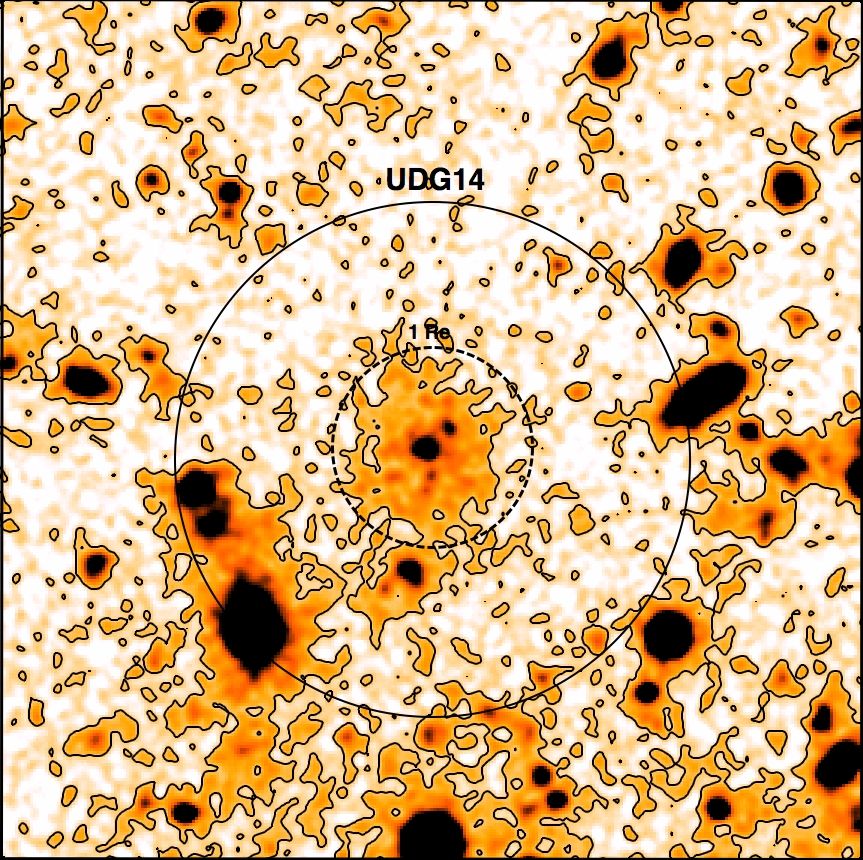} &
    \includegraphics[width=7cm]{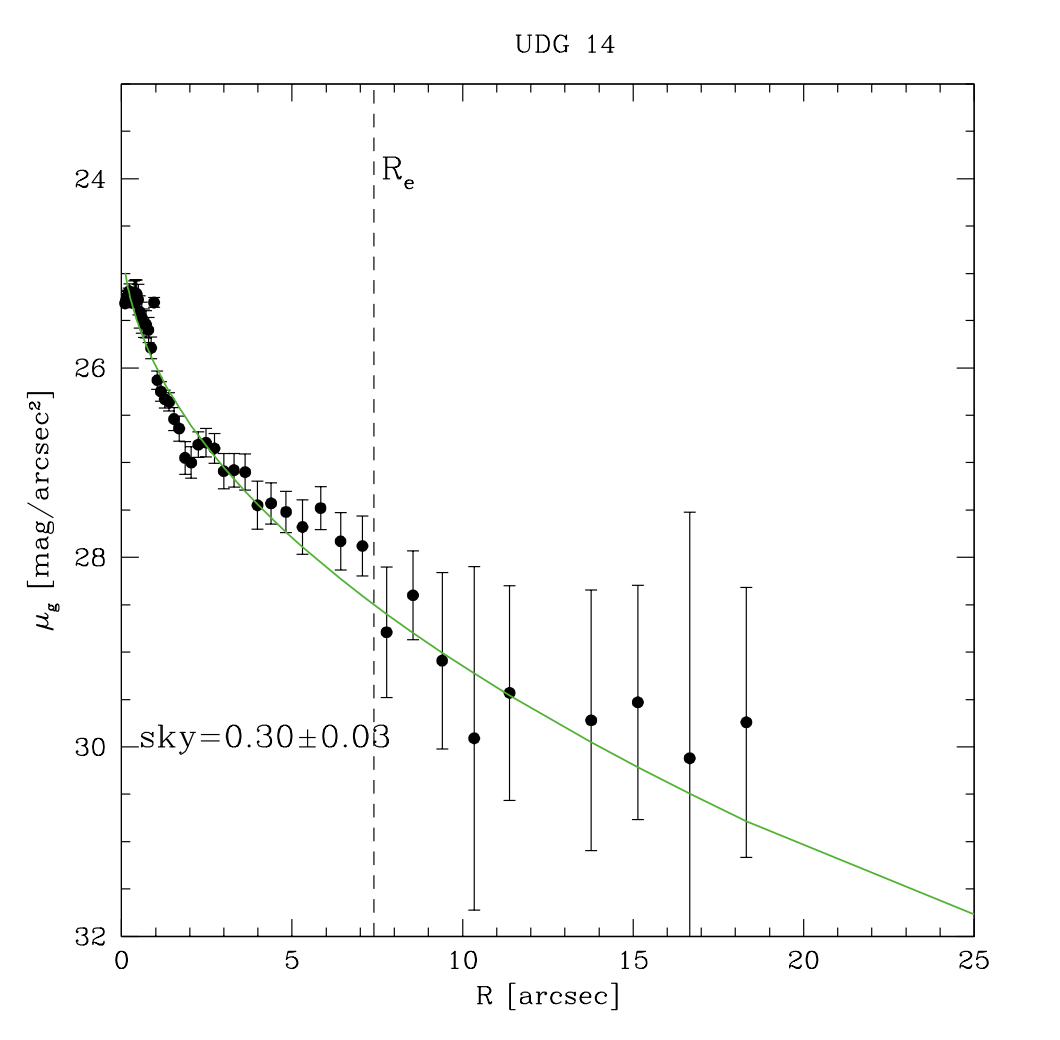} \\
    \includegraphics[width=7cm]{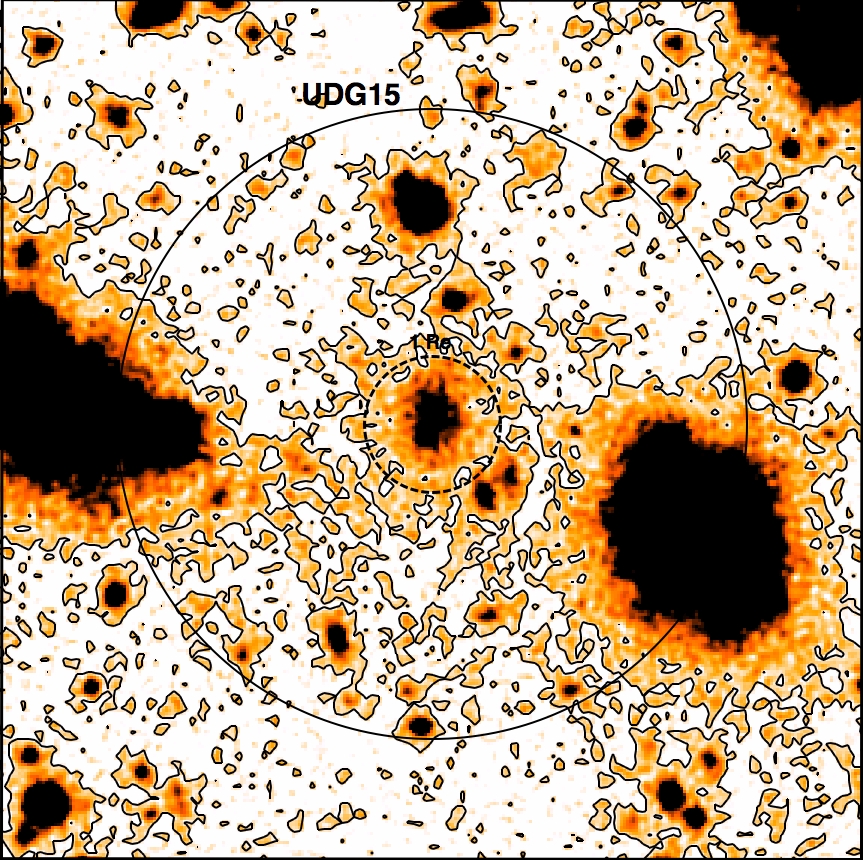} &
    \includegraphics[width=7cm]{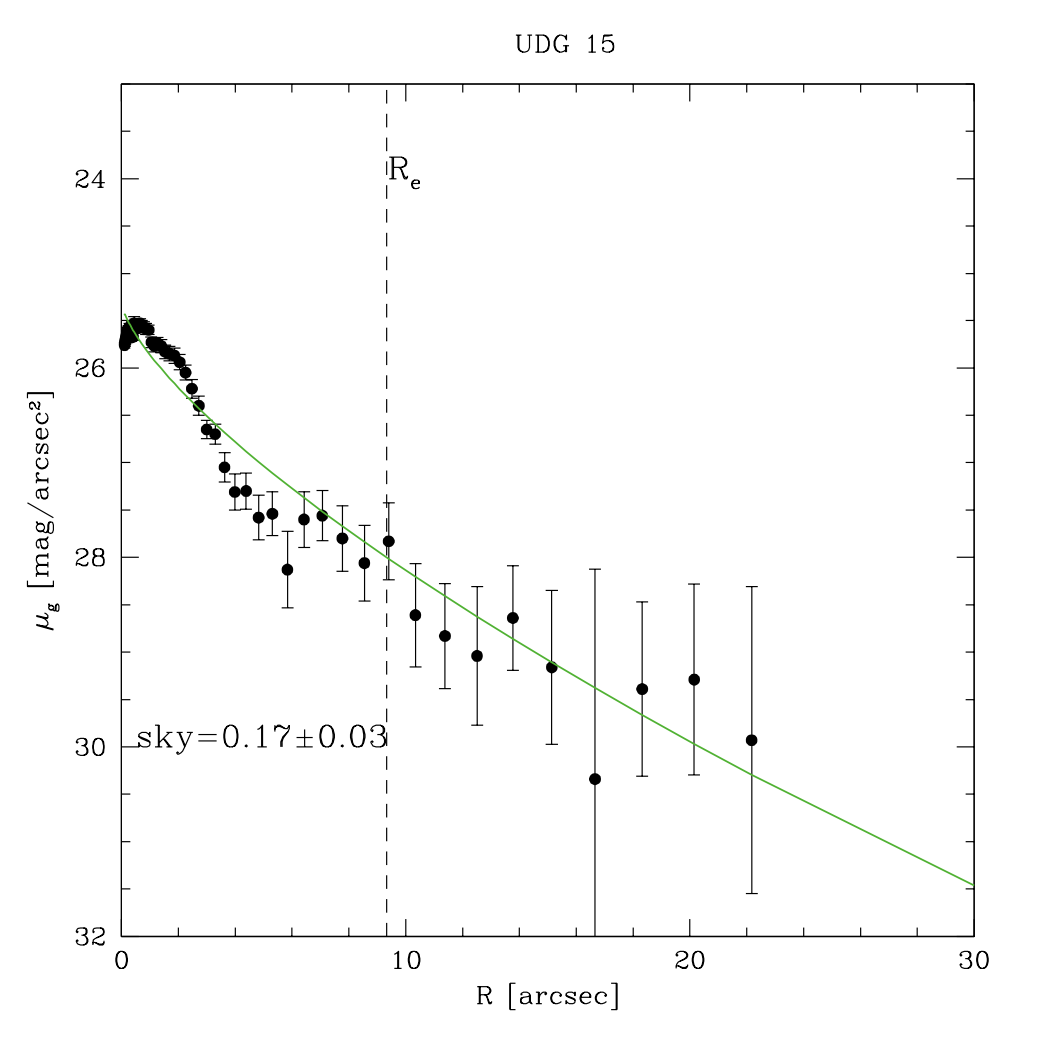}\\
   \end{tabular}
    \caption{Light distribution of the UDG candidates in the $g$ band. 
    Left panel - Image thumbnail centred on the UDG of $1\times 1$ arcmin ($\sim 14.9$ kpc). 
    The black dashed circle corresponds to $1 R_e$, listed in Tab.~\ref{tab:UDGsample}. 
    The black solid circle marks the outermost radius where flux blends into the sky level. This corresponds to the outermost data-point shown in the plot on the right panel.
    The black solid contours correspond at $\mu_{g}=29$~mag/arcsec$^2$.
	Right panel - Azimuthally averaged surface brightness profile (black points) of the UDG shown in the left panel, 
	as function of the semi-major axis radius, and the best fit model (green solid line). The vertical dashed line indicates the effective radius derived from the best fit model.
	The local sky level (in counts) is reported in the lower-left corner of the plot.
}
    \label{fig:UDG_1}
\end{figure*}

\begin{figure*}
\centering
\begin{tabular}{cc}

    \includegraphics[width=7cm]{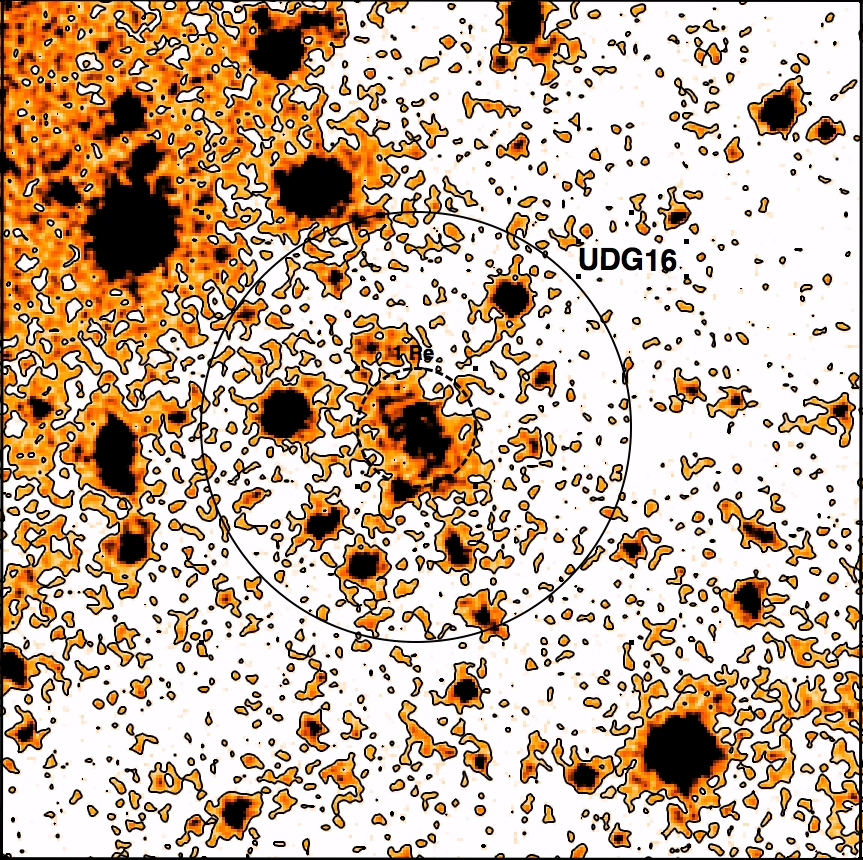} &
    \includegraphics[width=7cm]{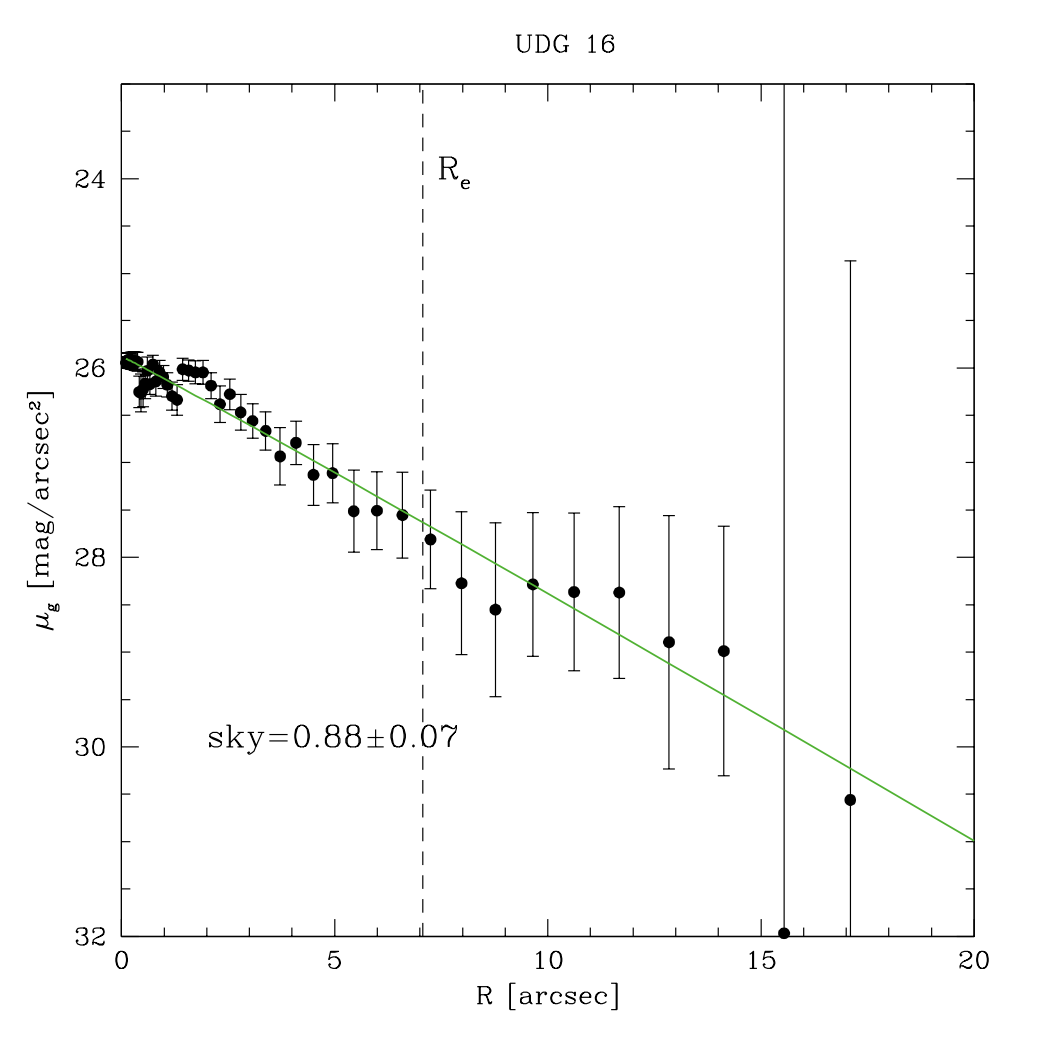}\\
    \includegraphics[width=7cm]{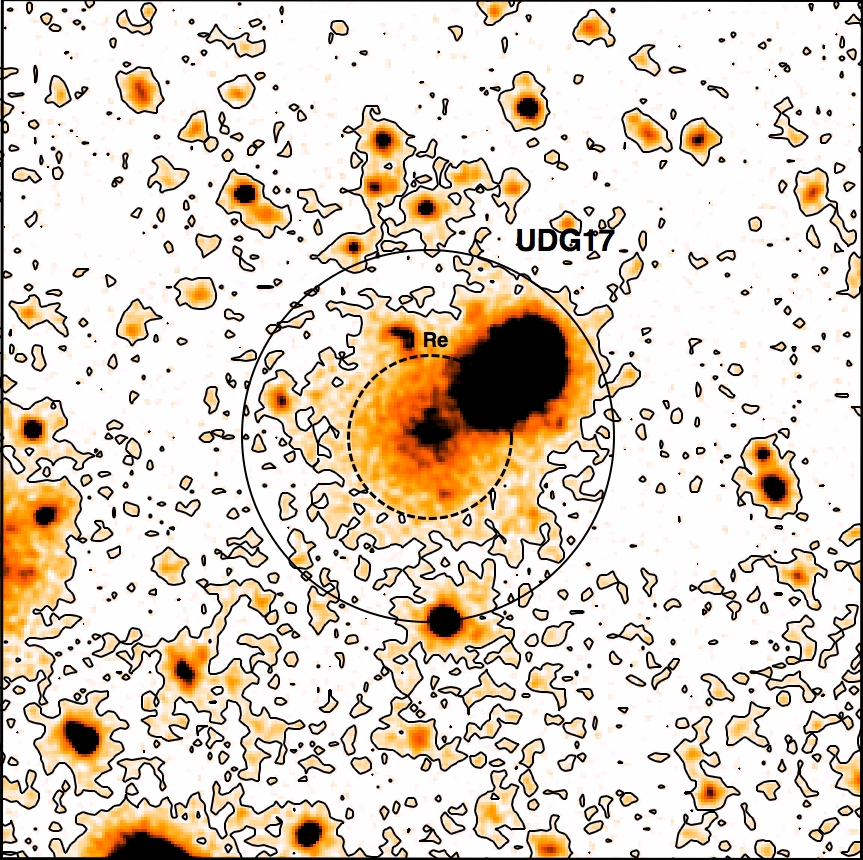} &
    \includegraphics[width=7cm]{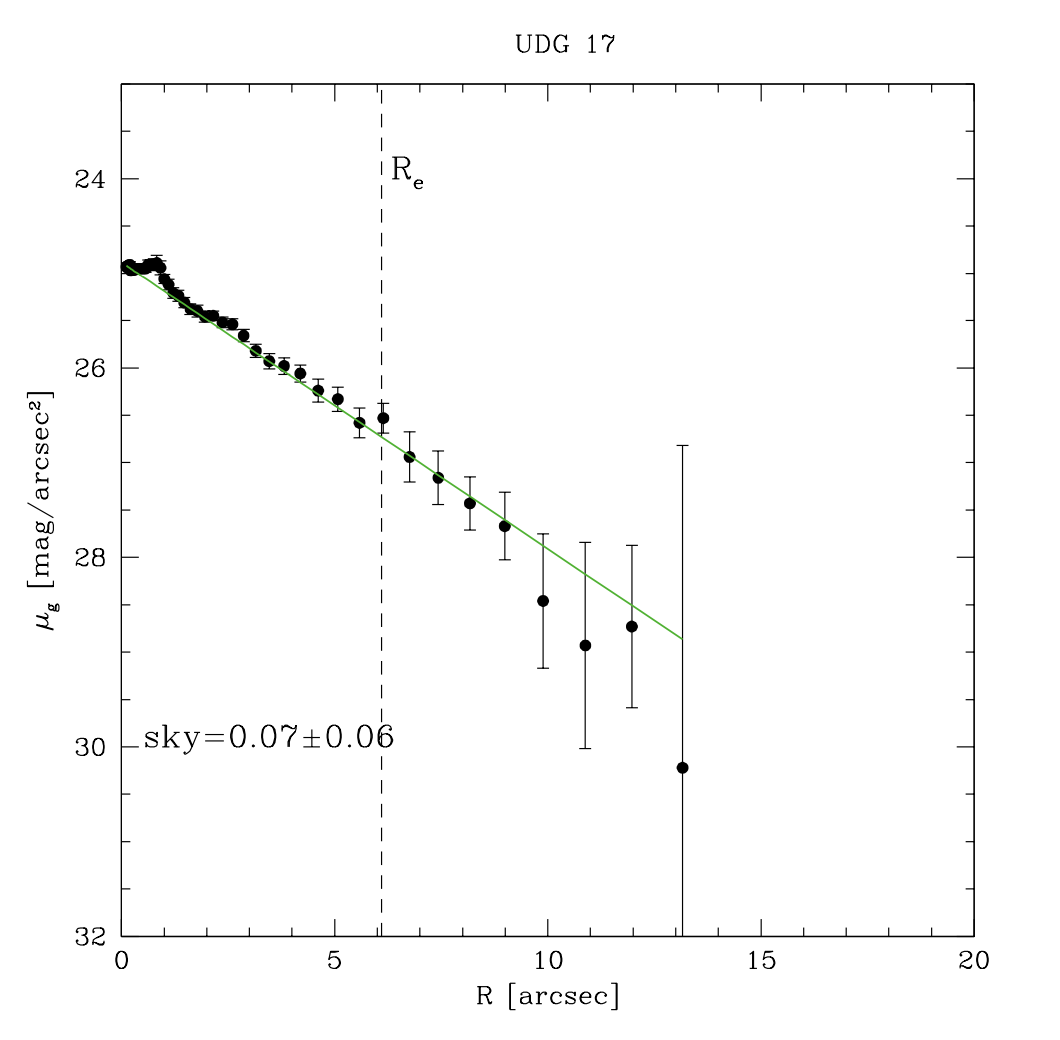} \\
    \includegraphics[width=7cm]{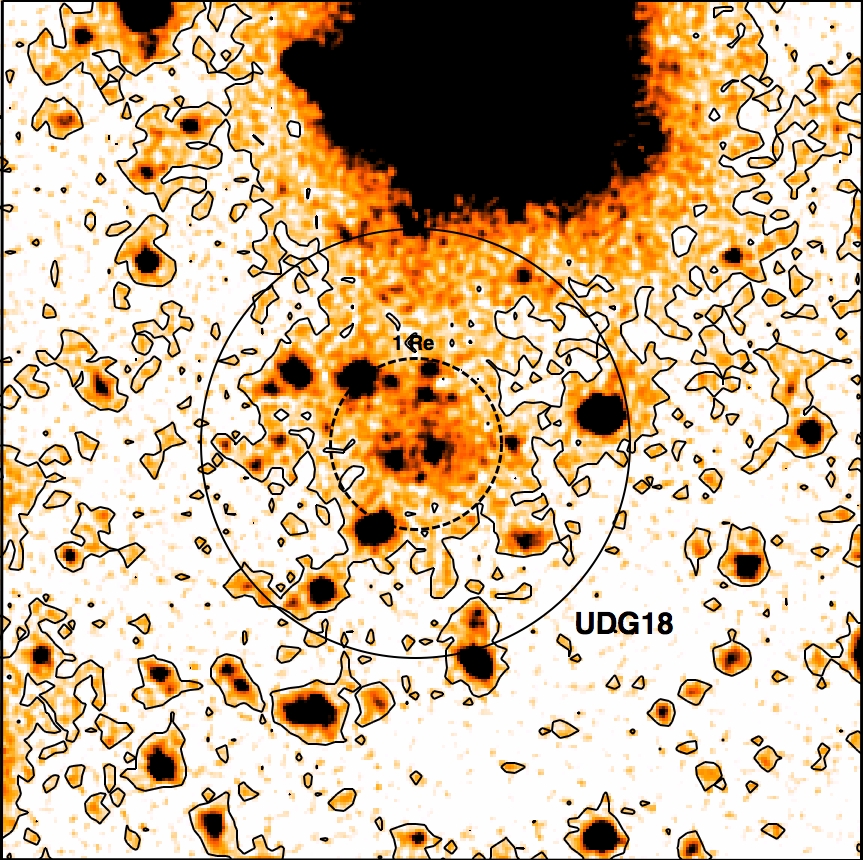} &
    \includegraphics[width=7cm]{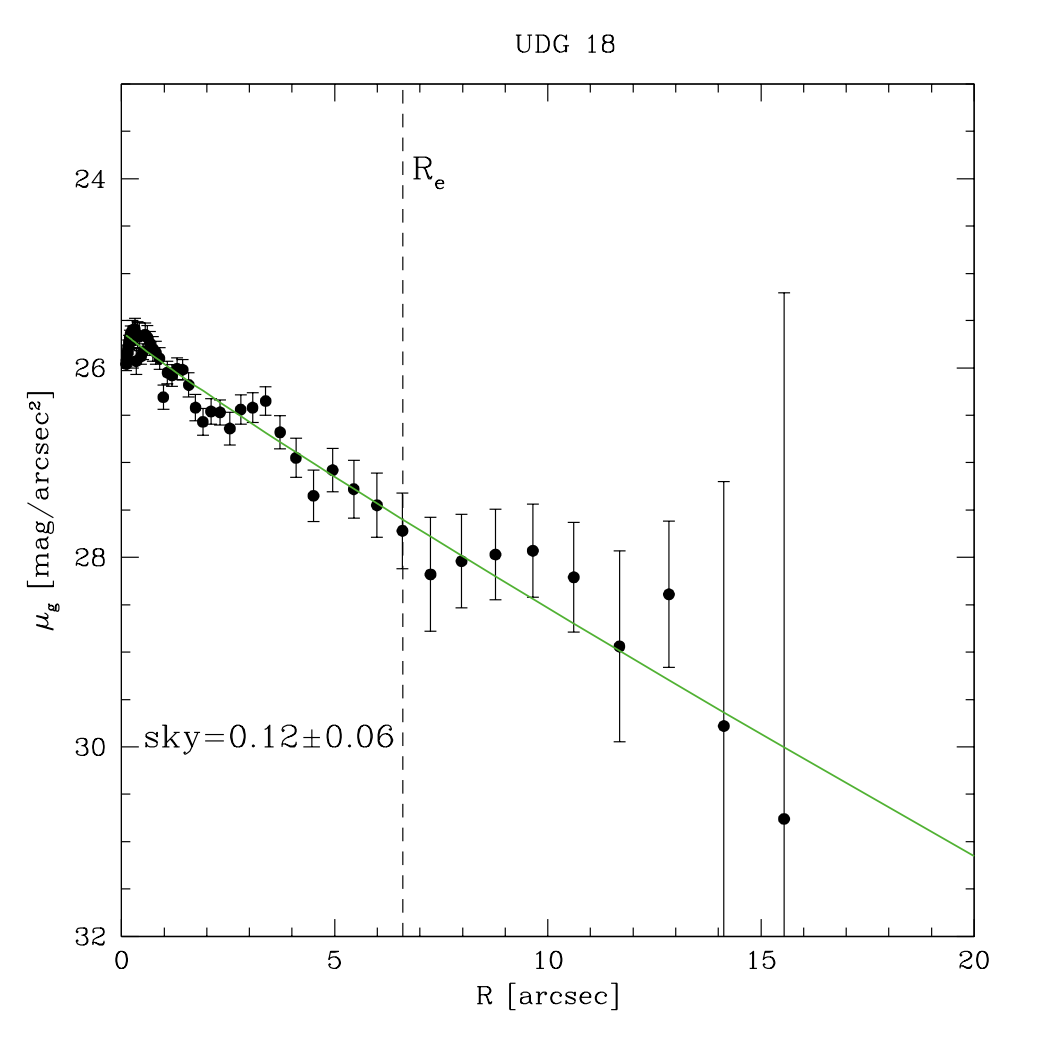}\\
   \end{tabular}
    \caption{Same as Fig.~\ref{fig:UDG_1} for UDG~16, UDG~17 and UDG~18.}
    \label{fig:UDG_2}
\end{figure*}

\begin{figure*}
\centering
\begin{tabular}{cc}
    \includegraphics[width=7cm]{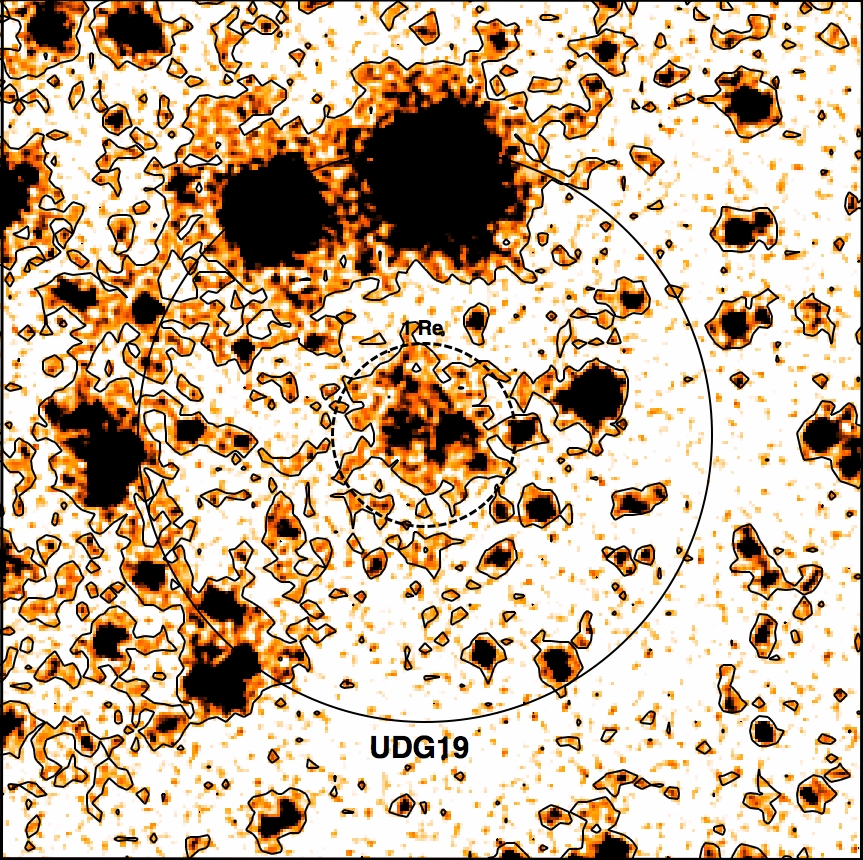} &
    \includegraphics[width=7cm]{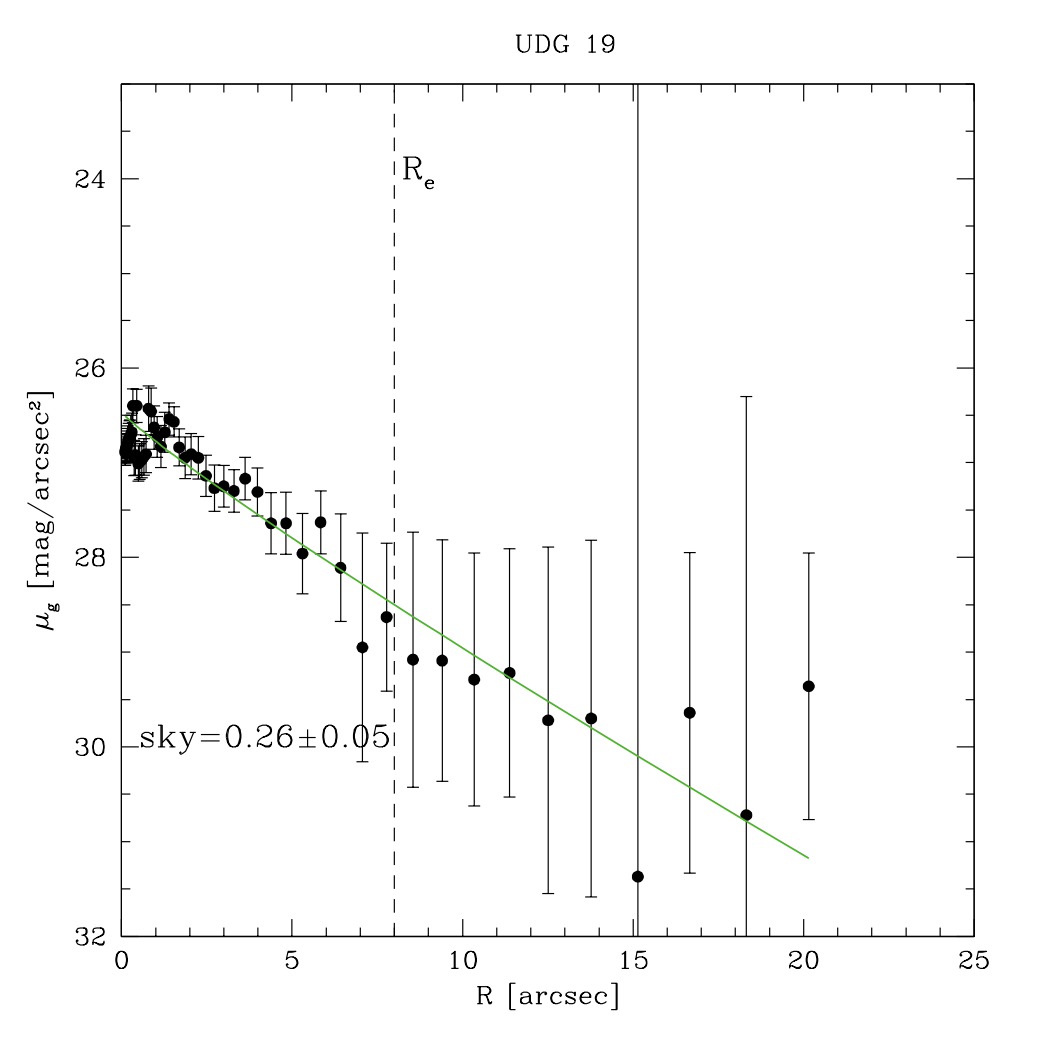} \\
    \includegraphics[width=7cm]{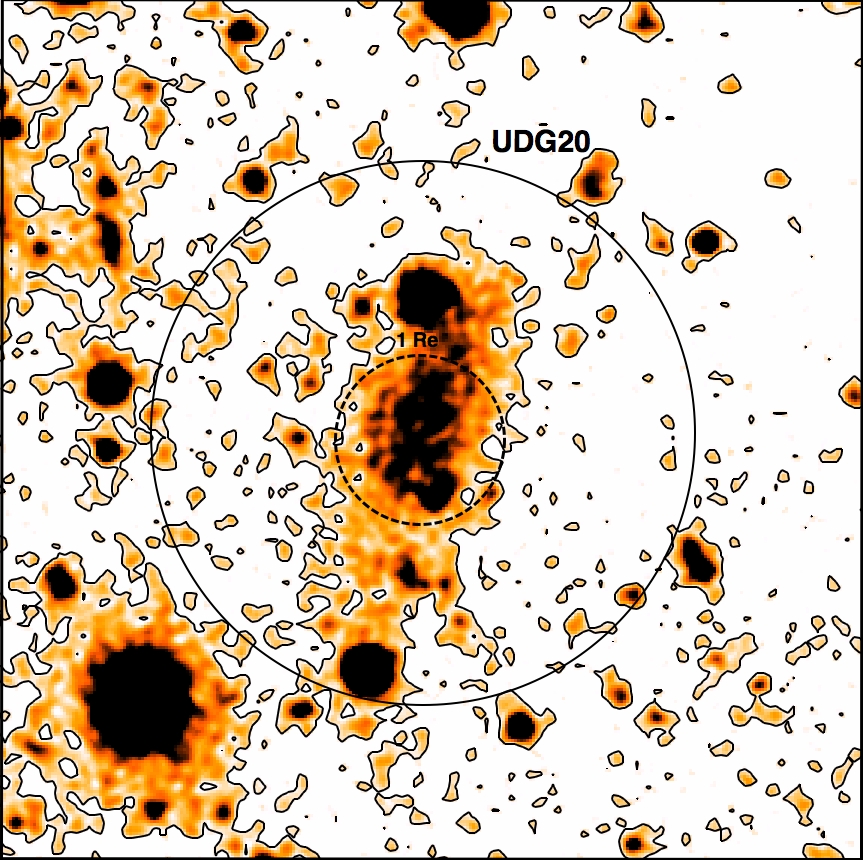} &
    \includegraphics[width=7cm]{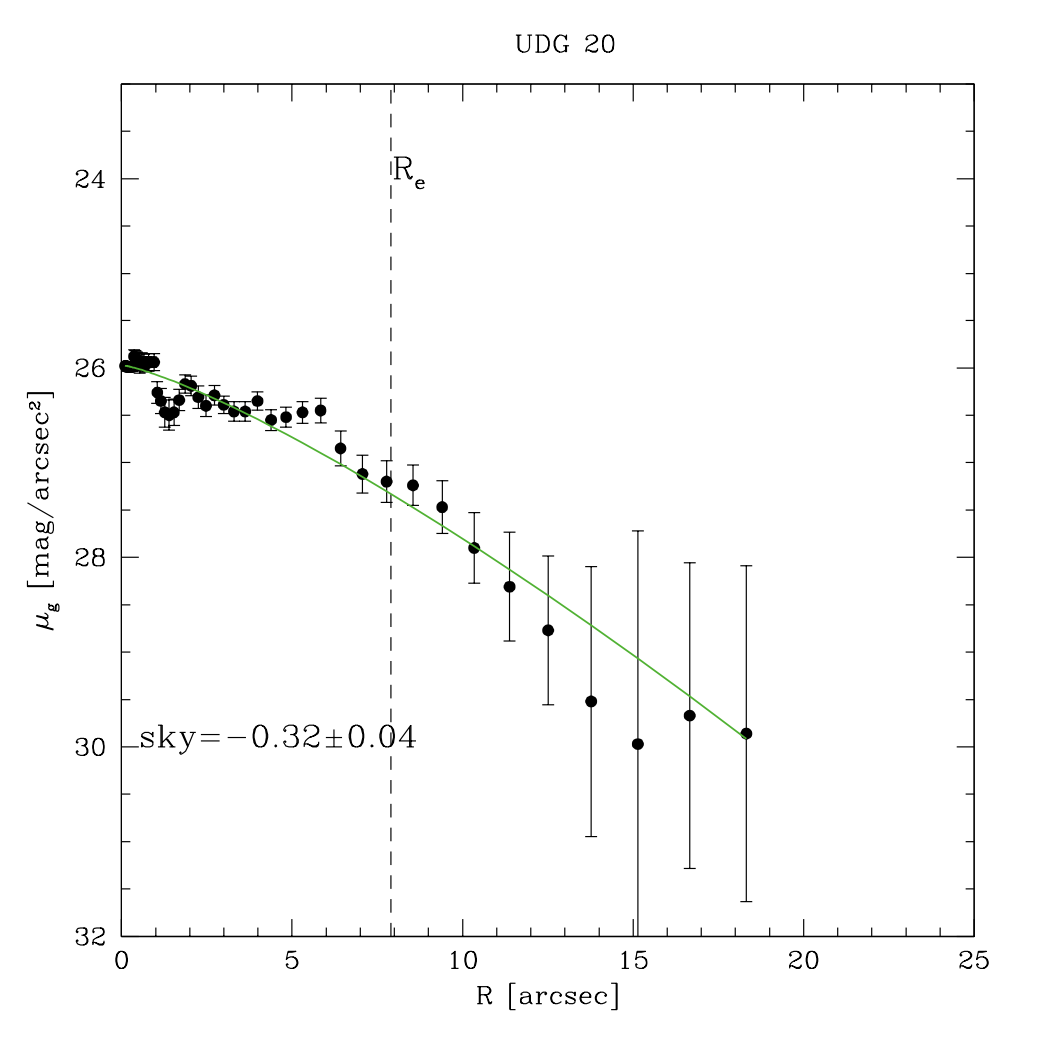} \\
    \includegraphics[width=7cm]{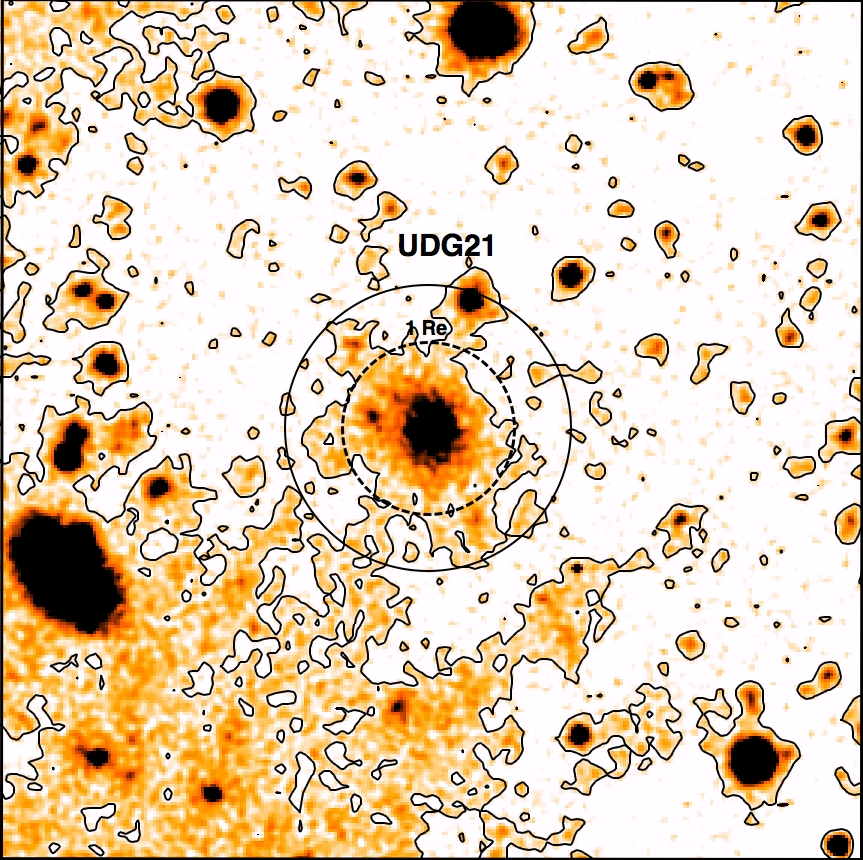} &
    \includegraphics[width=7cm]{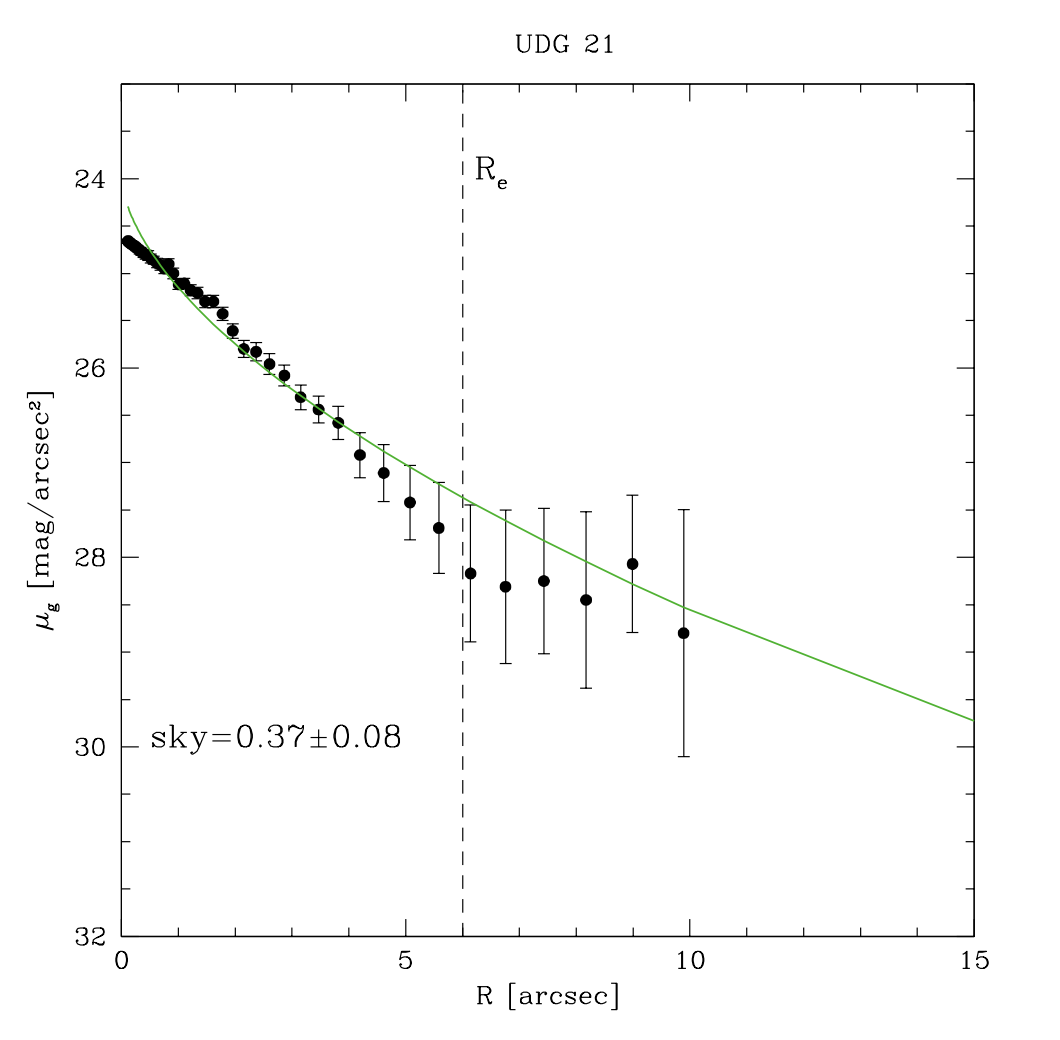} \\
   \end{tabular}
    \caption{Same as Fig.~\ref{fig:UDG_1} for UDG~19, UDG~20 and UDG~21.}
    \label{fig:UDG_3}
\end{figure*}

\begin{figure*}
\centering
\begin{tabular}{cc}
    \includegraphics[width=7cm]{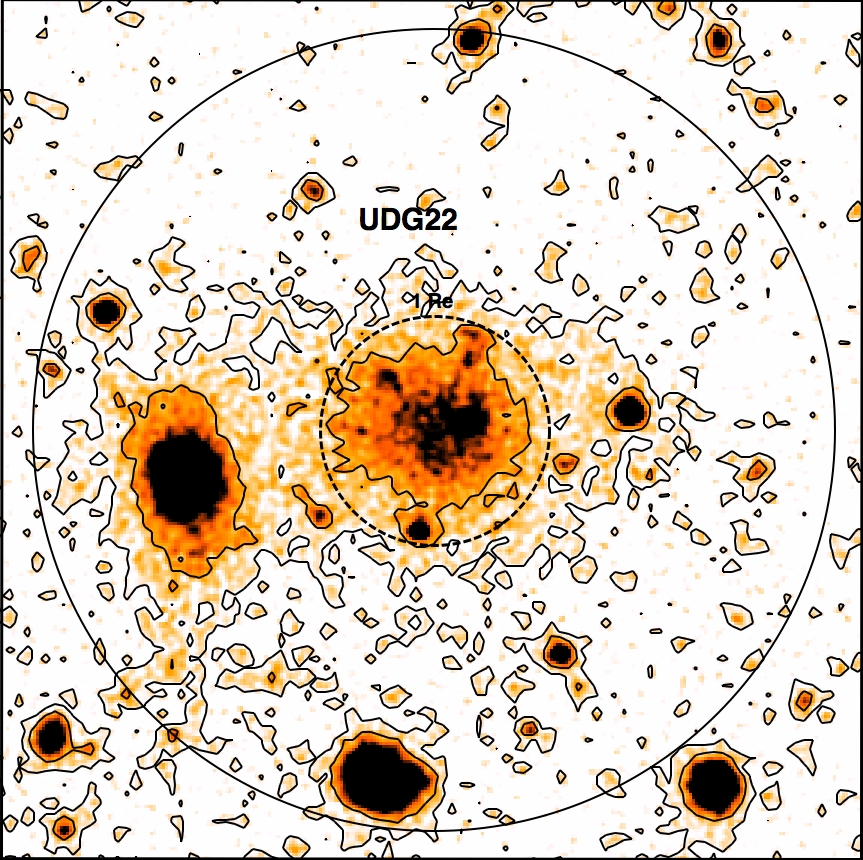} &
    \includegraphics[width=7cm]{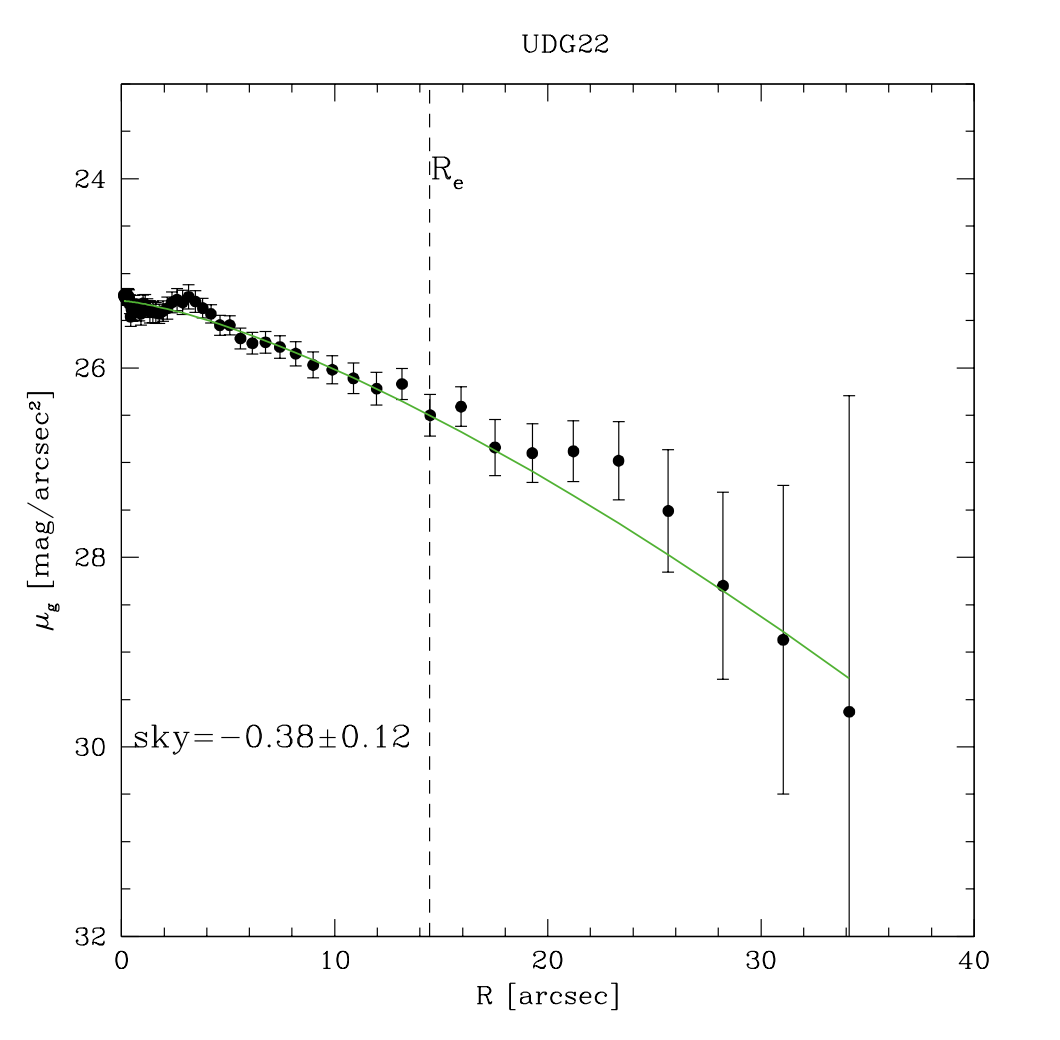} \\
    \includegraphics[width=7cm]{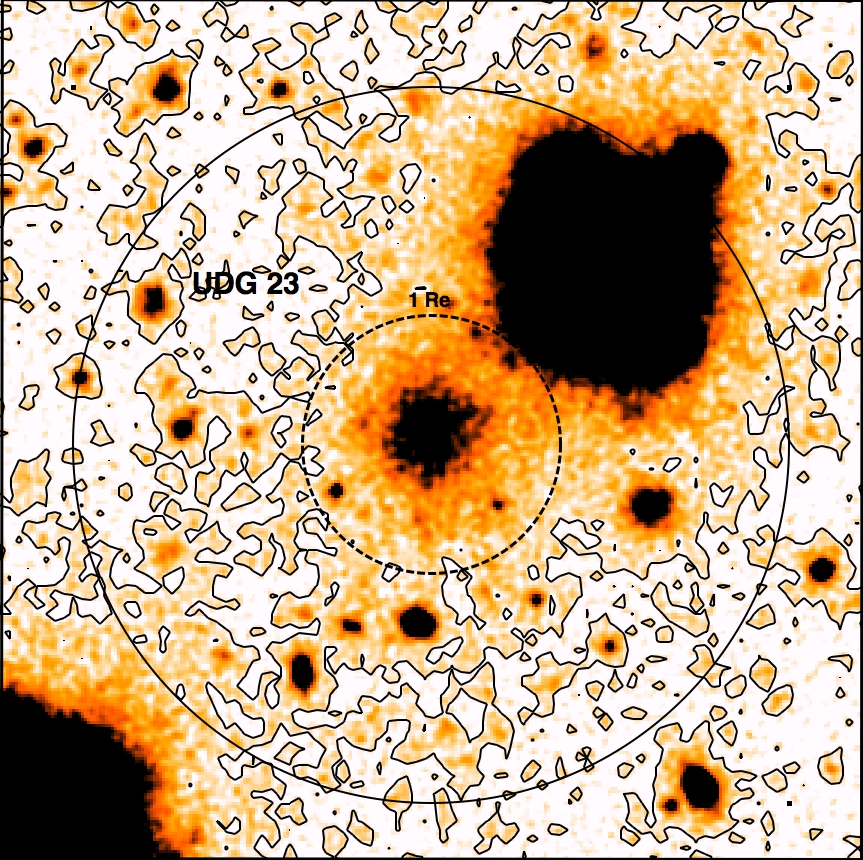} &
    \includegraphics[width=7cm]{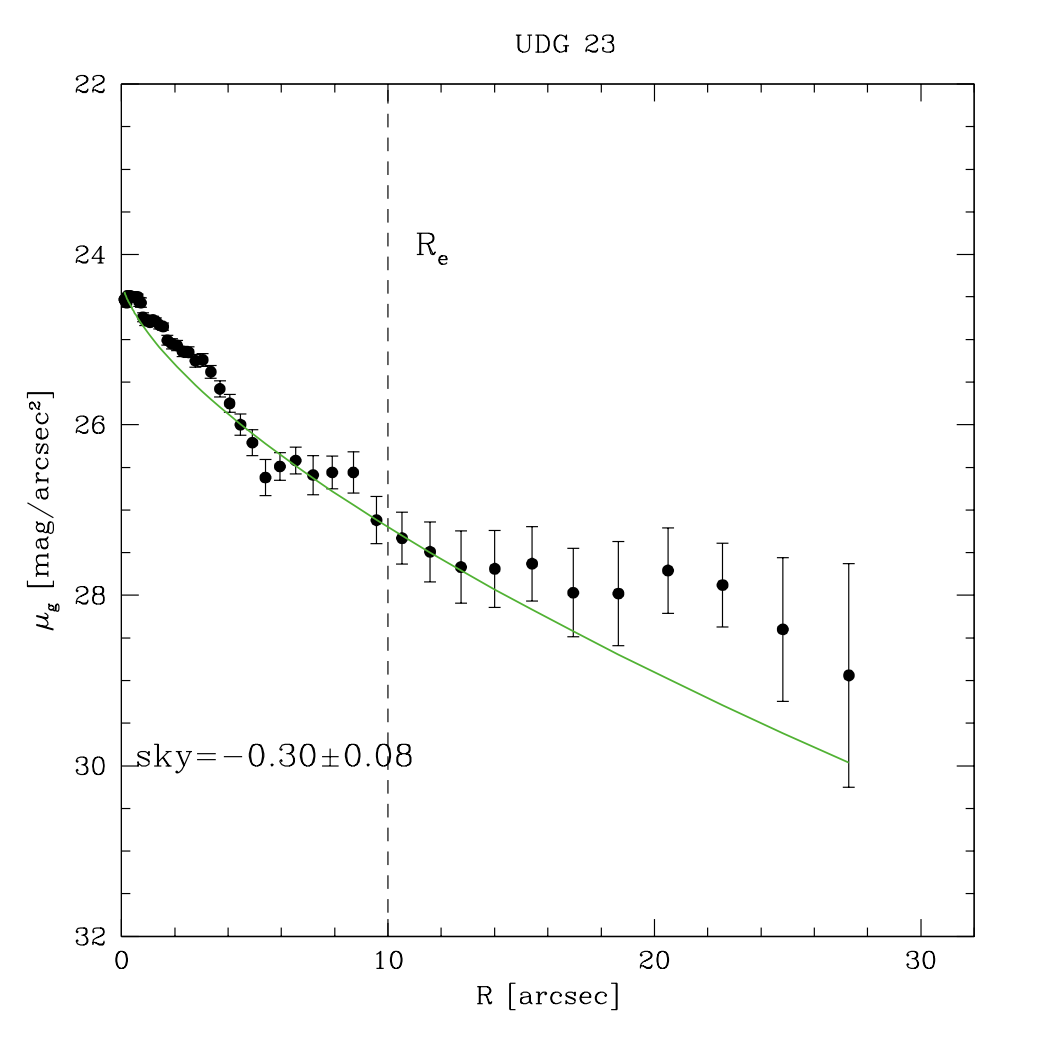} \\
   \end{tabular}
    \caption{Same as Fig.~\ref{fig:UDG_1} for UDG~22 and UDG~23. For UDG22, we have also included the contour
    surface brightness level corresponding at $\mu_e=26.50$~mag/arcsec$^2$, see also Tab.~\ref{tab:UDGsample}.
    }
    \label{fig:UDG_4}
\end{figure*}

\begin{figure*}
\centering
\begin{tabular}{cc}
    \includegraphics[width=7cm]{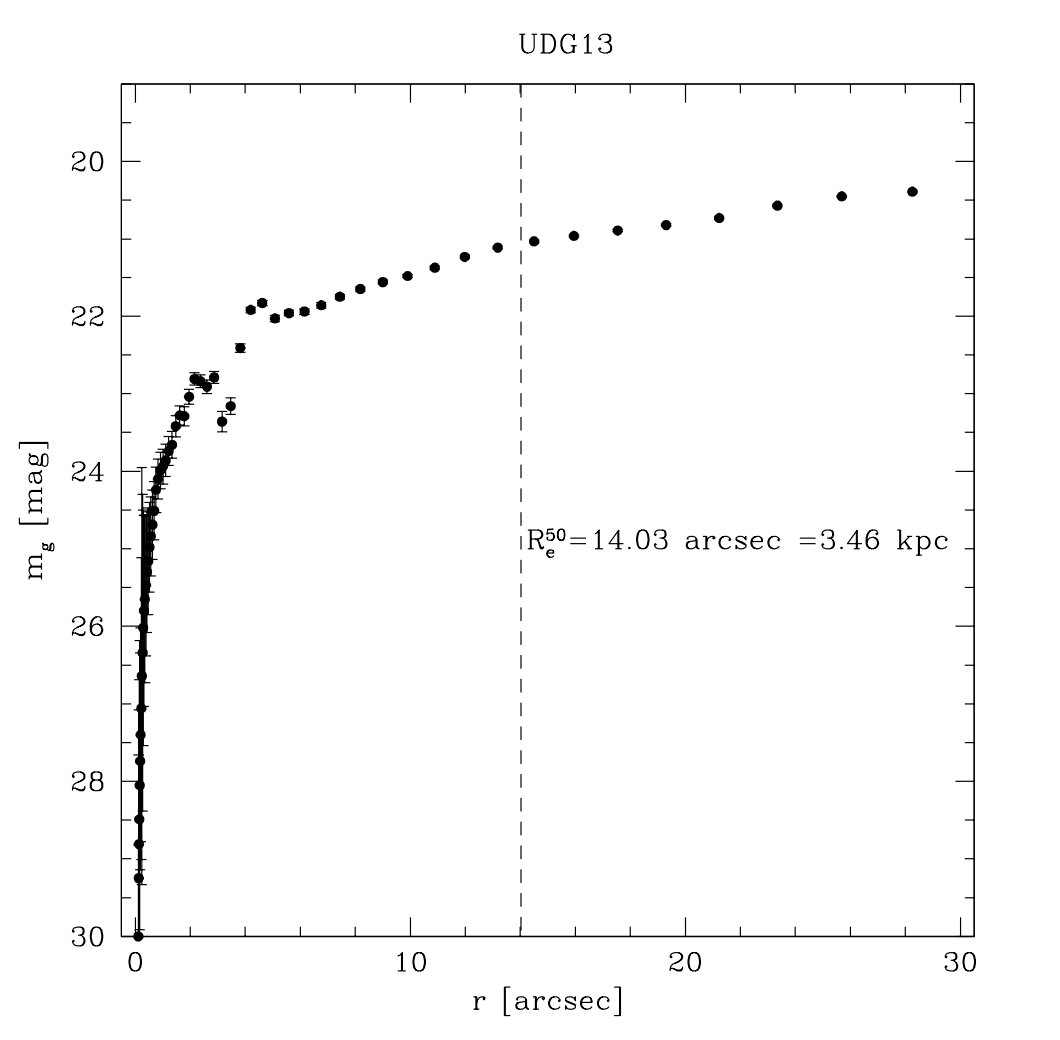} &
    \includegraphics[width=7cm]{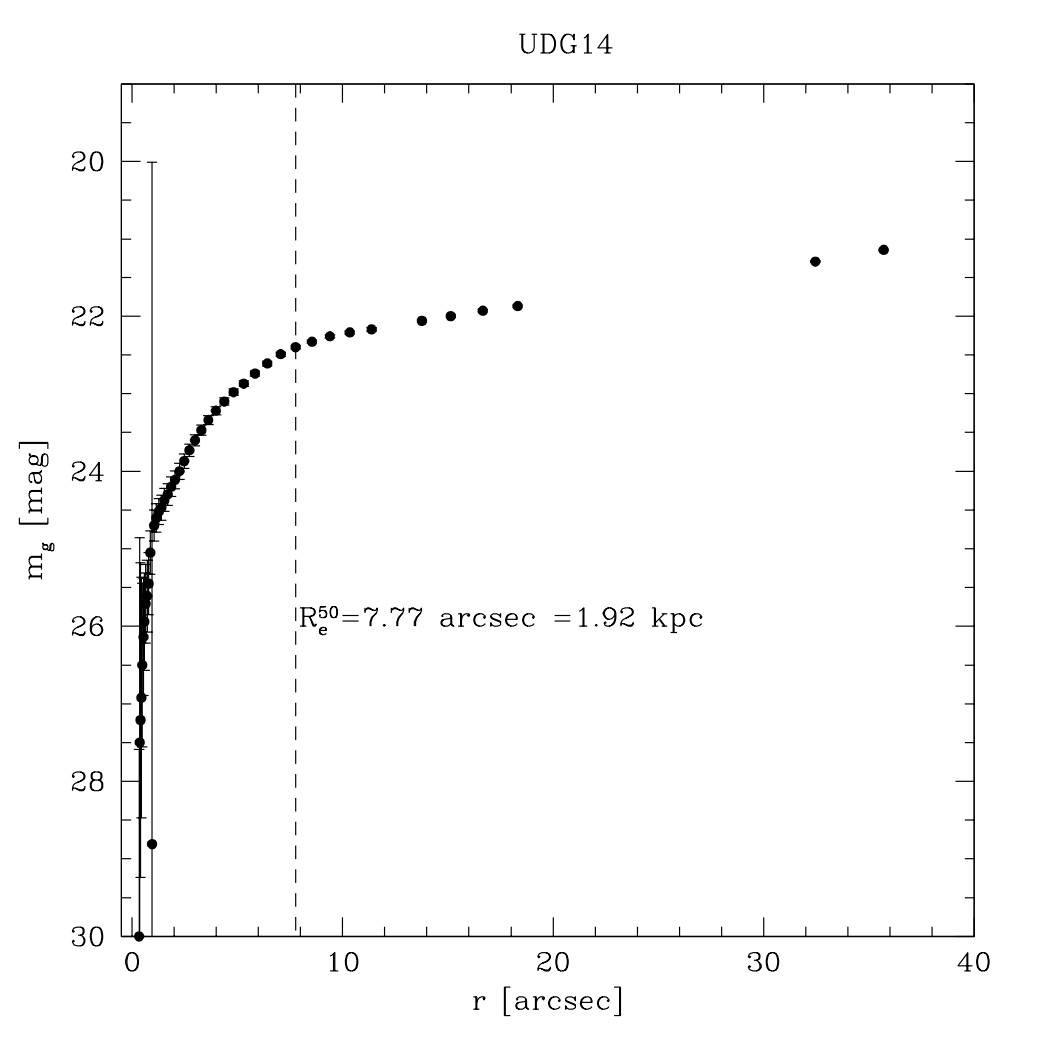} \\
    \includegraphics[width=7cm]{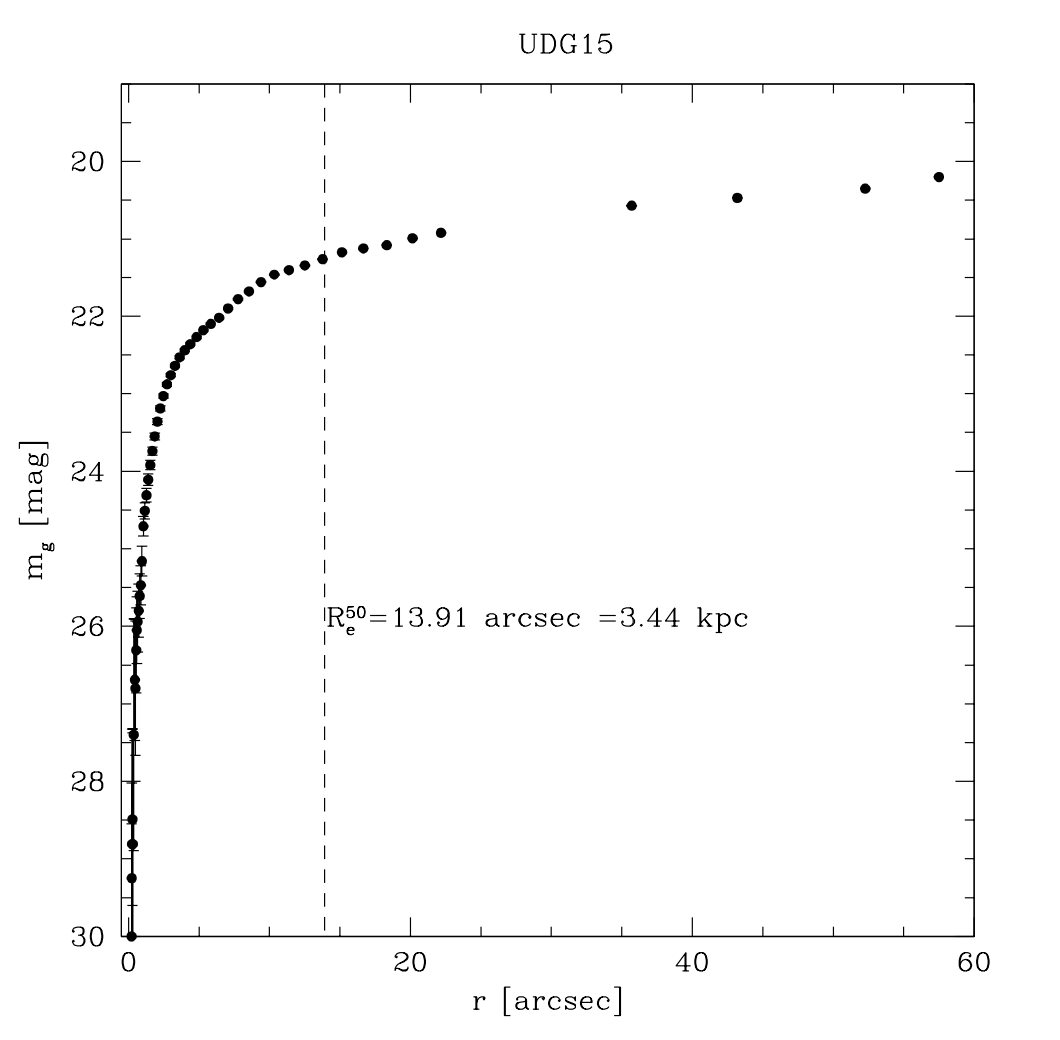} &
    \includegraphics[width=7cm]{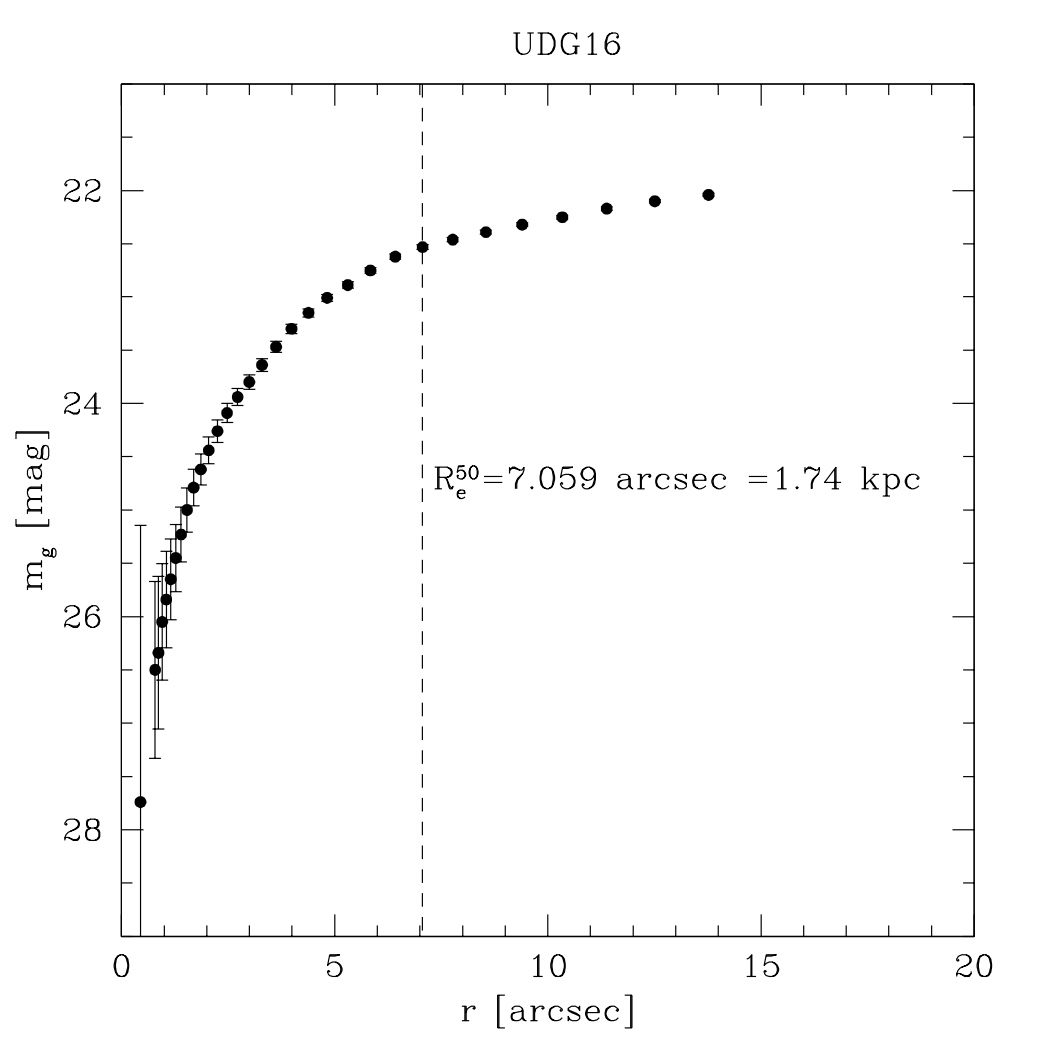} \\
    \includegraphics[width=7cm]{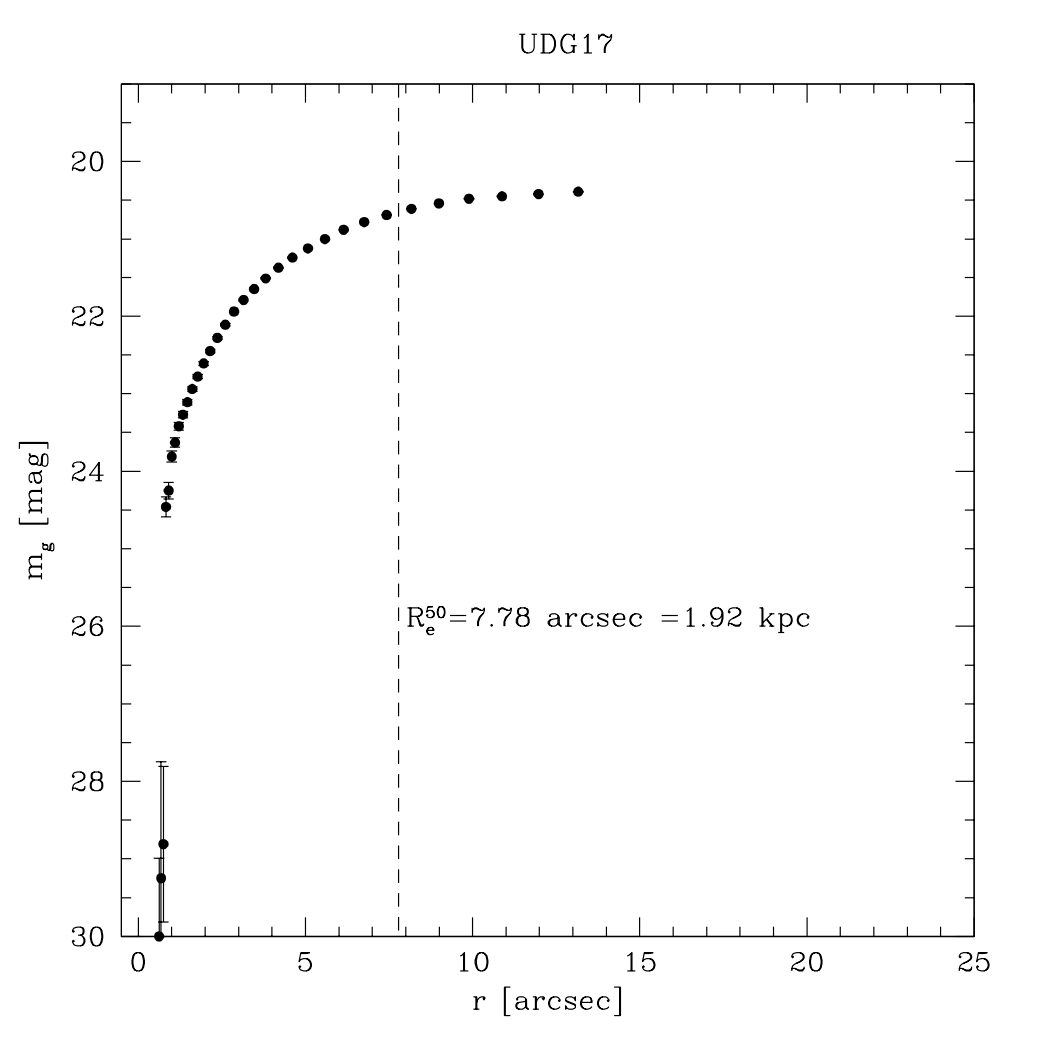} &
    \includegraphics[width=7cm]{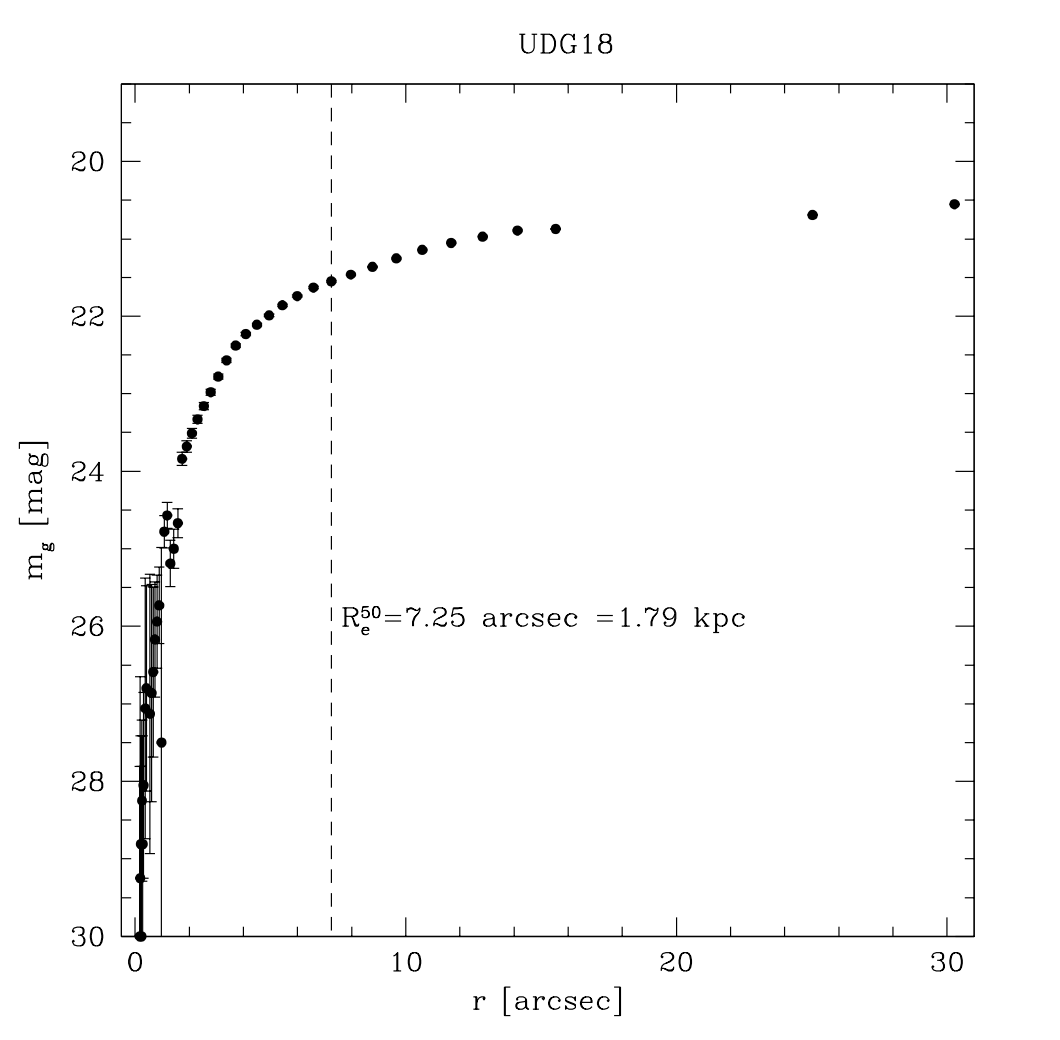} \\
   \end{tabular}
    \caption{Growth curves for UDG candidates in the $g$ band. Integrated magnitudes are computed from the
    integrated fluxes given by the isophote fits (see Sec.~\ref{sec:detection} for details). The vertical dashed line indicates the semi-major half-light radius $R_e^{50}$.}
    \label{fig:growUDG_1}
\end{figure*}

\begin{figure*}
\centering
\begin{tabular}{cc}
    \includegraphics[width=7cm]{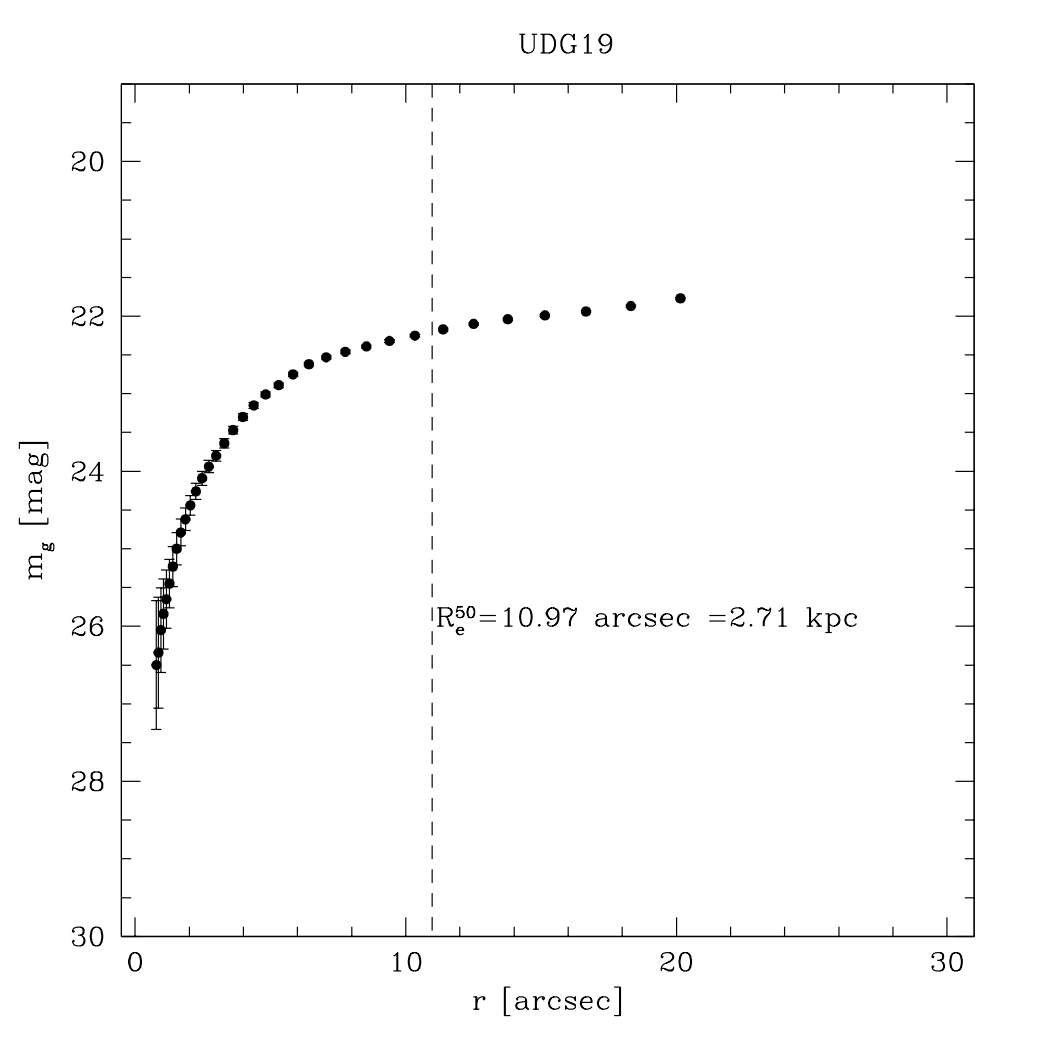} &
    \includegraphics[width=7cm]{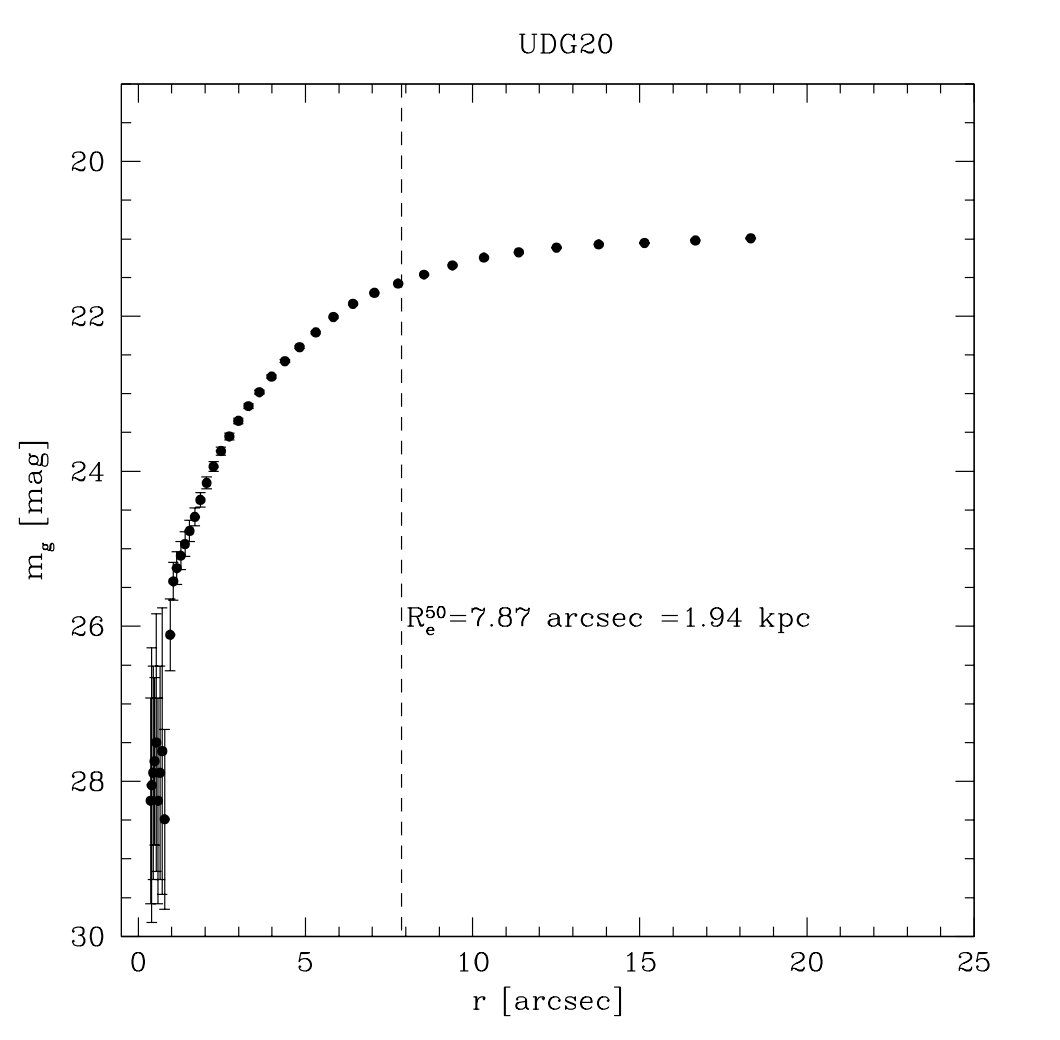} \\
    \includegraphics[width=7cm]{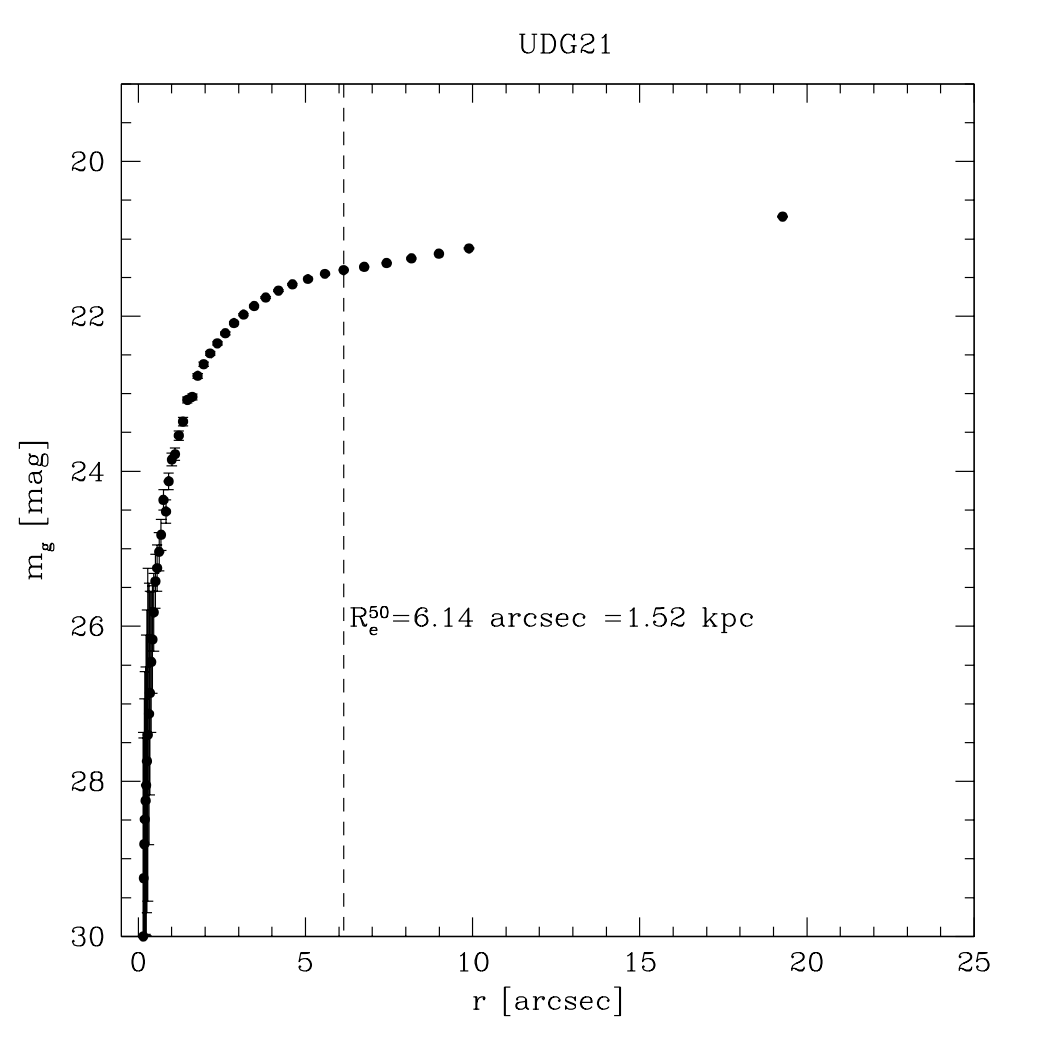} &
    \includegraphics[width=7cm]{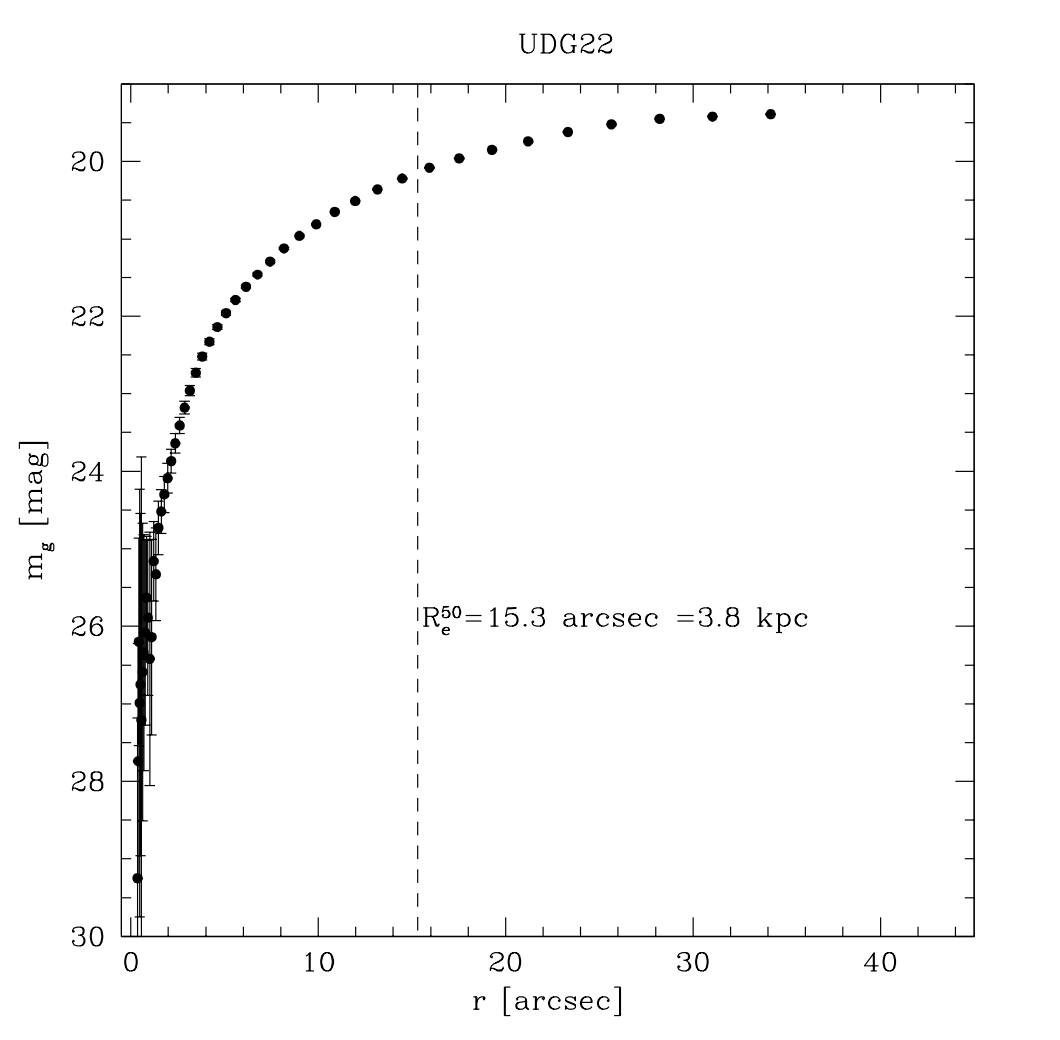} \\
    \includegraphics[width=7cm]{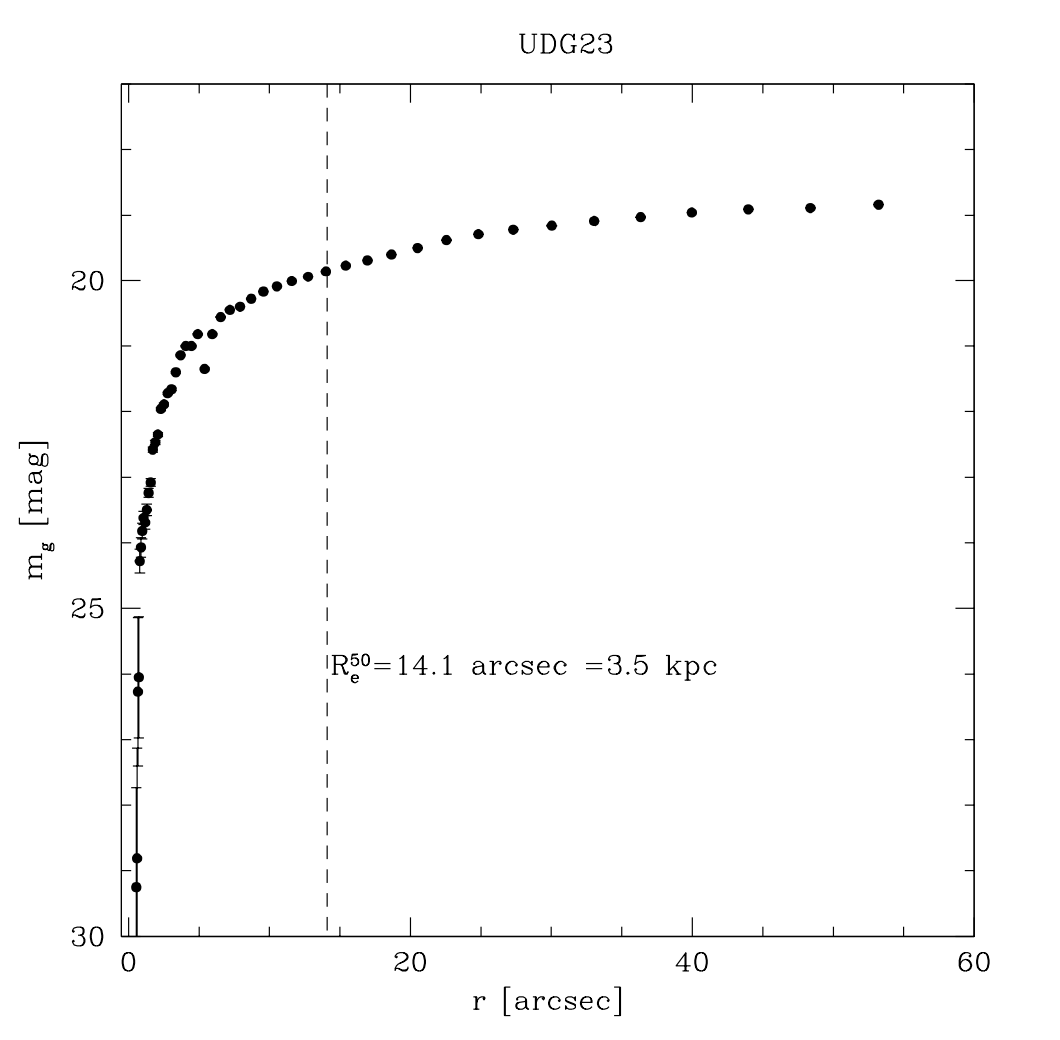} 
   \end{tabular}
    \caption{Same as Fig. \ref{fig:growUDG_1} for UDG 19, UDG 20, UDG 21, UDG 22, and UDG 23.}
    \label{fig:growUDG_2}
\end{figure*}

Here we present the individual images for the new UDG candidates, side by side with their surface brightness profiles in the $g$-band (Fig.~\ref{fig:UDG_1} to Fig.~\ref{fig:UDG_4}). 
The best fit of a single Sérsic profile is overlaid to each profile. 
In Fig.~\ref{fig:growUDG_1} and Fig.~\ref{fig:growUDG_2} we show the growth curves for each UDG, derived from the isophote fitting. In each panel we have included the effective radius derived as the distance from the galaxy centre that includes half of the total flux. This value can be easily compared with the $R_e$ derived from the 1D fit of the surface brightness profile, listed in Tab.~\ref{tab:UDGsample} and also shown in Fig.~\ref{fig:UDG_1} to Fig.~\ref{fig:UDG_4}.

%------------------

\section{Images and surface brightness profiles of 8 new  LSB galaxies}\label{sec:LSBimages}

\begin{figure*}
\centering
\begin{tabular}{cc}
    \includegraphics[width=7cm]{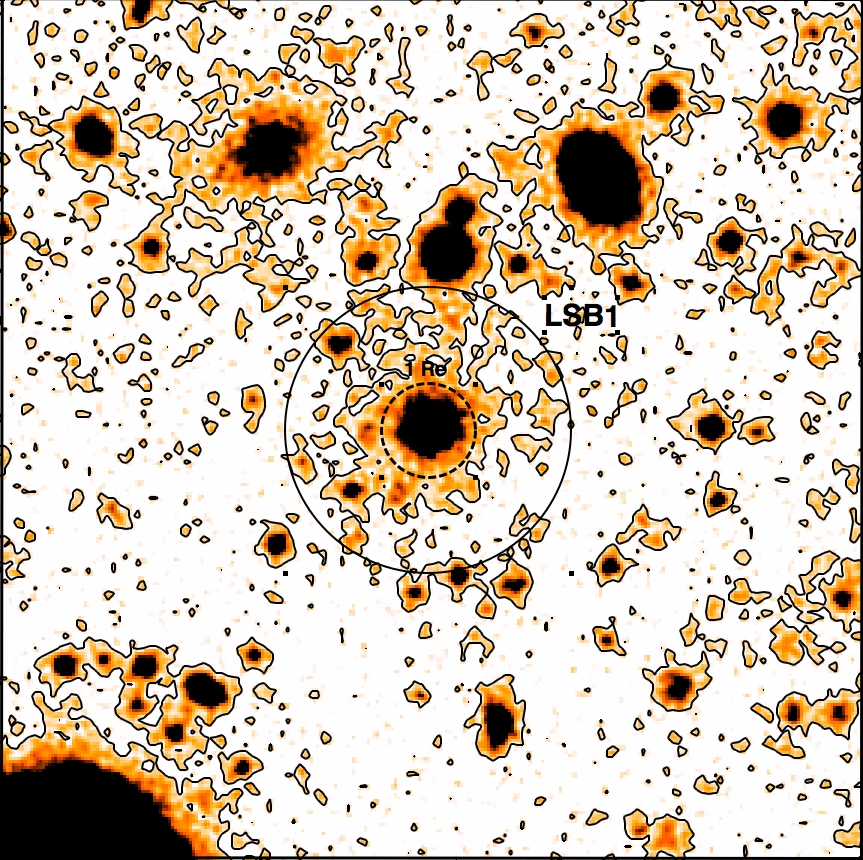} &
    \includegraphics[width=7cm]{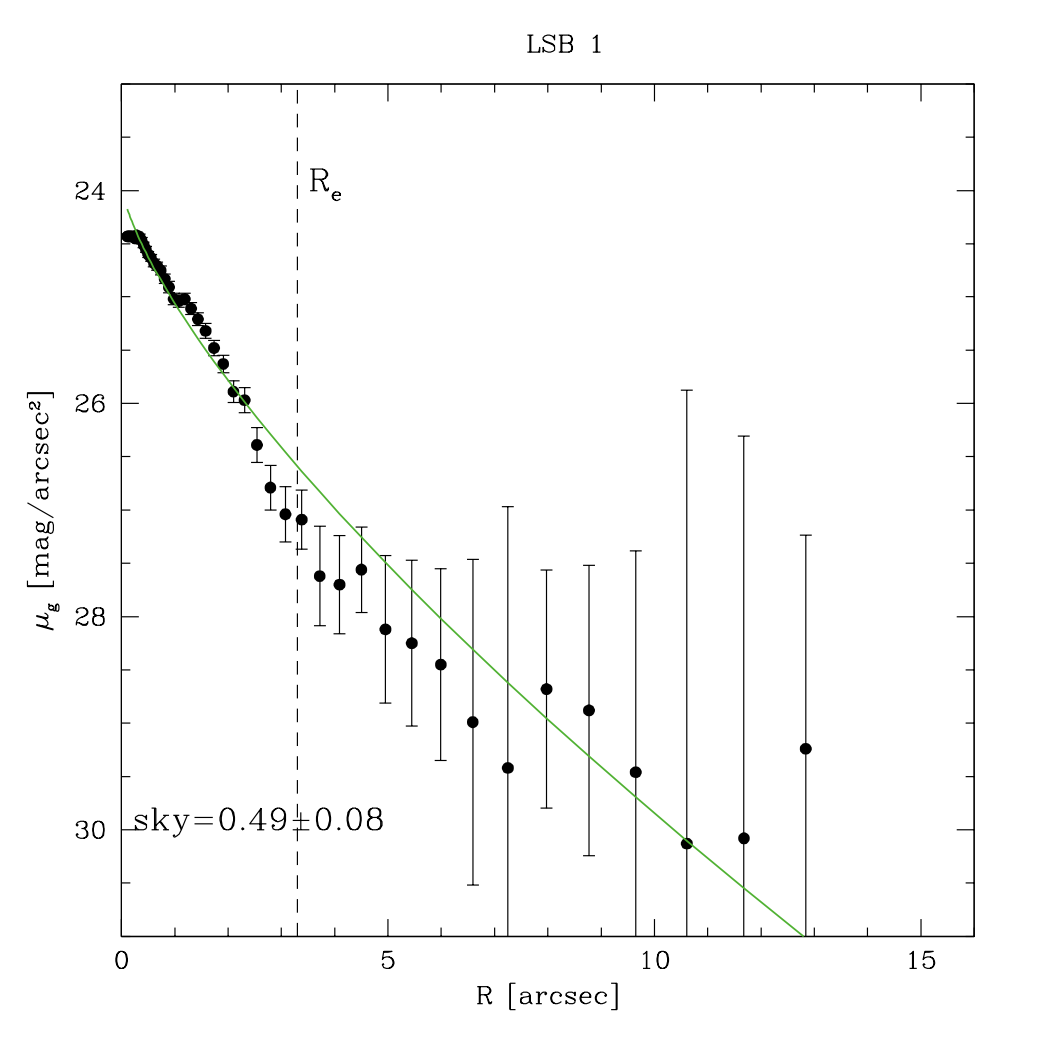} \\
    \includegraphics[width=7cm]{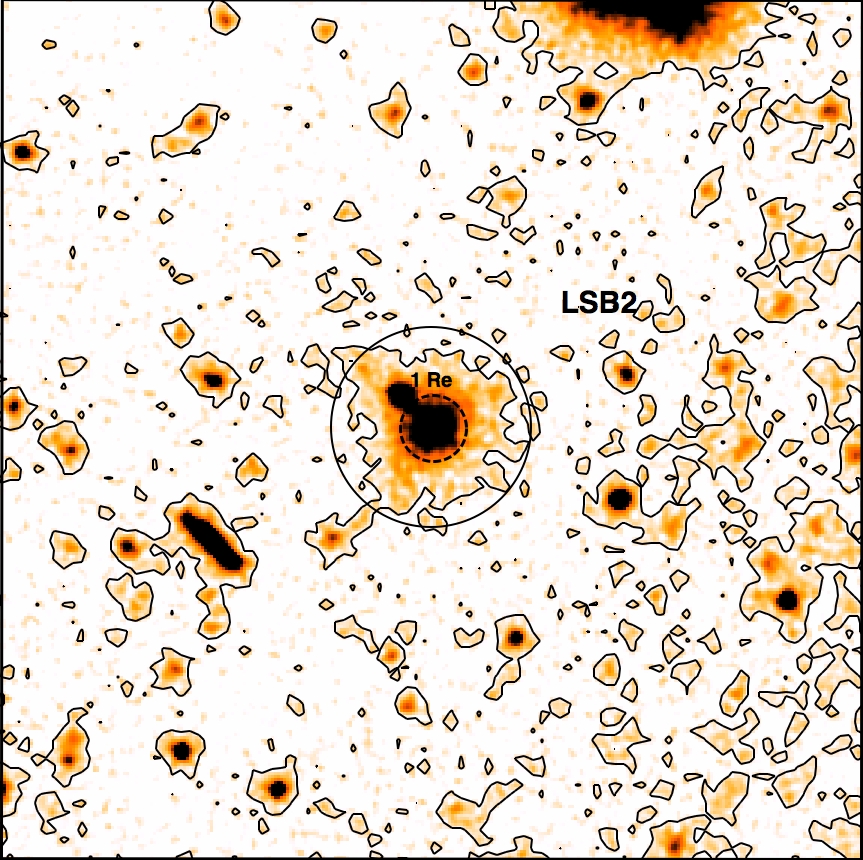} &
    \includegraphics[width=7cm]{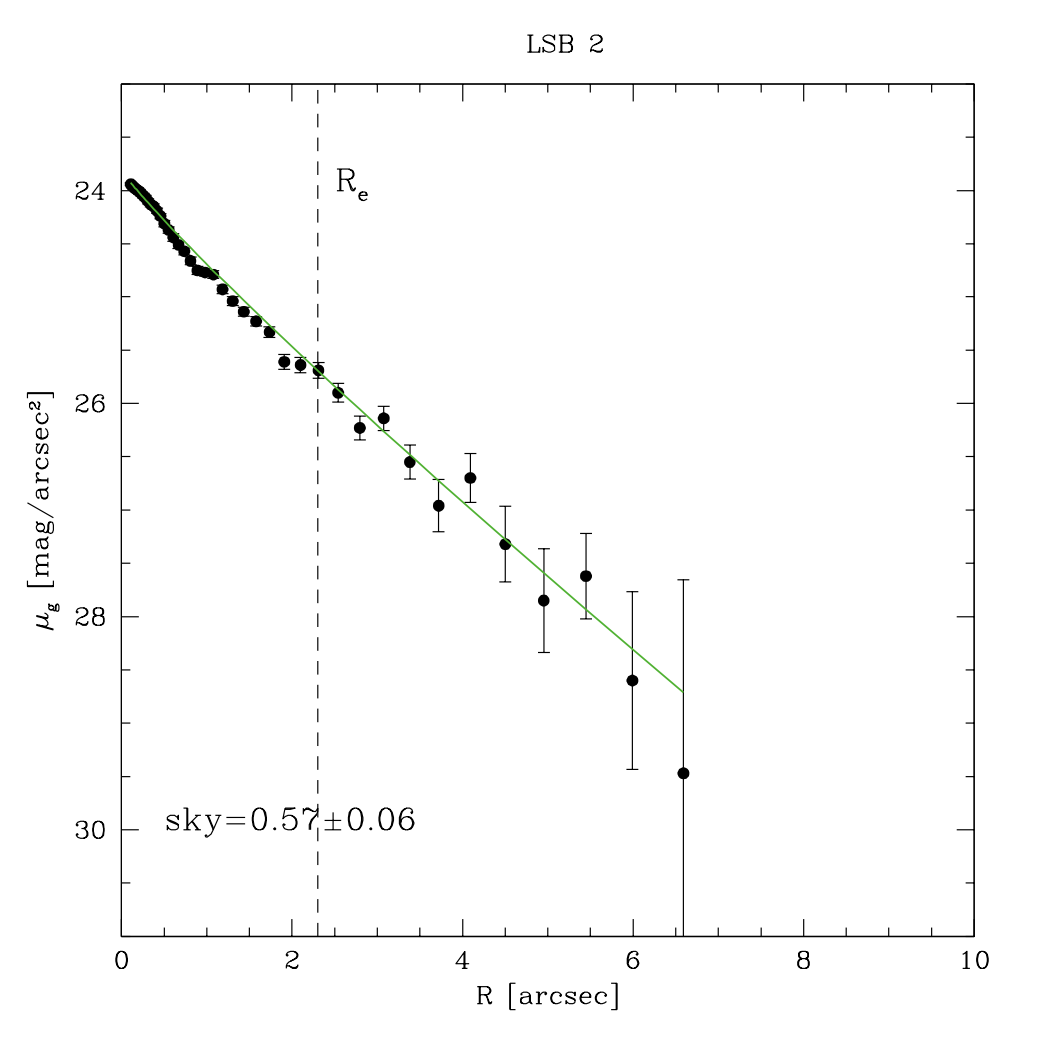} \\
    \includegraphics[width=7cm]{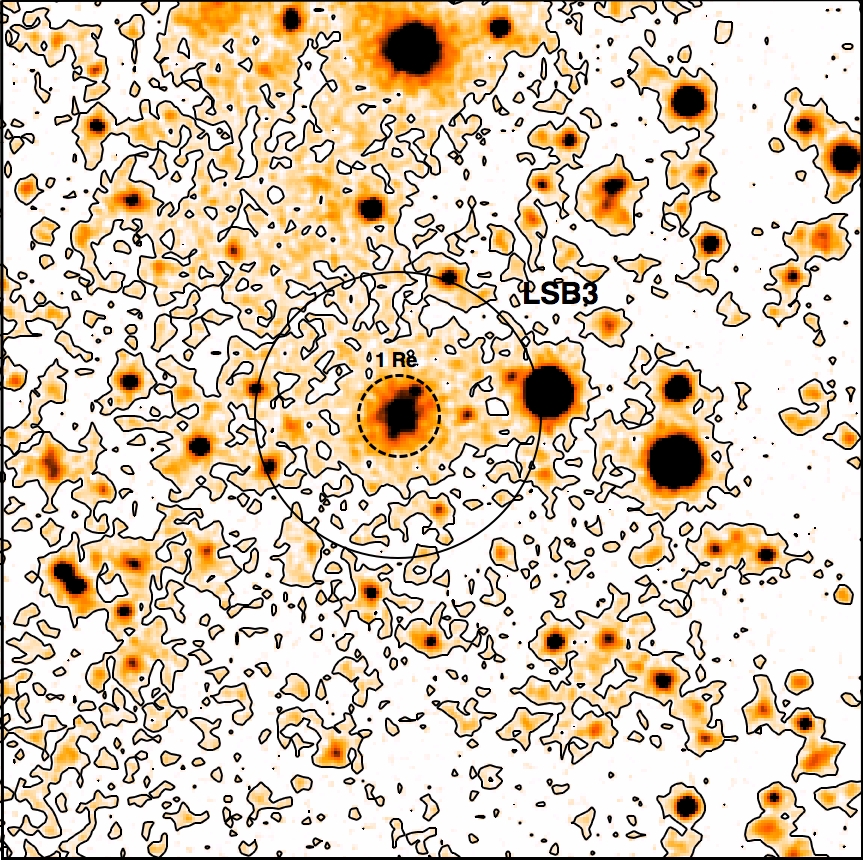} &
    \includegraphics[width=7cm]{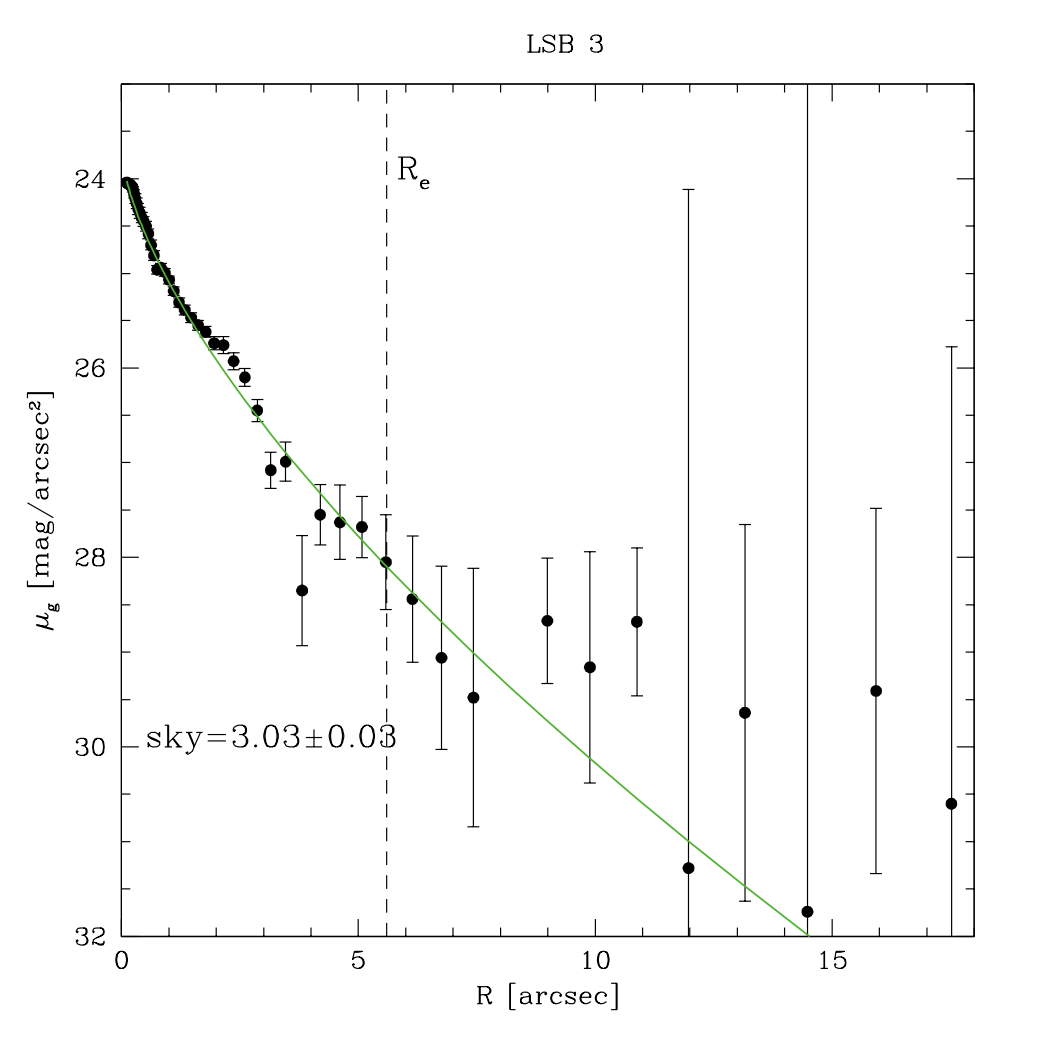} \\
   \end{tabular}
    \caption{Same as Fig.~\ref{fig:UDG_1}, for the  LSB galaxies LSB~1, LSB~2 and LSB~3.}
    \label{fig:LSB_1}
\end{figure*}

\begin{figure*}
\centering
\begin{tabular}{cc}
    \includegraphics[width=7cm]{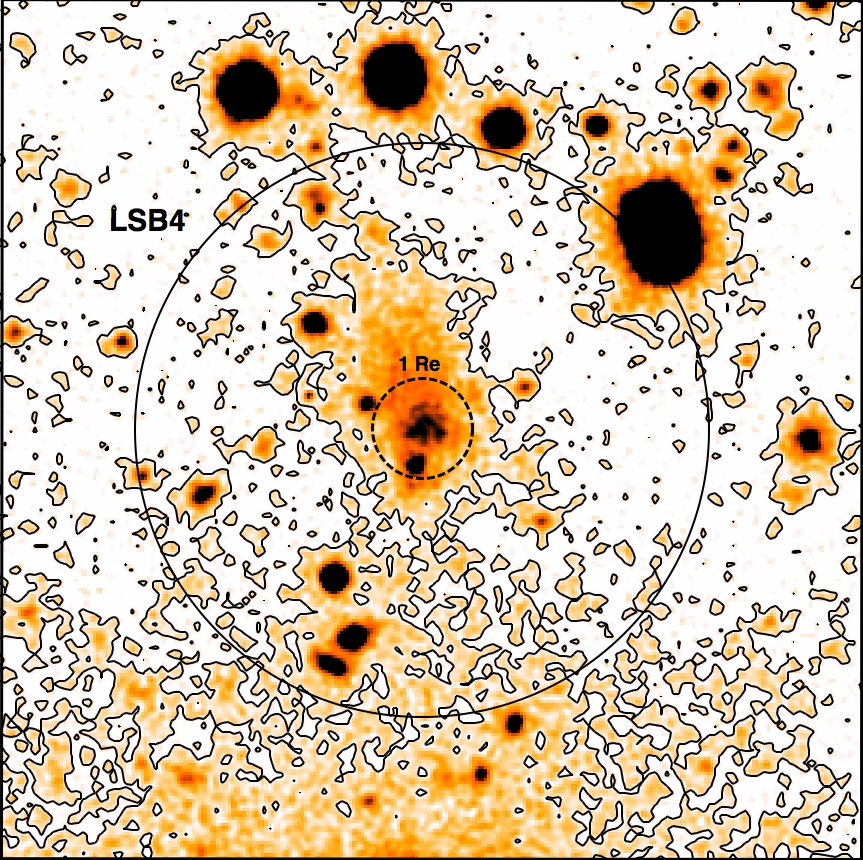} &
    \includegraphics[width=7cm]{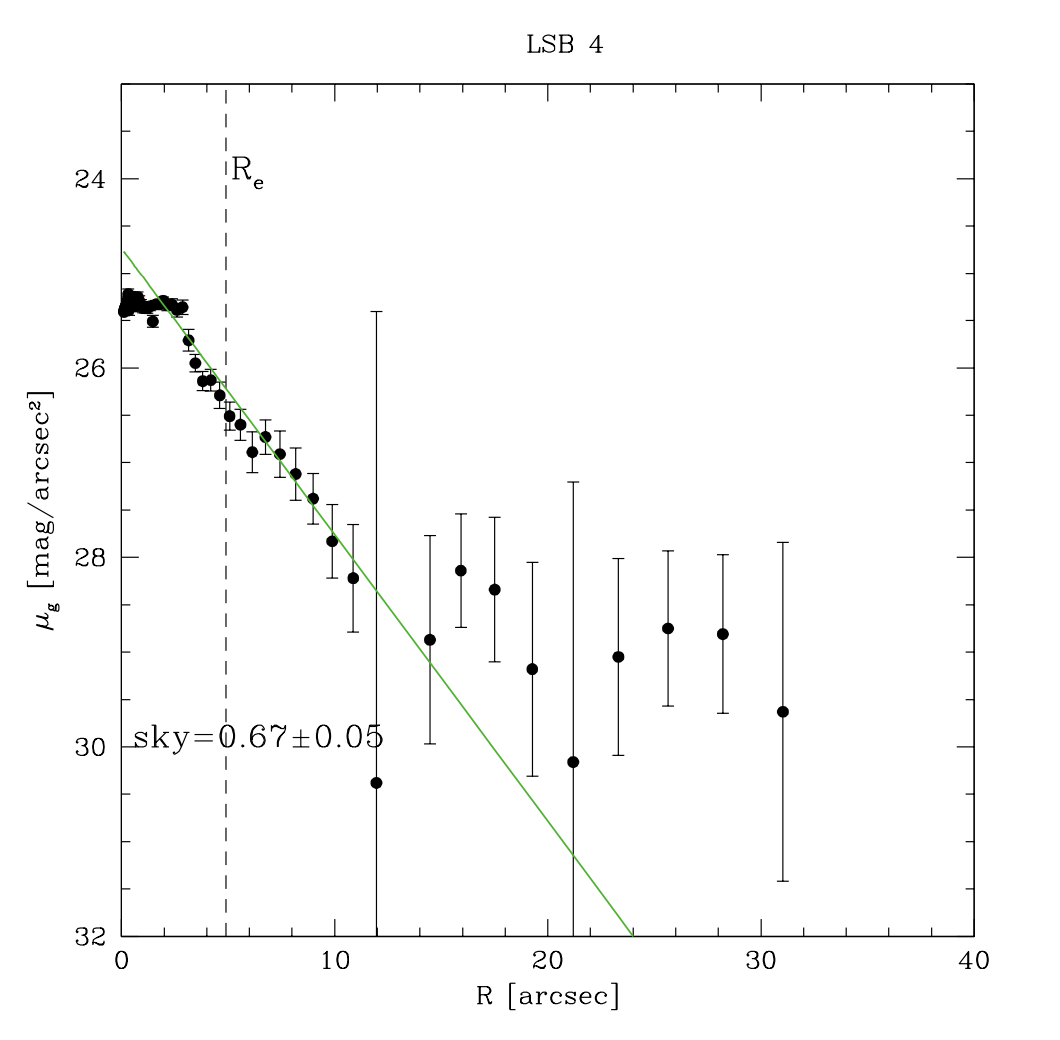} \\
    \includegraphics[width=7cm]{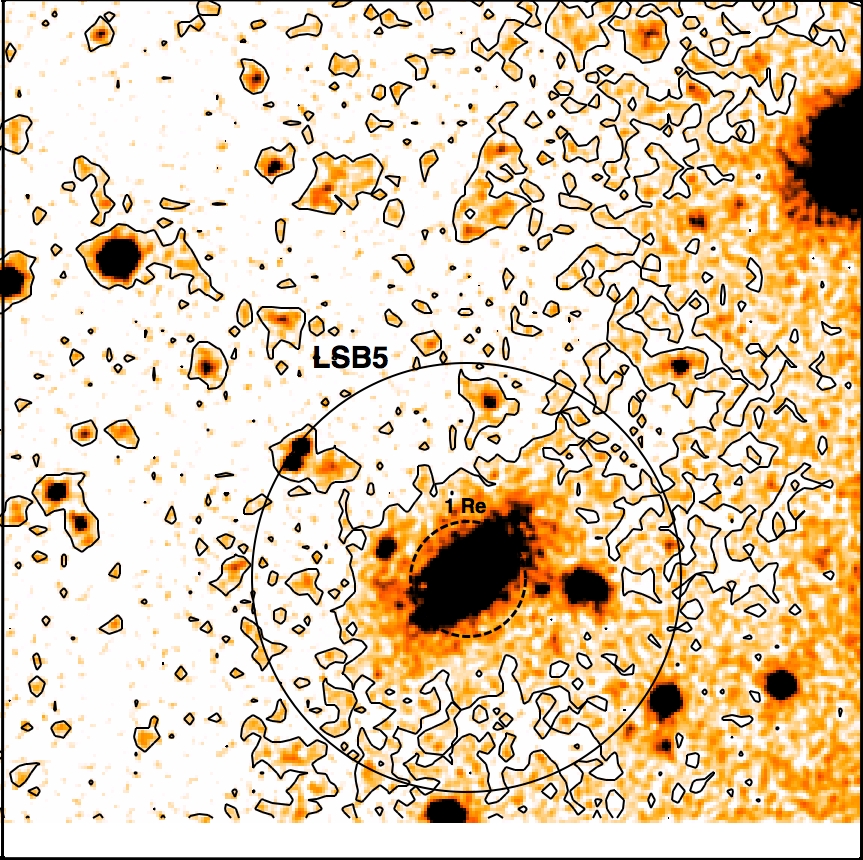} &
    \includegraphics[width=7cm]{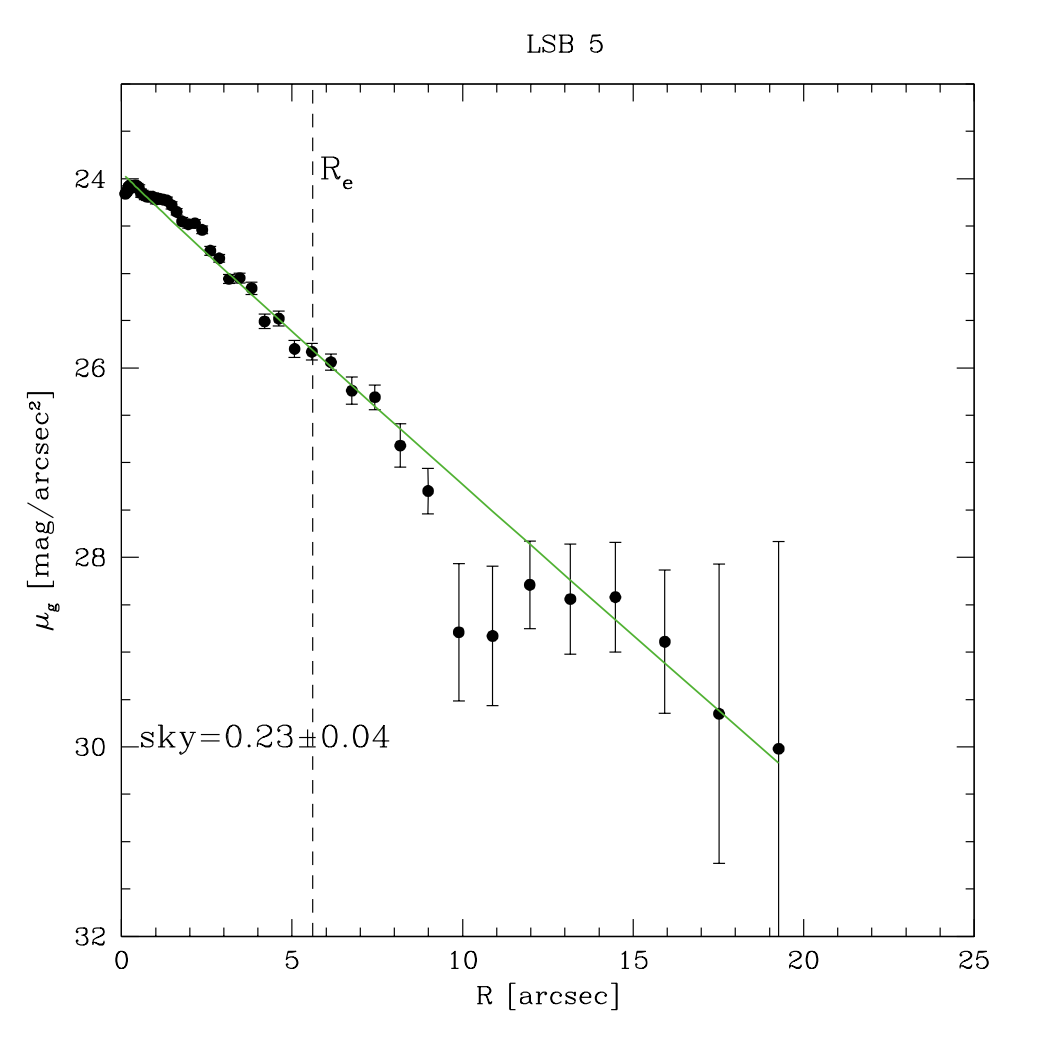} \\
    \includegraphics[width=7cm]{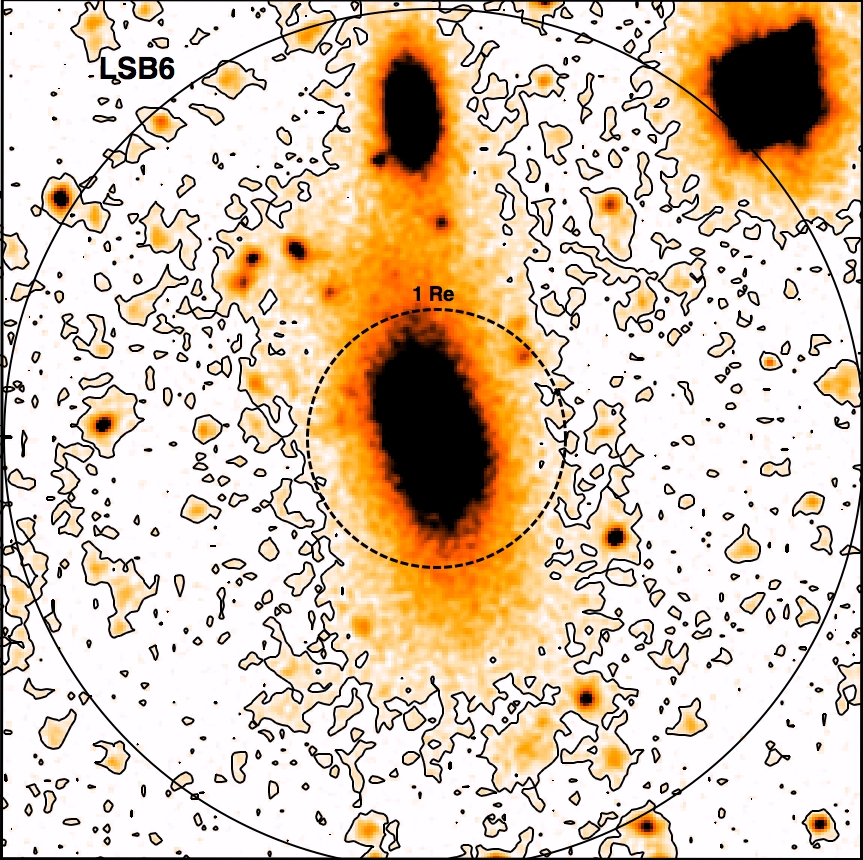} &
    \includegraphics[width=7cm]{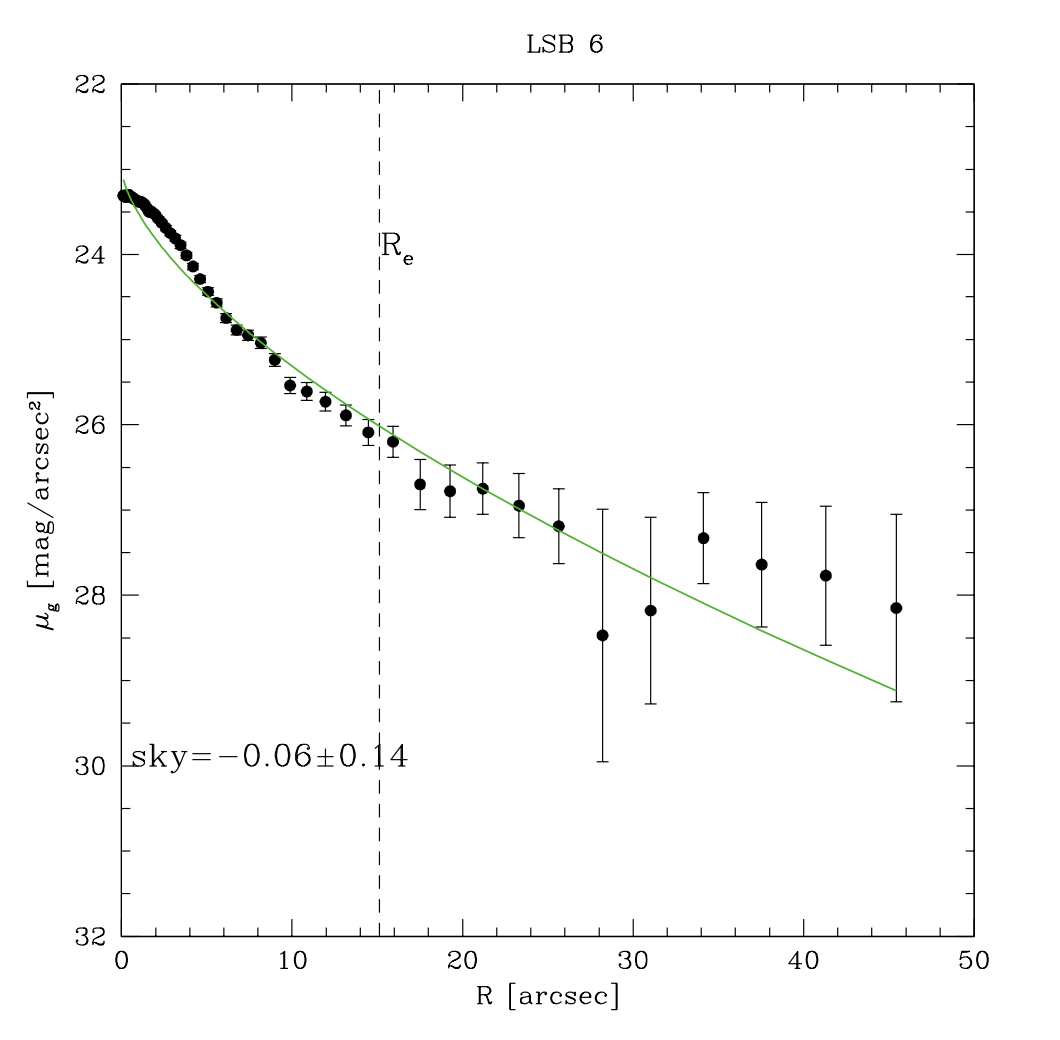} \\
   \end{tabular}
    \caption{Same as Fig.~\ref{fig:LSB_1} for LSB~4, LSB~5 and LSB~6.}
    \label{fig:LSB_2}
\end{figure*}

\begin{figure*}
\centering
\begin{tabular}{cc}
    \includegraphics[width=7cm]{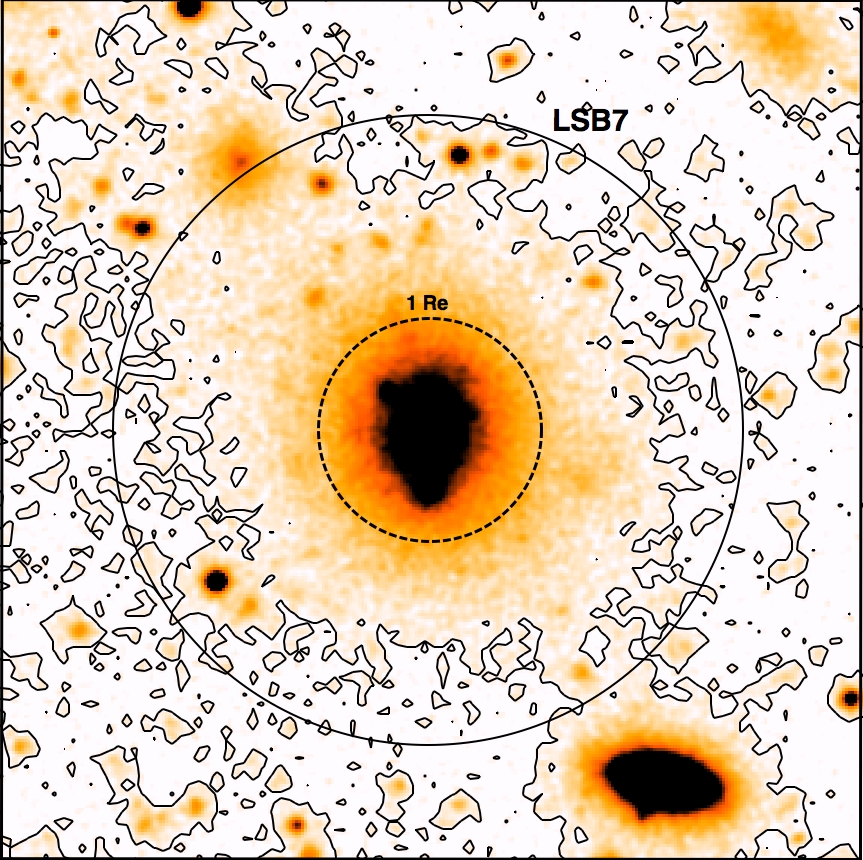} &
    \includegraphics[width=7cm]{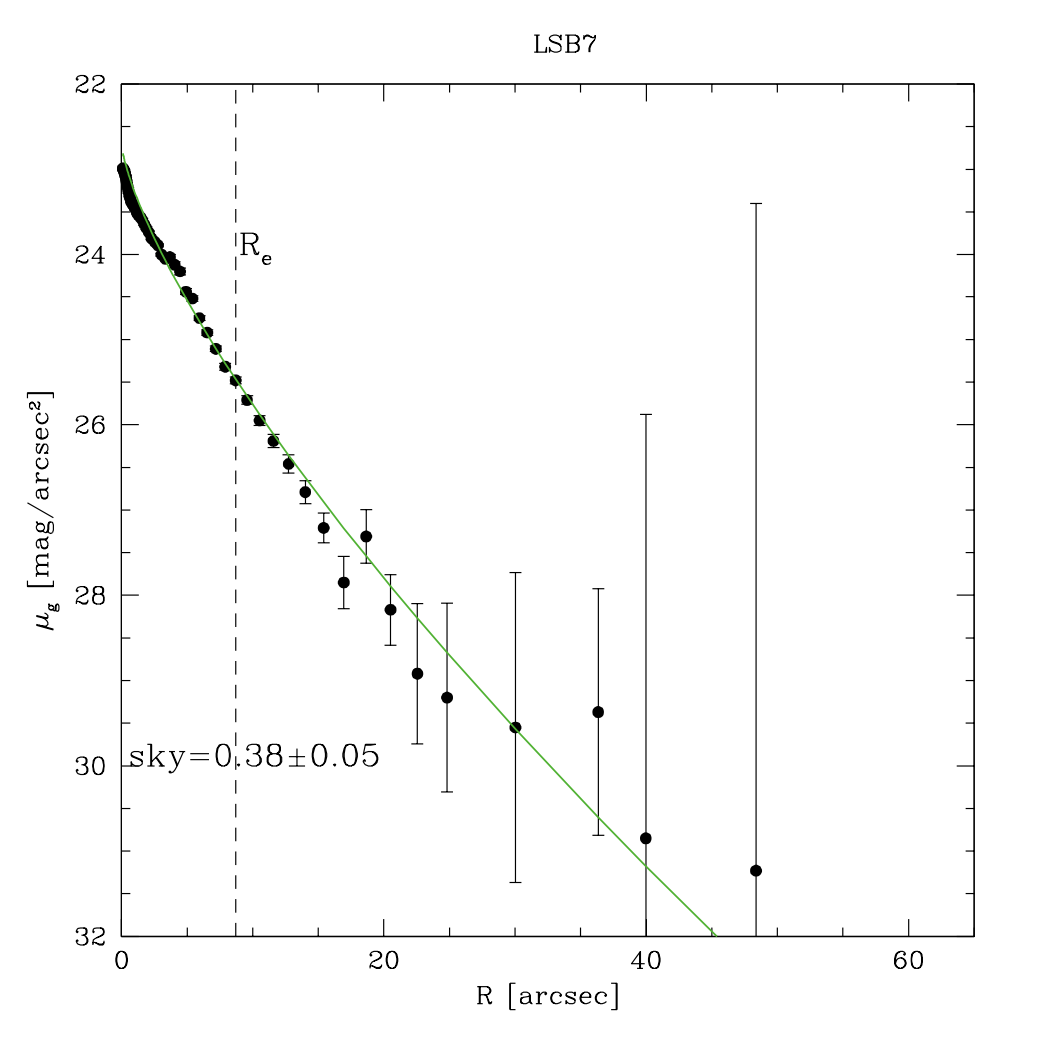} \\
    \includegraphics[width=7cm]{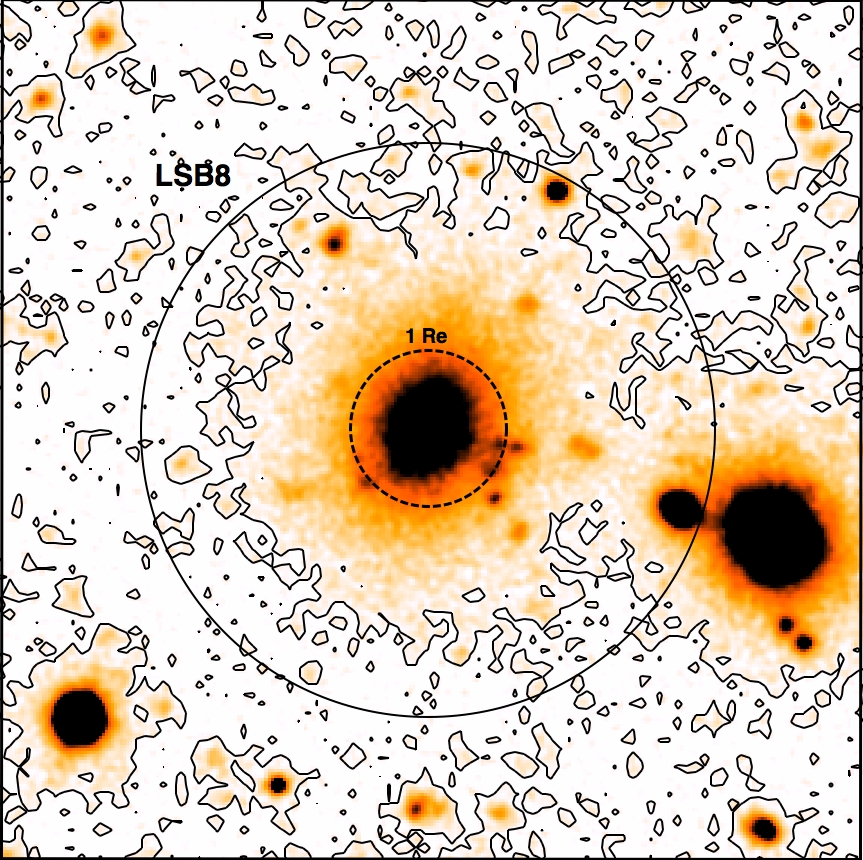} &
    \includegraphics[width=7cm]{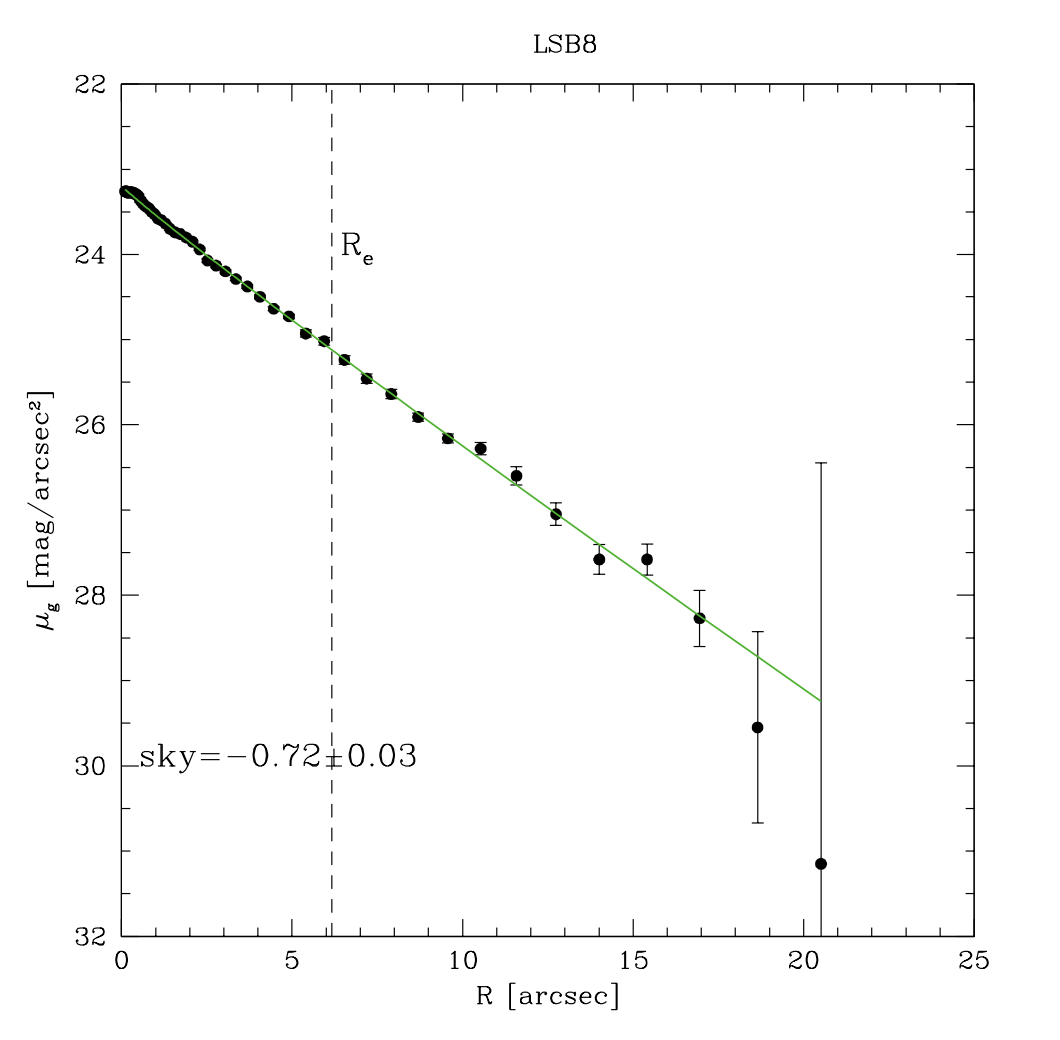} \\
   \end{tabular}
    \caption{Same as Fig.~\ref{fig:LSB_1} LSB~7  and LSB~8.}
    \label{fig:LSB_3}
\end{figure*}

\begin{figure*}
\centering
\begin{tabular}{cc}
    \includegraphics[width=7cm]{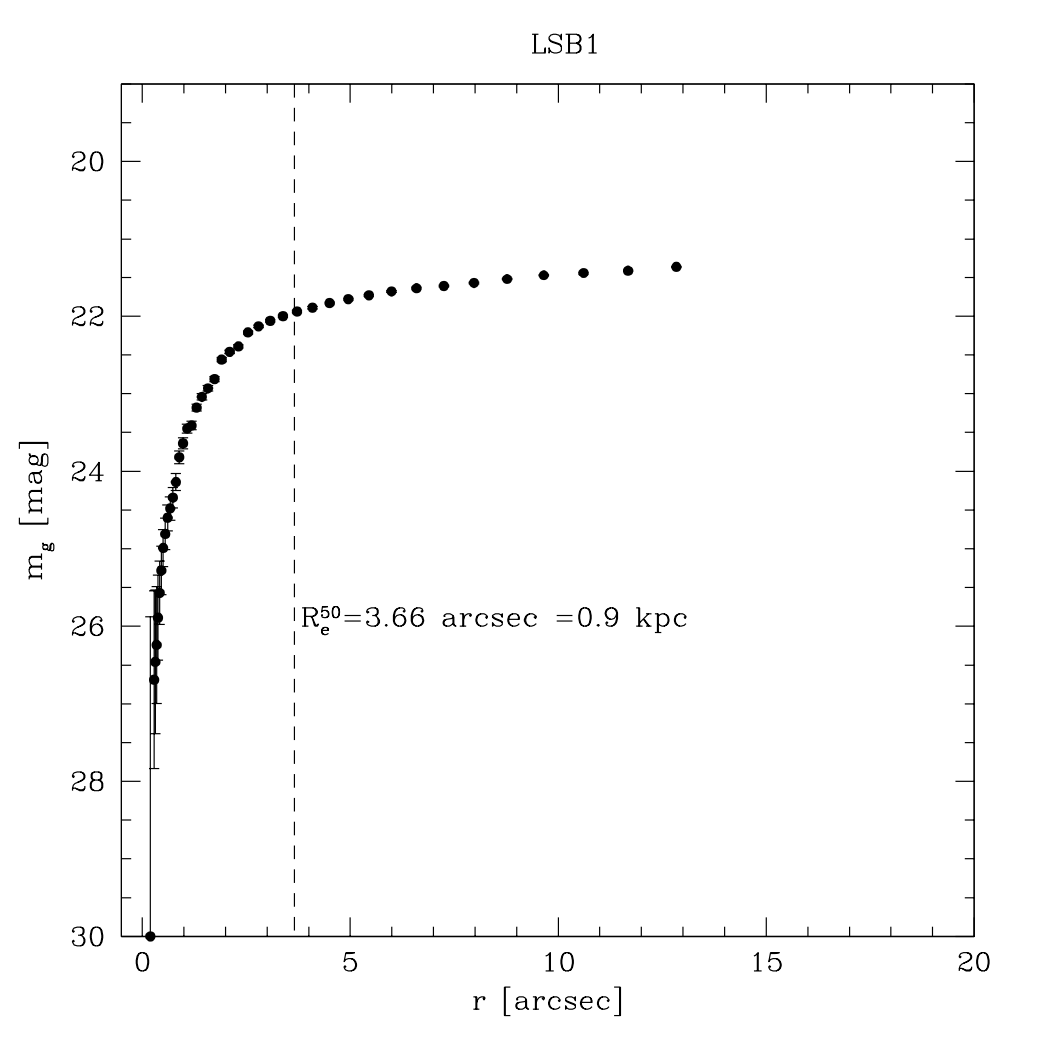} &
    \includegraphics[width=7cm]{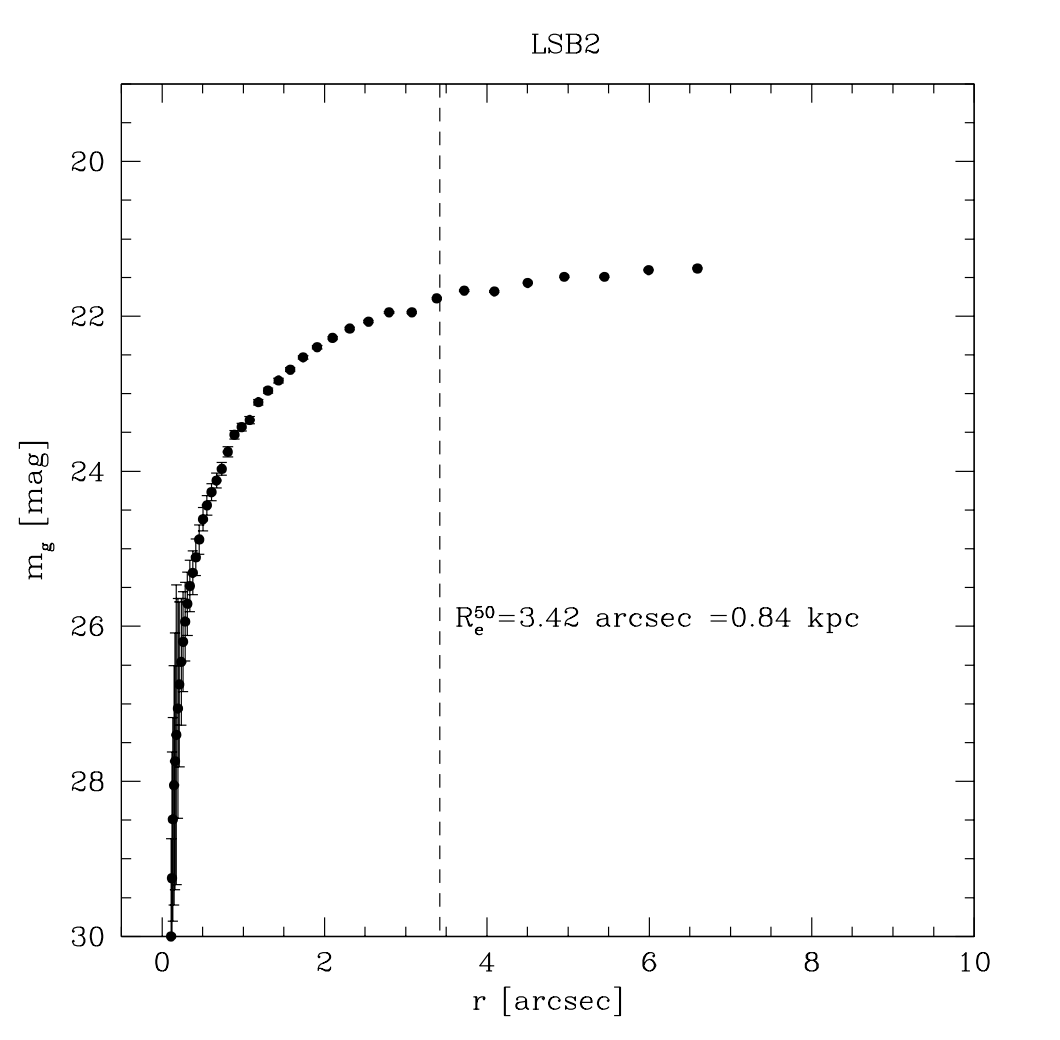} \\
    \includegraphics[width=7cm]{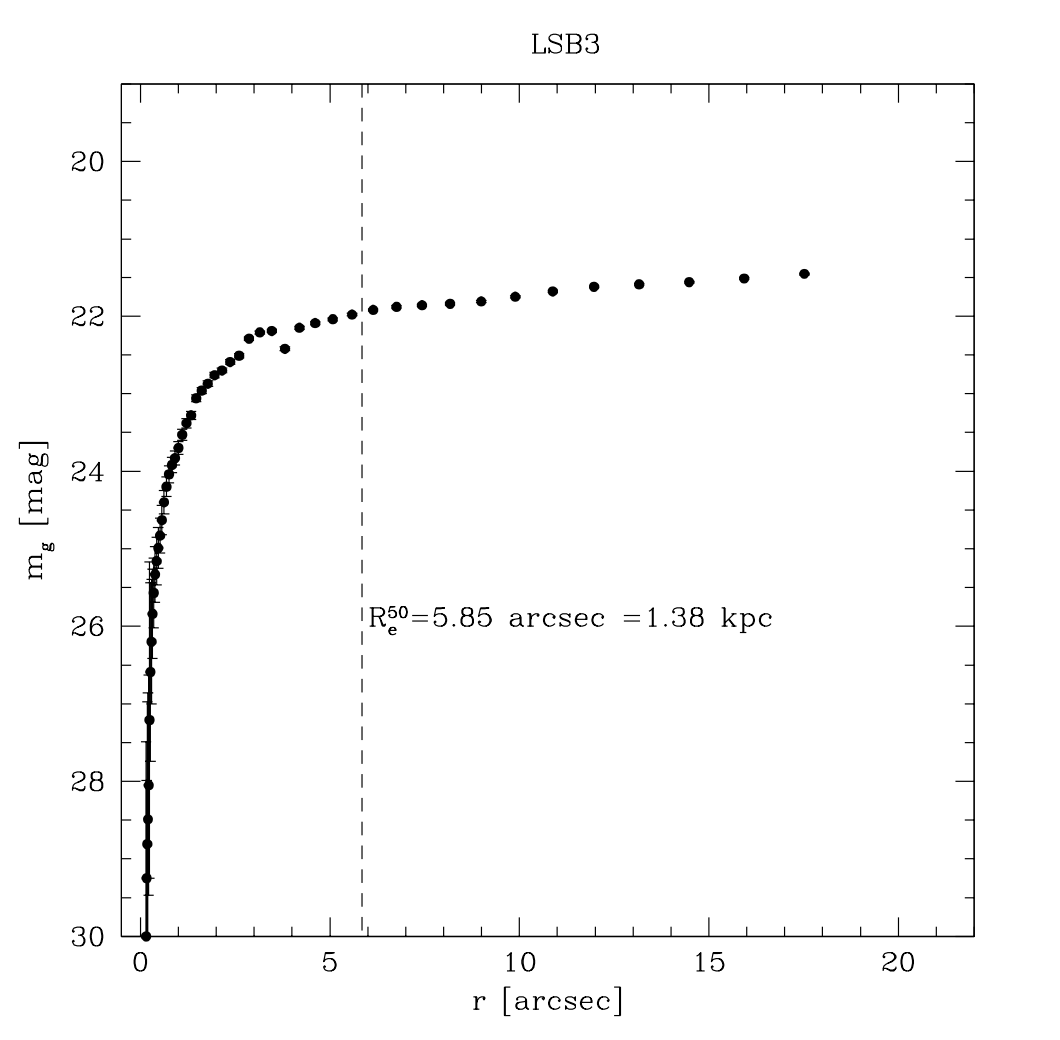} &
    \includegraphics[width=7cm]{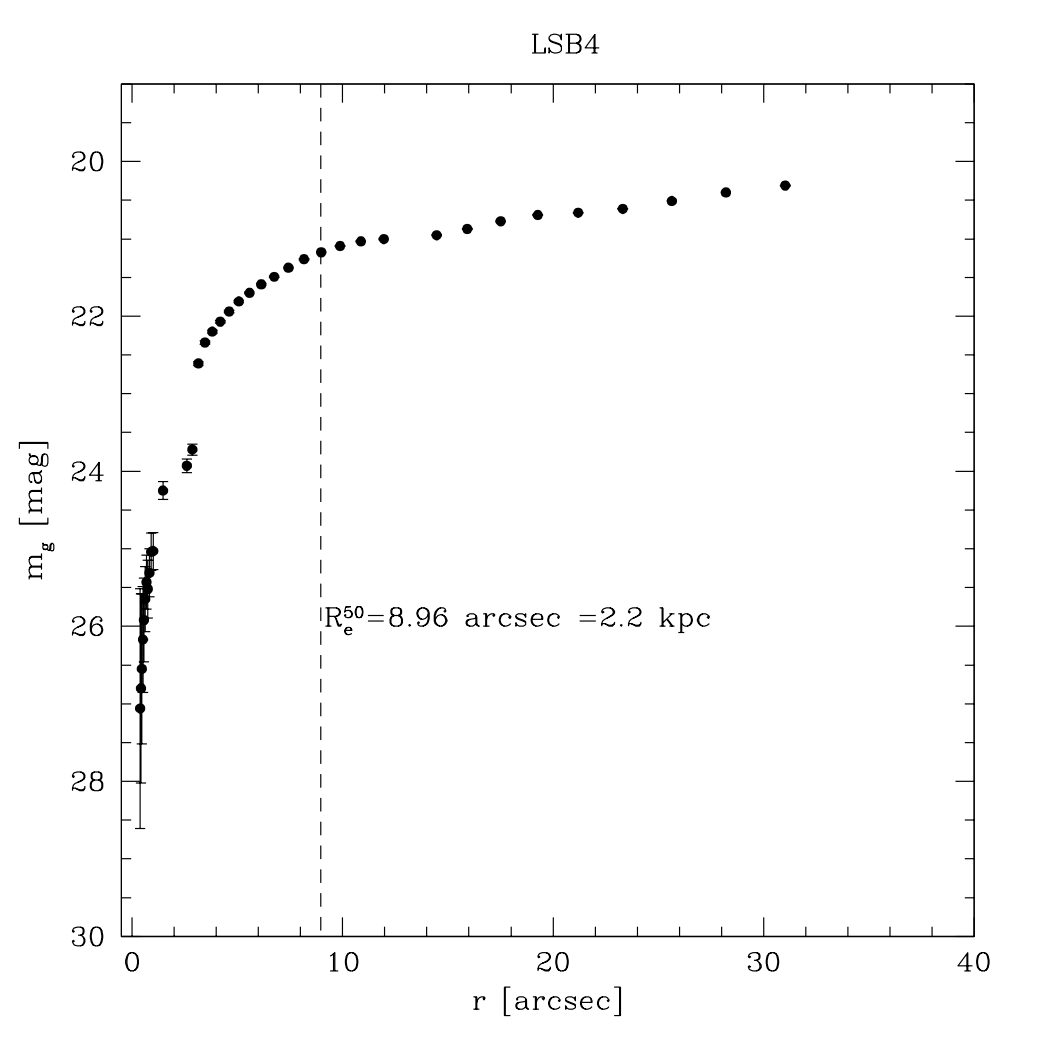} \\
    \includegraphics[width=7cm]{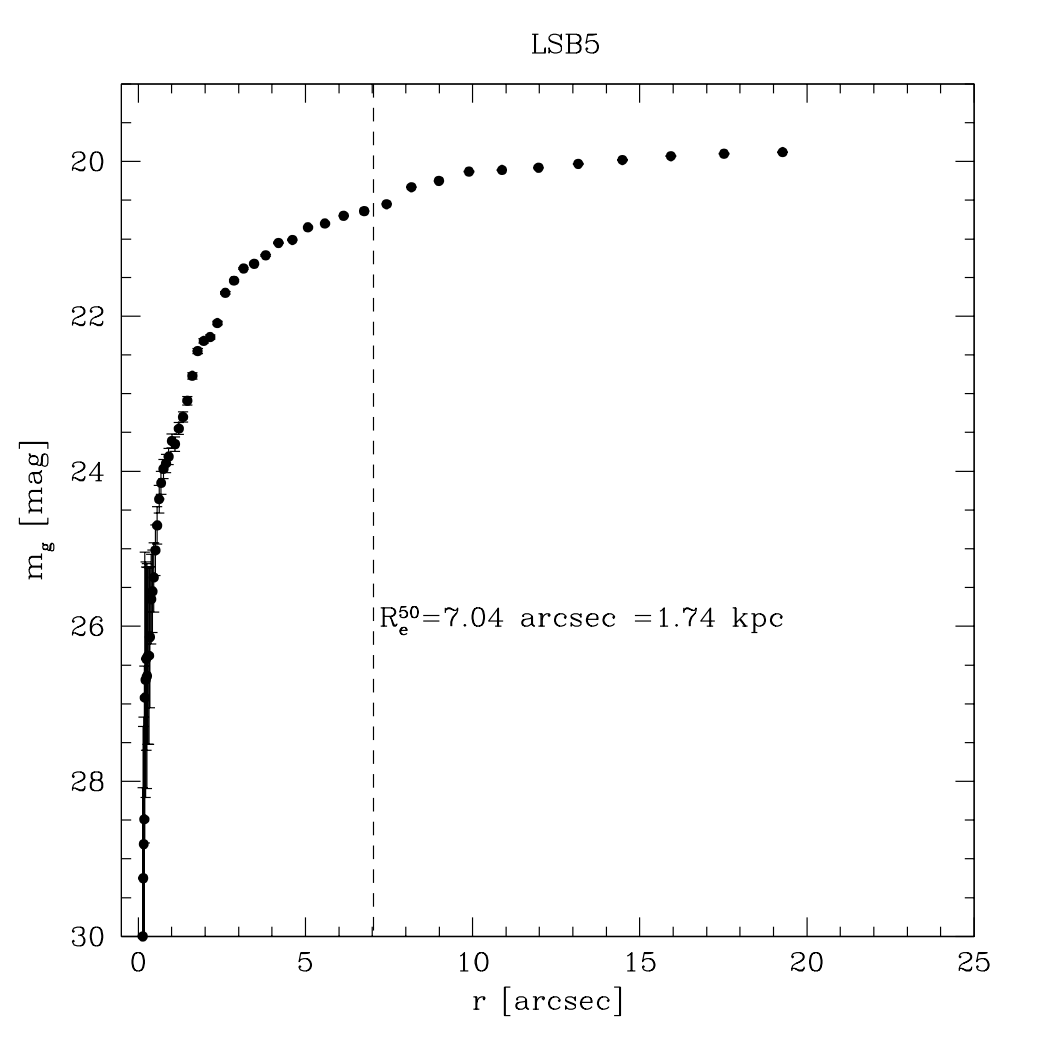} &
    \includegraphics[width=7cm]{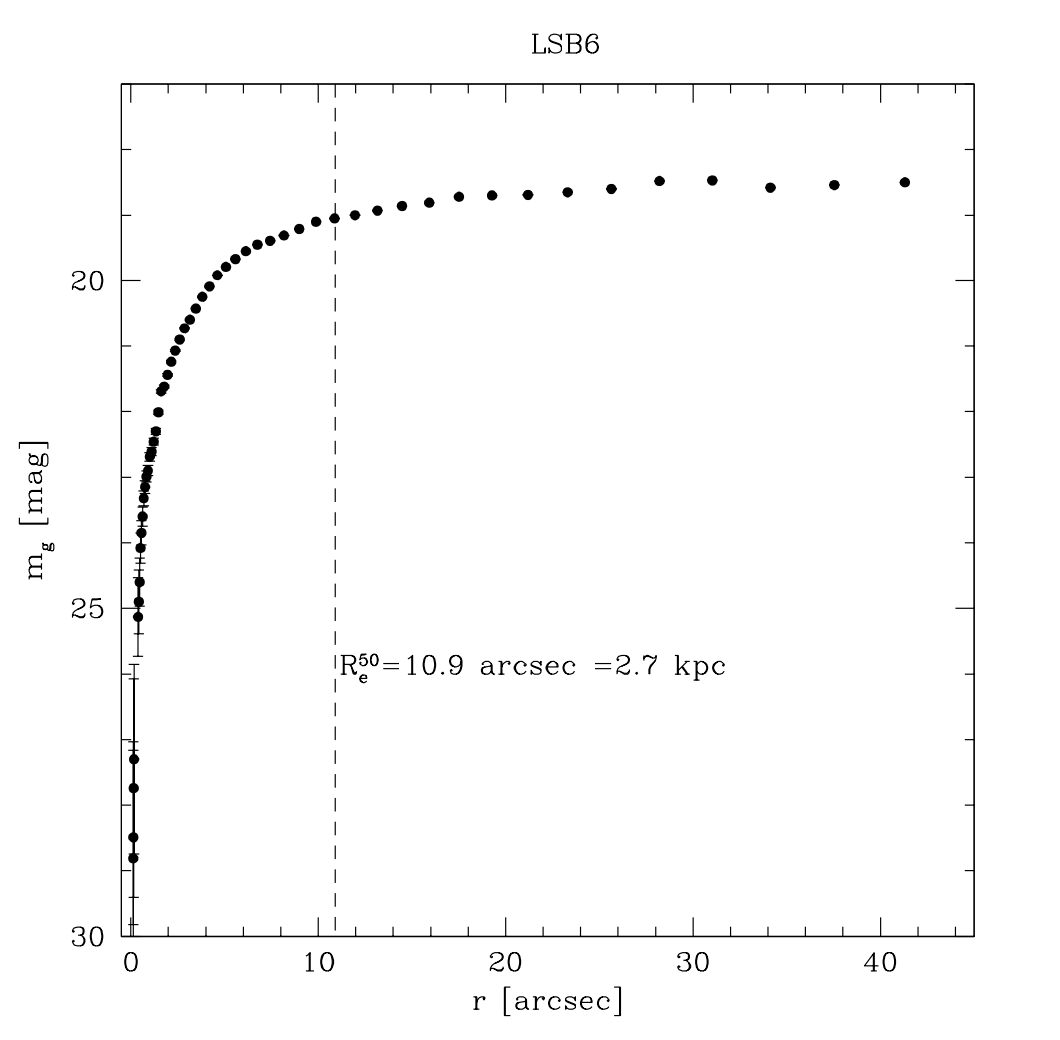} \\
   \end{tabular}
    \caption{Same as Fig. \ref{fig:growUDG_1} for LSB 1, LSB 2, LSB 3, LSB 4, LSB 5, and LSB 6.}
    \label{fig:growLSB_1}
\end{figure*}

\begin{figure*}
\centering
\begin{tabular}{cc}
    \includegraphics[width=7cm]{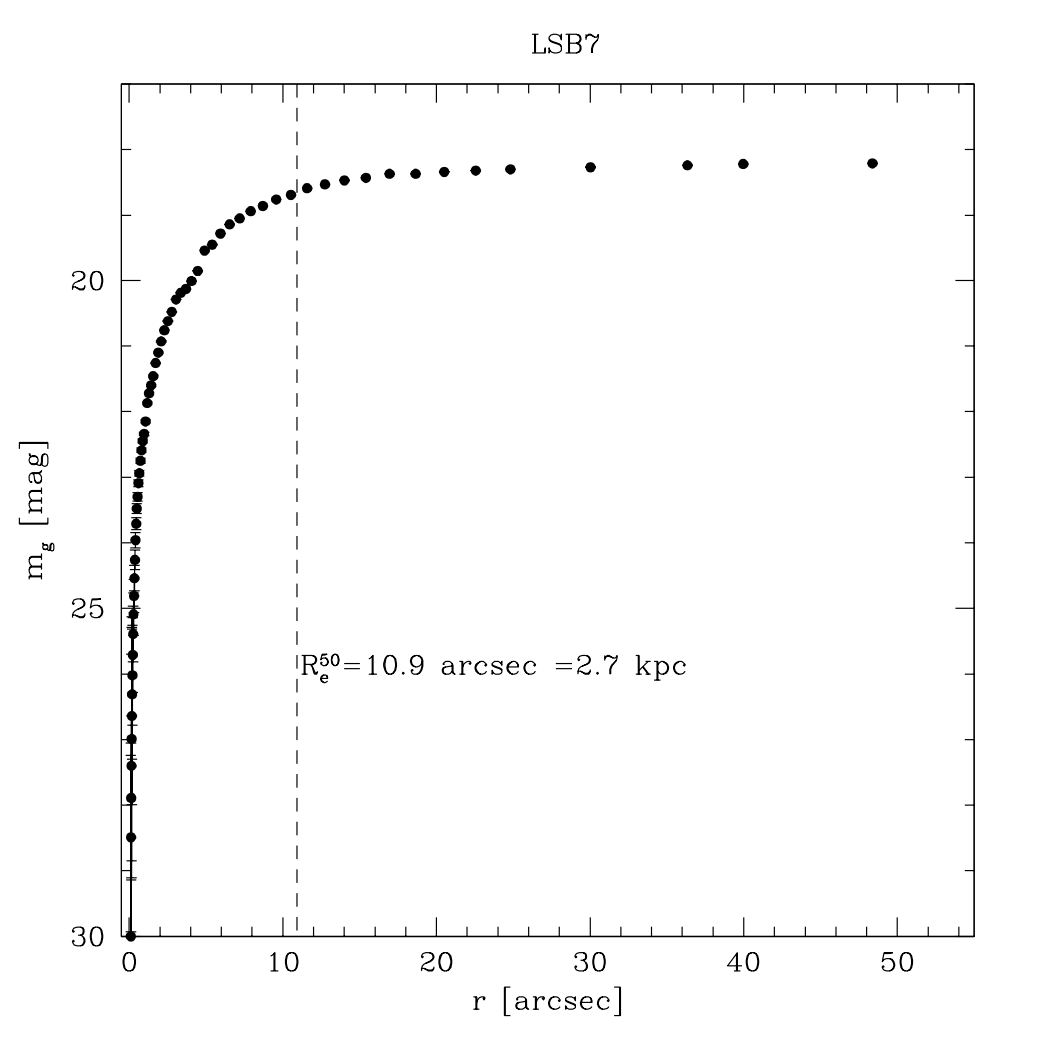} &
    \includegraphics[width=7cm]{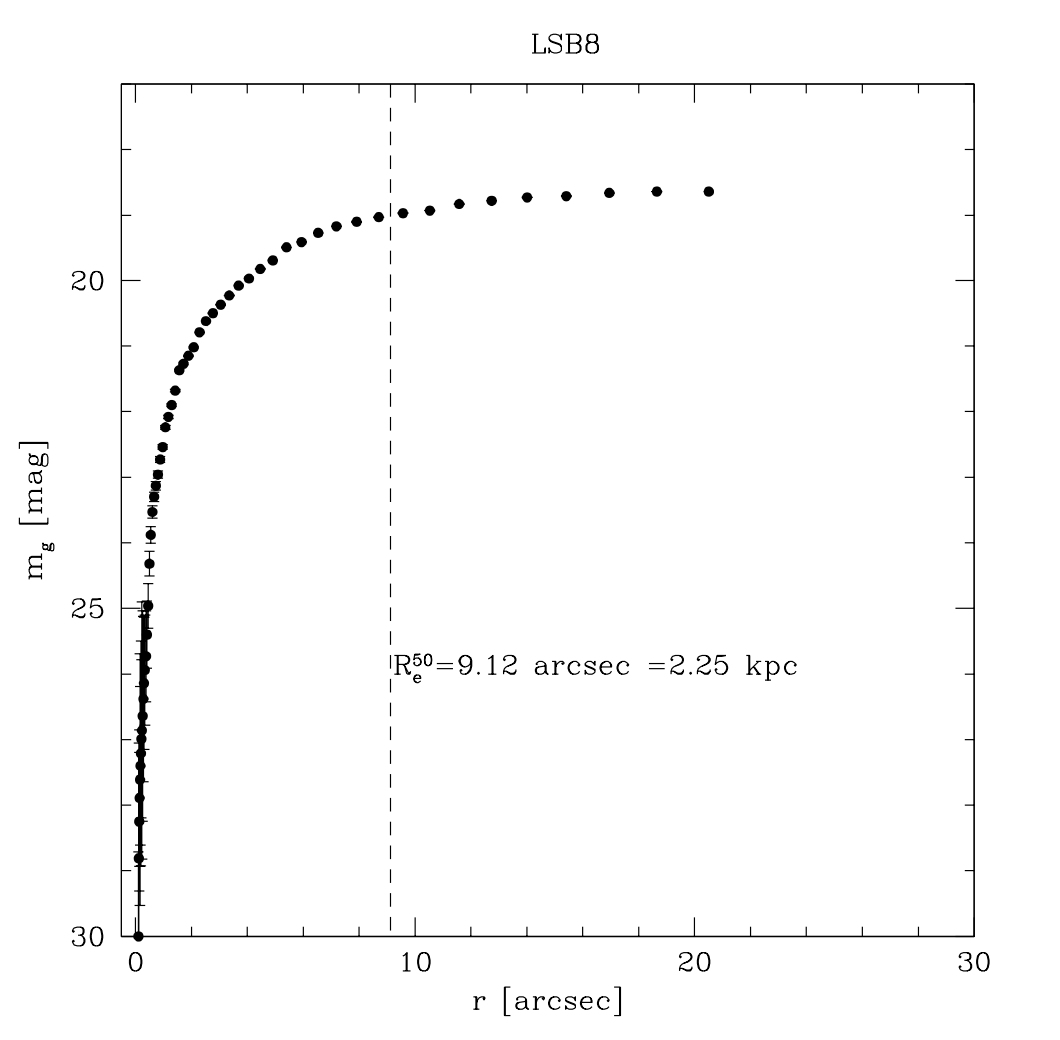} \\
   \end{tabular}
    \caption{Same as Fig. \ref{fig:growUDG_1} for LSB 7 and LSB 8.}
    \label{fig:growLSB_2}
\end{figure*}

Here we present the individual images for the 8 new  LSB galaxies (see Sec.~\ref{sec:detection}), 
side by side with their surface brightness profiles in the $g$-band (Fig.~\ref{fig:LSB_1} to 
Fig.~\ref{fig:LSB_3}).
As done for the UDGs in the sample, in Fig.~\ref{fig:growLSB_1} and Fig.~\ref{fig:growLSB_2} we show the
growth curves for each LSB galaxy, derived from the isophote fitting, where we also indicate 
the effective radius derived as the distance from the galaxy centre that includes half of the total flux.

\end{appendix} 

\newpage

\end{document}